\documentclass[amsmath,amssymb]{svjour3}                     
\smartqed  
\usepackage{graphicx}
\usepackage{bm}

\newcommand{\pp}[1]{\phantom{#1}}
\newcommand{\be}{\begin{eqnarray}}
\newcommand{\ee}{\end{eqnarray}}
\newcommand{\ve}{\varepsilon}
\newcommand{\ii}{\iota}

\journalname{Quantum Information Processing}
\begin{document}

\title{Two-spinors, oscillator algebras, and qubits:\\
Aspects of manifestly covariant approach to relativistic quantum information}


\author{
Marek Czachor 
}


\institute{M. Czachor \at
              Katedra Fizyki Teoretycznej i Informatyki Kwantowej\\
Politechnika Gda\'nska, 80-952 Gda\'nsk, Poland,\\
Krajowe Centrum Informatyki Kwantowej, 81-824 Sopot, Poland\\
and\\
Centrum Leo Apostel (CLEA)\\
Vrije Universiteit Brussel, 1050 Brussels, Belgium \\
              \email{mczachor@pg.gda.pl}           
}

\date{Received: date / Accepted: date}

\maketitle

\begin{abstract}
The first part of the paper reviews applications of 2-spinor methods to relativistic qubits (analogies between tetrads in Minkowski space and 2-qubit states, qubits defined by means of null directions and their role for elimination of the Peres-Scudo-Terno phenomenon, advantages and disadvantages of relativistic polarization operators defined by the Pauli-Lubanski vector, manifestly covariant approach to unitary representations of inhomogeneous SL(2,C)). The second part deals with electromagnetic fields quantized by means of harmonic oscillator Lie algebras (not necessarily taken in irreducible representations). As opposed to non-relativistic singlets one has to distinguish between maximally symmetric and EPR states. The distinction is one of the sources of `strange' relativistic properties of EPR correlations. As an example, EPR averages are explicitly computed for linear polarizations in states that are antisymmetric in both helicities and momenta. The result takes the familiar form $\pm p\cos 2(\alpha-\beta)$ independently of the choice of representation of harmonic oscillator algebra. Parameter $p$ is determined by spectral properties of detectors and the choice of EPR state, but is unrelated to detector efficiencies. Brief analysis of entanglement with vacuum and vacuum violation of Bell's inequality is given. The effects are related to inequivalent notions of vacuum states. Technical appendices discuss details of the representation I employ in field quantization. In particular, M-shaped delta-sequences are used to define Dirac deltas regular at zero.
\end{abstract}

\section{Introduction}

Einstein-Podolsky-Rosen (EPR) paradox \cite{EPR} and Bell's theorem \cite{Bell} --- cornerstones of quantum information and cryptography --- deal with relativistic issues of locality, but methodology of proofs is non-relativistic. In the mid-1980s some authors, including myself, tried to amend the inconsistency and reformulated Bell's theorem in relativistic formalisms (relativistic quantum mechanics \cite{MC84}, algebraic field theory \cite{SW0}). First relativistic results were not entirely trivial (violation of Bell's inequality by vacuum fluctuations, non-invariance of EPR correlations for spins) but did not attract much attention. The turning point for the emerging field of relativistic quantum information was the paper by Peres, Scudo and Terno \cite{PST} on non-invariance of entropy defined in terms of `reduced spin density matrices'. The line of research following from \cite{PST} culminated in the review paper \cite{PT} and is still continued by various authors.

The goals of the present paper are, in a sense, complementary to those of \cite{PT}. I want to concentrate on manifestly covariant 2-spinor approach to storage and communication qubits, technically and philosophically inspired by the works of Penrose \cite{PR,PR2}, but not widely known and not included in \cite{PT}. One of the unusual features of qubits formulated in a 2-spinor way, and going beyond the standard helicity formalism, is the fundamental role of projections of spin on {\it null\/} directions in space-time, even if massive particles are concerned. In particular, it is always possible to project the Pauli-Lubanski vector on principal null directions of SL(2,C) transformations. Decoherence of the Peres-Scudo-Terno type is then eliminated, at least for massive particles \cite{MCMW}.

The second class of results comes from application of null directions to electromagnetic fields. It can be shown that a class of twistor-type spin-frames has covariance properties leading to 4-potentials that do not change gauge after Lorentz transformations (i.e. $A_a(x)$ is a world-vector field). Simultaneously, the same twistor-like transformation implies that momentum-space amplitudes of $A_a(x)$ split into classes belonging to different representations of the Poincar\'e group: Spin-1 zero-mass representation which we regard as photons, and two additional scalar massless fields. Electromagnetic qubits are associated only with the spin-1 part.

The construction can be performed for both first- and second-quantized fields. We concentrate on quantized Maxwell fields, but quantization procedure is not the usual one. I believe the procedure I employ, based on reducible representations of harmonic-oscillator Lie algebras, is in many respects superior to the standard one, where irreducibility is implicitly assumed. Readers interested in more details of the relativistic formalism based on reducible representations are referred to \cite{MCKW}.

In order to analyze relativistic EPR correlations of photons we have to make the notion of an EPR state more precise. The EPR paradox is based, technically speaking, on maximal entanglement in at least two complementary bases --- I term this property the EPR condition.   In non-relativistic quantum mechanics EPR states are simultaneously maximally entangled in all bases, and maximally symmetric.  In the relativistic case we have a similar object,  the two-index antisymmetric spinor $\ve_{AB}$. The problem is that $\ve_{AB}$ is invariant with respect to spinor representations, while these are the unitary representations we have to work with. Scalar states were used in the EPR context by Caban \cite{Caban}, but it turns out that typical scalar states cannot be identified with EPR states, unless the two photons have the same momenta. The EPR condition is not SL(2,C) invariant if the two photons have different momenta, a fact responsible for relativistic non-invariance of EPR correlations for linear polarizations (and certain interferometric qubits).

Another conceptual difficulty with relativistic qubits is the choice of yes-no observables. For massive particles a natural definition is given by the Pauli-Lubanski vector (cf. \cite{MC-PRA}) but the resulting observable is not linear in number of particles. I show how to define an analogue of the Pauli-Lubanski vector which is linear in number of particles, but this is done in representations of harmonic oscillator Lie algebras that lead to a well behaved vacuum part of 4-momentum (this is one of the reasons why the reducible representations are here useful). Still, components of the Pauli-Lubanski vector commute in the massless case, so are useless for EPR correlations.

Since the field is quantized in a nonstandard way, one might ask to what extent the EPR averages for linear polarizations and interferometric observables are sensitive to modifications of field quantization paradigm. The answer is negative: In both reducible and irreducible representations of oscillator algebras the EPR averages possess the well known $\cos 2(\alpha-\beta)$ term, multiplied by $p$ parameterized  by sets of wave vectors analyzed by photon detectors.

The paper is organized as follows. In Sec.~2 I recall basic links between 2-spinors and null vectors, and stress certain similarity between tetrads in Minkowski space and 2-qubit states. This section simultaneously introduces abstract-index notation needed later in the paper. Sec.~3 deals with first-quantized Dirac fields. I analyze properties of the Pauli-Lubanski vector and show why it pays to work at a level more general than spin defined via helicity. Sec.~3.5 discusses an unorthodox approach to unitary representations of the Poincar\'e group, regarded as passive versions of active spinor representations. In Sec.~3.7 I concentrate on the Peres-Scudo-Terno problem and show that a formalism based on principal null directions plays in this context a role of error correction. Sec.~4 introduces basic 2-spinor concepts needed for massless fields. I emphasize the special role played by twistor-like spin-frames and, in Sec.~5, apply them to first-quantized electromagnetic qubits. In Sec.~6 I perform field quantization in terms of reducible representations of harmonic oscillator Lie algebras, and define polarization operators that are linear in numbers of particles. In Sec.~7 I concentrate on relativistic aspects of EPR states. EPR correlations are computed in Sec.~8. The main result shows that the reducible representations I advertise predict essentially the same formula for EPR averages as the usual formalism. This is perhaps the most important new result of the paper.
Sec.~9 discusses the issues of `entanglement with vacuum' and `vacuum violations' of Bell's inequality. I illustrate on simple examples the ideas behind these approaches, and show that they are related to different meanings of the notion of `vacuum'.
I end the paper with technical Appendices related to the choice of representations employed in my approach to field quantization.

\section{Two-spinors: Geometric and algebraic preliminaries}

Let us consider three components $(x,y,z)$ of a unit vector,
\be
x^2+y^2+z^2 &=& 1,\label{S}
\ee
taken in some basis in $\bm R^3$. (\ref{S}) describes a unit sphere whose points can be equivalently parametrized by means of stereographic projection on the plane $z=0$. A particularly useful form of this map
is obtained if the plane is parametrized by complex coordinates $\zeta$, linked to the point on the sphere by $\zeta =(x+iy)/(1-z)$. Let us note that the `north pole' $(0,0,1)$ corresponds to $\zeta=\infty$, whereas the `south pole' $(0,0,-1)$ implies $\zeta=0$. Yet another possible parametrization of the sphere is in terms of projective coordinates $(\xi,\eta)$ satisfying $\zeta=\xi/\eta$. Now the poles correspond to $(\xi,0)$ and $(0,\eta)$. The coordinates are projective since $(\xi,\eta)$ and $(\lambda\xi,\lambda\eta)$, for any $\lambda\neq 0$, define the same point on the sphere.

Let us now treat our sphere as a light cone in Minkowski space,
\be
1-x^2-y^2-z^2=0.
\ee
The four components $(1,x,y,z)$ of a null future-pointing world-vector are linked to projective coordinates by
\be
(1,x,y,z) &=& \Big(1,\frac{\xi\bar\eta+\eta\bar\xi}{\xi\bar\xi+\eta\bar\eta},-i\frac{\xi\bar\eta-\eta\bar\xi}{\xi\bar\xi+\eta\bar\eta},
\frac{\xi\bar\xi-\eta\bar\eta}{\xi\bar\xi+\eta\bar\eta}\Big).
\ee
Multiplying both sides by $(\xi\bar\xi+\eta\bar\eta)/\sqrt{2}$ we obtain coordinates
\be
(T,X,Y,Z) &=&
\frac{1}{\sqrt{2}}\Big(\xi\bar\xi+\eta\bar\eta,\xi\bar\eta+\eta\bar\xi,-i(\xi\bar\eta-\eta\bar\xi),\xi\bar\xi-\eta\bar\eta\Big)\label{TXYZ}
\ee
of a point belonging to the same null line in Minkowski space as the point corresponding to $(1,x,y,z)$. (\ref{TXYZ}) is future-pointing since $T> 0$ for a non-vanishing $(\xi,\eta)$. There is, of course, nothing fundamentally special about $1/\sqrt{2}$, but this choice is convenient, as we shall see later. Given a null future-pointing world-vector with components $(T,X,Y,Z)$, and solving the four equations (\ref{TXYZ}), one can determine  $\xi$ and $\eta$ up to a common phase factor.
$(T,X,Y,Z)$ is termed the flagpole of $(\xi,\eta)$. A flag plane can be defined by a vector tangent at $(x,y,x)$ to the sphere (\ref{S}). The flagpole and the flag plane determine $(\xi,\eta)$ up to a sign. This sign ambiguity is related to topological properties of the rotation roup.

I have purposefully stressed that $(T,X,Y,Z)$ are {\it components\/} of a Minkowski-space world-vector evaluated with respect to {\it some\/} basis. The pair $(\xi,\eta)$ can be associated with any basis in a 2-dimensional complex space, but there is no obvious link between the Minkowski space and this 2-dimensional linear space.
Putting it differently, if $\bm t$, $\bm x$, $\bm y$, $\bm z$, define a Minkowski tetrad the vector $X \bm x+Y \bm y+Z \bm z+T \bm t$ can be linked with a vector (a 2-spinor) $\xi \bm o+ \eta \bm \ii$, where $\bm o$ and $\bm \ii$ are basis vectors in $\bm C^2$, but there is no {\it a priori\/} link between $\bm t$, $\bm x$, $\bm y$, $\bm z$ and $\bm o$, $\bm \ii$. However, let us consider the tensor product of this spinor with its complex conjugate,
\be
(\xi \bm o+ \eta \bm \ii)\otimes(\bar\xi \bar{\bm o}+ \bar\eta \bar{\bm \ii})
&=&
\xi\bar\xi  \bm o\otimes\bar{\bm o}
+
\xi\bar\eta \bm o \otimes\bar{\bm \ii}
+
\eta\bar\xi  \bm \ii\otimes\bar{\bm o}
+
\eta \bar\eta \bm \ii\otimes\bar{\bm \ii}
\\
&=&
\underbrace{\frac{1}{\sqrt{2}}\Big(\xi\bar\eta+\eta\bar\xi\Big)}_X
\tilde{\bm x}
\underbrace{-
i\frac{1}{\sqrt{2}}\Big(\xi\bar\eta-\eta\bar\xi\Big)}_Y
\tilde{\bm y}
+
\underbrace{\frac{1}{\sqrt{2}}\Big(\xi\bar\xi-\eta\bar\eta\Big)}_Z
\tilde{\bm z}
\nonumber\\
&\pp=&
+
\underbrace{\frac{1}{\sqrt{2}}\Big(\xi\bar\xi+\eta\bar\eta\Big)}_T
\tilde{\bm t}
,
\ee
where
\be
\tilde{\bm t}
&=&
\frac{1}{\sqrt{2}}\Big(\bm o\otimes\bar{\bm o}+\bm \ii\otimes\bar{\bm \ii}\Big),\label{tilde t}\\
\tilde{\bm x}
&=&
\frac{1}{\sqrt{2}}\Big(\bm o\otimes\bar{\bm \ii}+\bm \ii\otimes\bar{\bm o}\Big),\label{tilde x}\\
\tilde{\bm y}
&=&
\frac{i}{\sqrt{2}}\Big(\bm o\otimes\bar{\bm \ii}-\bm \ii\otimes\bar{\bm o}\Big),\label{tilde y}\\
\tilde{\bm z}
&=&
\frac{1}{\sqrt{2}}\Big(\bm o\otimes\bar{\bm o}-\bm \ii\otimes\bar{\bm \ii}\Big),\label{tilde z}
\ee
and
\be
\bm o\otimes\bar{\bm o}
&=&
\frac{1}{\sqrt{2}}\Big(\tilde{\bm t}+\tilde{\bm z}\Big),\label{oo}\\
\bm o \otimes\bar{\bm \ii}
&=&
\frac{1}{\sqrt{2}}\Big(\tilde{\bm x}-i\tilde{\bm y}\Big)\label{oi},\\
\bm \ii\otimes\bar{\bm o}
&=&
\frac{1}{\sqrt{2}}\Big(\tilde{\bm x}+i\tilde{\bm y}\Big)\label{io},\\
\bm \ii\otimes\bar{\bm \ii}
&=&
\frac{1}{\sqrt{2}}\Big(\tilde{\bm t}-\tilde{\bm z}\Big).\label{ii}
\ee
A change of basis $\{\bm o,\bm\ii\}\to \{\bm o',\bm\ii'\}$ implies an associated change of components $(\xi,\eta)\to (\xi',\eta')$. If $(\xi,\eta)\to (\xi',\eta')$ is a linear transformation with unit determinant (an element of the group SL(2,C)) then the associated transformation
$(X,Y,Z,T)\to (X',Y',Z',T')$ turns out to be a proper isochronous Lorentz transformation. Since, $X\tilde{\bm x}+Y\tilde{\bm y}+Z\tilde{\bm z}+T\tilde{\bm t}=X'\tilde{\bm x}{'}+Y'\tilde{\bm y}{'}+Z'\tilde{\bm z}{'}
+T'\tilde{\bm t}{'}$, the map $\{\tilde{\bm x},\tilde{\bm y},\tilde{\bm z},\tilde{\bm t}\}\to
\{\tilde{\bm x}{'},\tilde{\bm y}{'},\tilde{\bm z}{'},\tilde{\bm t}{'}\}$ is a proper isochronous Lorentz transformation as well. Tensor products of 2-spinors with their complex conjugates transform under SL(2,C) transformations in the same way as world-vectors in Minkowski space under boosts and rotations. This leads us to the abstract-index formalism.

\subsection{Abstract-index formalism of Penrose}

Second-rank spinors $\{\tilde{\bm x},\tilde{\bm y},\tilde{\bm z},\tilde{\bm t}\}$ have transformation properties of a Minkowski tetrad.
Spinor approach to space-time geometry is based on the assumption that one can {\it identify\/} the two bases, $\{\tilde{\bm x},\tilde{\bm y},\tilde{\bm z},\tilde{\bm t}\}$ in $\bm C^2\otimes \bm C^2$ and
$\{{\bm x},{\bm y},{\bm z},{\bm t}\}$ in Minkowski space, and simply skip the tildas in (\ref{tilde t})--(\ref{tilde z}).
From this perspective 2-spinors are more fundamental than world-vectors in Minkowski space. The Minkowski space can be regarded as a structure derived from a more fundamental spinorial level.

This is the starting point of the abstract-index formalism of Penrose \cite{PR}, where (\ref{tilde t})--(\ref{tilde z}) would be written as
\be
t^a&=&
\frac{1}{\sqrt{2}}\Big(o^A \bar o^{A'}+\ii^A\bar \ii^{A'}\Big)=t^{AA'},\\
x^a
&=&
\frac{1}{\sqrt{2}}\Big(o^A \bar \ii^{A'}+\ii^A \bar o^{A'}\Big)=x^{AA'},\\
y^a
&=&
\frac{i}{\sqrt{2}}\Big(o^A \bar \ii^{A'}-\ii^A \bar o^{A'}\Big)=y^{AA'},\\
z^a&=&
\frac{1}{\sqrt{2}}\Big(o^A \bar o^{A'}-\ii^A\bar \ii^{A'}\Big)=z^{AA'}.
\ee
The indices are just labels (analogous to `Alice' and `Bob'; think of $t^{\rm alice}=t^{\rm Alice,Alice'}$) and do not take numerical values.
The null tetrad corresponding to (\ref{oo})--(\ref{ii}) is expressed in the abstract-index form as
\be
l^a &=& o^A \bar o^{A'},\\
m^a &=& o^A \bar \ii^{A'},\\
\bar m^a &=& \ii^A \bar o^{A'},\\
n^a &=& \ii^A \bar \ii^{A'}.
\ee
In this approach there are no operations that would involve (anti-)symmetrizations of primed and unprimed indices. In consequence, we do not loose generality by assuming that the primed labels occur to the right of the unprimed ones. In particular, we identify $\alpha_A\beta_{A'}$ with $\beta_{A'}\alpha_A$ for all 2-spinors $\alpha_A$ and $\beta_{A'}$, and one does not need to distinguish between hermiticity and reality. As an illustration, let us consider reality properties of the null tetrad:
\be
\overline{l^a} &=& \overline{o^A} \overline{\bar o^{A'}}=\bar o^{A'}o^A=o^A\bar o^{A'}=l^a ,\\
\overline{m^a} &=& \overline{o^A} \overline{\bar \ii^{A'}}=\bar o^{A'}\ii^A=\ii^A\bar o^{A'}=\bar m^a,
\ee
and so on. This formalism is very convenient and does not lead to ambiguities in practical computations but may seem somewhat counterintuitive, especially to those who are accustomed to the more traditional spinor notation employed in relativistic quantum mechanics. This is why in the next subsection we shall describe a variant of the Penrose formalism that can be directly translated into formulas we know from quantum mechanics textbooks.

But before we do that let us discuss relations between abstract and numerical indices.  Numerical indices, 0 and 1, are sometimes needed but then we denote them by upright boldface fonts. Accordingly, the symbol $\phi^A$ denotes a 2-spinor (`a spinor of Alice'), but $\phi^{\bf A}$ may equal $\phi^{0}$ or $\phi^{1}$. $\phi^A$ is basis independent, but $\phi^{\bf A}$ implicitly depends on a basis.
Now consider two 2-spinors, $\alpha^A=\alpha^0 o^A+\alpha^1 \ii^A=\tilde\alpha^0 \tilde o^A+\tilde\alpha^1 \tilde\ii^A$, $\beta^A=\beta^0 o^A+\beta^1 \ii^A=\tilde\beta^0 \tilde o^A+\tilde\beta^1 \tilde\ii^A$, where the two bases are related by $\tilde o^A=(So)^A$,  $\tilde\ii^A=(S\ii)^A$, $\det S=1$. The determinant
\be
\left|
\begin{array}{cc}
\alpha^0 & \alpha^1\\
\beta^0 & \beta^1\\
\end{array}
\right|
=
\left|
\begin{array}{cc}
\tilde\alpha^0 & \tilde\alpha^1\\
\tilde\beta^0 & \tilde\beta^1\\
\end{array}
\right|=\alpha^0\beta^1-\alpha^1\beta^0=\ve_{\bf AB}\alpha^{\bf A}\beta^{\bf B}=\alpha_{\bf B}\beta^{\bf B}=-
\alpha^{\bf A}\beta_{\bf A}\label{contr}
\ee
is independent of $S$ and defines
\be
\ve_{\bf AB}
=
\left(
\begin{array}{cc}
0 & 1\\
-1 & 0\\
\end{array}
\right).
\ee
The last two equalities in (\ref{contr}) show how to lower spinor indices in SL(2,C)-invariant manner: $\alpha_{\bf B}=\alpha^{\bf A}\ve_{\bf AB}$. The inverse rule is $\ve^{\bf AB}\alpha_{\bf B}=\alpha^{\bf A}$, with $\ve^{\bf AB}=\ve_{\bf AB}$.

The lower-index $\alpha_{\bf B}$ may be regarded as a component of a spinor $\alpha_{A}$ dual to $\alpha^{A}$, the duality being given by
$\alpha_{A}\beta^{A}=\alpha_{\bf A}\beta^{\bf A}$. The formulas $\alpha_{B}=\alpha^{A}\ve_{AB}$ and $\ve^{AB}\alpha_{B}=\alpha^{A}$ define at the abstract-index level the isomorphisms between modules of upper-index 2-spinor fields and their duals. Note that $\ve^{CB}\ve_{CA}=\ve{_A}{^B}$ acts by
$\ve{_A}{^B}\phi_B=\phi_A$, and
\be
\ve{_{\bf A}}{^{\bf B}}
=
\left(
\begin{array}{cc}
1 & 0\\
0 & 1\\
\end{array}
\right).
\ee
$\ve{_A}{^B}$ thus plays a role of spinorial Kronecker delta; at the abstract-index level $\ve{_A}{^B}$ is the isomorphism between `the spinor of Bob' and `the spinor of Alice', a map quite similar to the teleportation protocol \cite{MC-CQG}.

The basis $o^A$ and $\ii^A$, satisfying $o_A\ii^A=1$, is termed the spin-frame. Since $\phi_A\phi^A=0$ for any $\phi_A$, spin-frames possess a kind of gauge freedom: $(o_A+\lambda \ii_A)\ii^A=o_A(\ii^A+\mu o^A)=o_A\ii^A=1$. It is interesting that in electrodynamics this ambiguity indeed manifests itself in a form of gauge transformation associated with Lorentz transformations of four-potentials. Any spin-frame satisfies
\be
\ve_{AB} &=& o_A\ii_B-\ii_A o_B,\label{ve_A_B}\\
\ve{_A}{^B} &=& o_A\ii^B-\ii_A o^B.\label{ve_A^B}
\ee
If $g_{ab}$ is the Minkowski-space metric tensor then  $g{_a}{^b}$ must be the abstract-index form of Kronecker's delta in Minkowski space. Accordingly, $g{_a}{^b}=\ve{_A}{^B}\ve{_{A'}}{^{B'}}$ ($\overline{\ve{_A}{^B}}=\ve{_{A'}}{^{B'}}$). Lowering the indices we obtain the following three useful abstract-index forms of the metric:
\be
g_{ab} &=& \ve_{AB}\ve_{A'B'}\label{g1}\\
&=&
t_at_b-x_ax_b-y_ay_b-z_az_b\label{g2}\\
&=&
n_al_b+l_an_b-\bar m_am_b-m_a\bar m_b.\label{g3}
\ee
The reader may have noticed that in the Penrose formalism the Minkowski tetrad has all the properties of the Bell basis of two-qubit entangled states. The null basis $l=\bm o\otimes\bar{\bm o}=l^a$, $m=\bm o\otimes\bar{\bm \ii}=m^a$, $\bar m=\bm \ii\otimes\bar{\bm o}=\bar m^a$, $n=\bm \ii\otimes\bar{\bm \ii}=n^a$ is analogous to the basis of product states. There exists an unexplored `duality' between space-time geometry and quantum information theory. Some preliminary considerations in this spirit can be found in \cite{MC-CQG}, where analogies between teleportation protocols and metric tensors of Lorentzian manifolds were discussed. Lorentzian techniques, with applications to multi-qubit entanglement, can be found also in \cite{Levay,Frumosu,Jaeger,Jaeger2}.

\subsection{Abstract-index analogues of Pauli matrices: Infeld-van der Waerden tensors}

Let us return to the formula (\ref{TXYZ}), but written as
\be
X
&=&(X^0,X^1,X^2,X^3) \nonumber\\
&=&
\frac{1}{\sqrt{2}}\Big(\phi^0\bar\phi^{0'}+\phi^1\bar\phi^{1'},\phi^0\bar\phi^{1'}+\phi^1\bar\phi^{0'},-i(\phi^0\bar\phi^{1'}-\phi^1\bar\phi^{0'}),
\phi^0\bar\phi^{0'}-\phi^1\bar\phi^{1'}\Big)\label{TXYZ1}\\
&=&
(g{^0}{_{\bf A}}{_{{\bf A}'}},g{^1}{_{\bf A}}{_{{\bf A}'}},g{^2}{_{\bf A}}{_{{\bf A}'}},g{^3}{_{\bf A}}{_{{\bf A}'}})\phi^{\bf A}\bar\phi^{{\bf A}'},\\
X^{\bf a}
&=&
g{^{\bf a}}{_{\bf A}}{_{{\bf A}'}}\phi^{\bf A}\bar\phi^{{\bf A}'}.
\ee
The coefficients $g{^{\bf a}}{_{\bf A}}{_{{\bf A}'}}$ are known as Infeld-van der Waerden symbols, and can be grouped into four matrices
\be
g{^0}{_{\bf A}}{_{{\bf A}'}}
&=&
\left(
\begin{array}{cc}
g{^0}{_{0}}{_{0'}} & g{^0}{_{0}}{_{1'}}\\
g{^0}{_{1}}{_{0'}} & g{^0}{_{1}}{_{1'}}
\end{array}
\right)
=
\frac{1}{\sqrt{2}}
\left(
\begin{array}{cc}
1 & 0\\
0 & 1
\end{array}
\right)=g{_0}{^{\bf A}}{^{{\bf A}'}},\\
g{^1}{_{\bf A}}{_{{\bf A}'}}
&=&
\left(
\begin{array}{cc}
g{^1}{_{0}}{_{0'}} & g{^1}{_{0}}{_{1'}}\\
g{^1}{_{1}}{_{0'}} & g{^1}{_{1}}{_{1'}}
\end{array}
\right)
=
\frac{1}{\sqrt{2}}
\left(
\begin{array}{cc}
0 & 1\\
1 & 0
\end{array}
\right)=g{_1}{^{\bf A}}{^{{\bf A}'}},\\
g{^2}{_{\bf A}}{_{{\bf A}'}}
&=&
\left(
\begin{array}{cc}
g{^2}{_{0}}{_{0'}} & g{^2}{_{0}}{_{1'}}\\
g{^2}{_{1}}{_{0'}} & g{^2}{_{1}}{_{1'}}
\end{array}
\right)
=
\frac{1}{\sqrt{2}}
\left(
\begin{array}{cc}
0 & -i\\
i & 0
\end{array}
\right)=-g{_2}{^{\bf A}}{^{{\bf A}'}},\\
g{^3}{_{\bf A}}{_{{\bf A}'}}
&=&
\left(
\begin{array}{cc}
g{^3}{_{0}}{_{0'}} & g{^3}{_{0}}{_{1'}}\\
g{^3}{_{1}}{_{0'}} & g{^3}{_{1}}{_{1'}}
\end{array}
\right)
=
\frac{1}{\sqrt{2}}
\left(
\begin{array}{cc}
1 & 0\\
0 & -1
\end{array}
\right)=g{_3}{^{\bf A}}{^{{\bf A}'}}.
\ee
The numerical indices are lowered or raised by means of the epsilons and $g^{\bf ab}=g_{\bf ab}={\rm diag}(1,-1,-1,-1)$.
In quantum mechanics textbooks one works with matrix 4-vectors $\sigma=(1,\sigma_x,\sigma_y,\sigma_z)$ and  $\tilde\sigma=(1,-\sigma_x,-\sigma_y,-\sigma_z)=\ve\bar\sigma\ve^T$. Infeld-van der Waerden symbols, with appropriately raised or lowered indices,  play precisely the same role, but make the formalism more flexible and prepared for more advanced applications.

An important relation between the four types of coefficients,
\be
g_{\bf ab}
&=&
g{_{\bf a}}{^{\bf A}}{^{{\bf A}'}}g{_{\bf b}}{^{\bf B}}{^{{\bf B}'}}\ve_{\bf AB}\ve_{{\bf A}'{\bf B}'},
\ee
can be rewritten in several {\it equivalent\/} ways, each of them revealing another aspect of their mutual relations:
\be
g_{\bf ab}g{^{\bf a}}{_{\bf A}}{_{{\bf A}'}}g{^{\bf b}}{_{\bf B}}{_{{\bf B}'}}
&=&
\ve_{\bf AB}\ve_{{\bf A}'{\bf B}'},\\
g{^{\bf a}}{_{\bf A}}{_{{\bf A}'}}g{_{\bf a}}{^{\bf B}}{^{{\bf B}'}}
&=&
\ve{_{\bf A}}{^{\bf B}}\ve{_{{\bf A}'}}{^{{\bf B}'}},
\\
g{_{\bf a}}{^{\bf A}}{^{{\bf A}'}}g{^{\bf b}}{_{\bf A}}{_{{\bf A}'}}
&=&
g{_{\bf a}}{^{\bf b}},\\
g{_{\bf a}}{_{\bf A}}{_{{\bf A}'}}g{_{\bf b}}{^{\bf B}}{^{{\bf A}'}}
+
g{_{\bf b}}{_{\bf A}}{_{{\bf A}'}}g{_{\bf a}}{^{\bf B}}{^{{\bf A}'}}
&=&
\ve{_{\bf A}}{^{\bf B}}g_{\bf ab}.\label{Cl}
\ee
Eq.~(\ref{Cl}) is the 2-spinor form of the anticommutator known from the algebra of Dirac matrices.
The above formulas involve boldface (numerical) indices but can be regarded as components of appropriate abstract-index relations, say,
\be
g_{ab}
&=&
g{_{a}}{^{A}}{^{{A}'}}g{_{b}}{^{B}}{^{{B}'}}\ve_{AB}\ve_{{A}'{B}'},\label{43}
\ee
and the like. Infeld-van der Waerden {\it tensors\/} $g{_{a}}{^{A}}{^{{A}'}}$, of a mixed world-vector and 2-spinor type, play the roles of the isomorphisms allowing to identify world-vector indices with pairs of the spinor ones: $X_a=g{_{a}}{^{A}}{^{{A}'}}X_{AA'}$, and so forth with tensors of any rank.

Let us note that, in general, $g{_{a}}{^{A}}{^{{A}'}}$ may possess a nontrivial dependence on the point $x$ of an appropriate Lorentzian manifold. The identification $X_a=X_{AA'}$ implicitly assumes that in all differential equations one encounters in a given theory one can freely commute
$g{_{a}}{^{A}}{^{{A}'}}$ with derivatives. The latter means that one works with covariant derivatives satisfying $\nabla_a g{_{b}}{^{C}}{^{{C}'}}=0$ (plus analogous equations obtained by raising and lowering the indices). In most applications to Minkowski space we do not have to worry about such subtleties.

In the Penrose formalism no special role is given to Infeld-van der Waerden tensors --- they are just regarded as one of the possible instances of the abstract-index rule $g{_a}{^b}=g{_a}{^{BB'}}$. However, it is sometimes useful to work with Infeld-van der Waerden tensors occurring explicitly, if we need to translate some formulas into their more standard shapes. As an exercise, let us show several equivalent abstract-index versions of (\ref{43}):
\be
g_{ab}
&=&
g{_{a}}{^{C}}{^{{C}'}}g{_{b}}{^{D}}{^{{D}'}}\ve_{CD}\ve_{{C}'{D}'}
=
g{_{a}}{^{c}}g{_{b}}{^{d}}g_{cd}
=
g{_{a}}{^{c}}g{_{b}}{^{d}}g_{CC'DD'}
=
g{_{a}}{^{CC'}}g{_{b}}{^{DD'}}g_{cDD'}\nonumber\\
&=&
g{_{AA'}}{^{c}}g{_{BB'}}{^{d}}g_{cd}
=
g{_{AA'}}{^{c}}g{_{BB'}}{^{d}}g_{CC'd}
=
\ve{_{A}}{^{C}}\ve{_{A'}}{^{C'}}
\ve{_{B}}{^{D}}\ve{_{B'}}{^{D'}}
\ve{_{C}}{_{D}}\ve{_{C'}}{_{D'}}
\nonumber\\
&=&
g{_{AB'}}{^{CD'}}
g{_{BA'}}{^{DC'}}
\ve{_{C}}{_{D}}\ve{_{C'}}{_{D'}}=g{_{AB'}}{^{CD'}}
g{_{BA'}}{^{DC'}}
g_{cd}=\ve{_{A}}{_{B}}\ve{_{A'}}{_{B'}}.
\nonumber
\ee
Flexibility of switching between forms is indeed immense. We shall use this freedom in derivation of generators of certain important representations of SL(2,C).

\subsection{Tetrad and diad notation}

Another notational element we will need is based on the following observation. Let us note that, by definition of the relation between abstract and numerical indices, any world-vector satisfies
\be
X_a &=& g{_a}{^b}X_b=g{_a}{^{\bf b}}X_{\bf b}=g{_a}{^{0}}X_{0}+g{_a}{^{1}}X_{1}+g{_a}{^{2}}X_{2}+g{_a}{^{3}}X_{3},
\ee
so that the four {\it world-vectors\/} $g{_a}{^{\bf b}}$ play a role of a basis. Actually, this is a Minkowski tetrad, since
\be
g{_a}{^{\bf b}}g{^a}{^{\bf c}}
&=&
g{_{\bf a}}{^{\bf b}}g{^{\bf a}}{^{\bf c}}=g{^{\bf b}}{^{\bf c}}.
\ee
Analogously,
\be
g{^{a}}{^{b}}
&=&
g{_c}{^{a}}g{^c}{^{b}}
=
g{_{\bf c}}{^{a}}g{^{\bf c}}{^{b}}
=
g^{\bf cd}g{_{\bf c}}{^{a}}g{_{\bf d}}{^{b}}
\\
&=&
g{_{0}}{^{a}}g{_{0}}{^{b}}
-
g{_{1}}{^{a}}g{_{1}}{^{b}}
-
g{_{2}}{^{a}}g{_{2}}{^{b}}
-
g{_{3}}{^{a}}g{_{3}}{^{b}}
\ee
(compare Eq.~(\ref{g2})). A similar construction can be performed
with 2-spinors, starting with
\be
\phi_A
&=&
\ve{_A}{^B}\phi_B=\ve{_A}{^{\bf B}}\phi_{\bf B},
\ee
showing that $\ve{_A}{^{\bf B}}$ is a diad of spinorial basis vectors. In particular, since $\ve^{01}=1$ we find $\ve{_A}{^{0}}\ve{^A}{^{1}}=
\ve{_{\bf A}}{^{0}}\ve{^{\bf A}}{^{1}}=1$, so that $o_A=\ve{_A}{^{0}}$, $\ii_A=\ve{_A}{^{1}}$ form a spin-frame.

\subsection{Simplest spinor representations of SL(2,C) and their generators}

SL(2,C) transformations act by $\phi_A\to \tilde\phi_A=\Lambda{_A}{^B}\phi_B$, $\phi^A\to \tilde\phi^A=\phi^B\Lambda^{-1}{_B}{^A}$,
$\phi_{A'}\to \tilde\phi_{A'}=\Lambda{_{A'}}{^{B'}}\phi_{B'}$, $\phi^{A'}\to \tilde\phi^{A'}=\phi^{B'}\Lambda^{-1}{_{B'}}{^{A'}}$. Here $\Lambda{_{A}}{^{B}}$ and $\Lambda{_{A'}}{^{B'}}=\overline{\Lambda{_{A}}{^{B}}}$ are the two inequivalent 2-dimensional representations of $\Lambda\in$~SL(2,C). Let us note that transformation properties of lower- and upper-index spinors do not have to be separately postulated. Indeed, for SL(2,C) transformations we have
$\Lambda{_A}{^C}\Lambda{_B}{^D}\ve{_C}{_D}=\ve{_A}{_B}$. Raising appropriate indices and employing antisymmetry of the epsilon, we can transform this formula as follows
$-\Lambda{_A}{^C}\Lambda{^B}{_C}=\ve{_A}{^B}$, which shows that $-\Lambda{^B}{_C}=\Lambda^{-1}{_C}{^B}$.

Representations $(\frac{1}{2},\frac{1}{2})$ (in Minkowski space), $(\frac{1}{2},0)$ (unprimed spinors), and $(0,\frac{1}{2})$ (primed spinors) of an element $\Lambda\in$~SL(2,C) are linked to the Lie-algebra of generators by:
$\Lambda{_r}{^s}=\exp\frac{-i}{2}{}y^{ab}\sigma_{ab}{_r}{^s}$, $\Lambda{_R}{^S}=\exp\frac{-i}{2}{}y^{ab}\sigma_{ab}{_R}{^S}$, and
$\Lambda{_{R'}}{^{S'}}=\exp\frac{-i}{2}{}y^{ab}\sigma_{ab}{_{R'}}{^{S'}}$. ${}y^{ab}$ is the antisymmetric tensor whose six independent components correspond to boosts ($y^{0\bf n}$) and rotations ($y^{\bf mn}$), ${\bf n,m}=1,2,3$. Accordingly, the generators can be obtained from
\be
\sigma_{ab}{_{\dots}}{^{\dots}}
&=&i\frac{\partial \Lambda{_{\dots}}{^{\dots}}}{\partial {}y^{ab}}\Big|_{{}y^{ab}=0}.
\ee
Differentiating at ${}y^{ab}=0$ both sides of $\Lambda{_A}{^C}\Lambda{_B}{^D}\ve{_C}{_D}=\ve{_A}{_B}$ we obtain
$\sigma_{ab}{_A}{_B}=\sigma_{ab}{_B}{_A}$. Similarly, differentiating $\Lambda{_{r}}{^{s}}=\Lambda{_{R}}{^{S}}\Lambda{_{R'}}{^{S'}}$, we find
\be
\sigma{_{ab}}{_r}{^s}
&=&
\sigma_{ab}{_R}{^S}\ve{_{R'}}{^{S'}}
+
\ve{_R}{^S}\sigma_{ab}{_{R'}}{^{S'}}
\label{gen rs}.
\ee
Components $\sigma{_{\bf ab}}{_{\bf r}}{^{\bf s}}$ of the Minkowski-space generators  can be collected into six matrices, the ones corresponding to boosts being symmetric, as opposed to the antisymmetric generators of rotations. Their manifestly covariant abstract-index form,
\be
\sigma{_{ab}}{_r}{^s}
&=&
i\,(g_{ar}g_b{^s} - g_{br}g_a{^s})
\ee
can be rewritten in all the possible equivalent ways by means of the tricks I have described above. In particular, we find
\be
\sigma{_{ab}}{_r}{^s}
&=&
i\,(g_{aRR'}g_b{^{SS'}} - g_{bRR'}g_a{^{SS'}}).
\label{gen rs'}
\ee
Since any symmetric spinor $\phi_{AB}=\phi_{BA}$ satisfies $\phi{_{A}}{^{A}}=\ve^{AB}\phi_{AB}=\ve^{BA}\phi_{BA}=-\ve^{AB}\phi_{AB}=0$,
comparing (\ref{gen rs}) with (\ref{gen rs'}) we get
\be
\sigma_{ab}{_R}{^S}
&=&
\frac{i}{2}(g_{aRX'}g_b{^{SX'}} - g_{bRX'}g_a{^{SX'}})
,
\\
\sigma_{ab}{_{R'}}{^{S'}}
&=&
\frac{i}{2}(g_{aXR'}g_b{^{XS'}} - g_{bXR'}g_a{^{XS'}})
.
\ee
Alternatively, after some simplifications,
\be
\sigma_{ab}{_R}{^S}
&=&
\frac{i}{2}\ve_{A'B'}(\ve_{AR}\ve{_{B}}{^{S}} + \ve{_{B}}{_{R}}\ve{_{A}}{^{S}})
\label{sigma RS},\\
\sigma_{ab}{_{R'}}{^{S'}}
&=&
\frac{i}{2}\ve_{AB}(\ve_{A'R'}\ve{_{B'}}{^{S'}} + \ve_{B'R'}\ve{_{A'}}{^{S'}})
\label{sigma R'S'}.
\ee
We will encounter several types of spinor fields, transforming {\it non-unitarily\/} (unitary representations will appear later) and denoted as follows
\be
\Lambda\phi{_{A_1\dots A_nA'_1\dots A'_m}}(x) &=& \Lambda{_{A_1}}{^{B_1}}\dots \Lambda{_{A'_m}}{^{B'_m}}
\phi{_{B_1\dots B_nB'_1\dots B'_m}}(\Lambda^{-1}x),\nonumber\\
&\pp=&
\textrm{(fields on Minkowski space)},\\
\Lambda\phi{_{A_1\dots A_nA'_1\dots A'_m}}(\bm p) &=& \Lambda{_{A_1}}{^{B_1}}\dots \Lambda{_{A'_m}}{^{B'_m}}\phi{_{B_1\dots B_nB'_1\dots B'_m}}(\bm{\Lambda^{-1}p}), \nonumber\\
&\pp=&\textrm{(fields on mass-$m$ hyperboloid)}.
\ee
$\bm{\Lambda^{-1}p}$ denotes the spacelike part
$\big(({\Lambda^{-1}p})^1,({\Lambda^{-1}p})^2,({\Lambda^{-1}p})^3\big)$.
Recall that $\Lambda p_a=\Lambda{_a}{^b}p_b$, $\Lambda p^a=p^b\Lambda^{-1}{_b}{^a}$, so that $\Lambda^{-1} p^a=p^b\Lambda{_b}{^a}$.

Generators of these representations,
\be
J{_{ab}}{_{A_1}}{^{B_1}}\dots {_{A'_m}}{^{B'_m}}
&=&
L_{ab}\ve{_{A_1}}{^{B_1}}\dots \ve{_{A'_m}}{^{B'_m}}
+
\sigma{_{ab}}{_{A_1}}{^{B_1}}\dots {_{A'_m}}{^{B'_m}},
\ee
split into orbital parts $L_{ab}$ and the spin parts
\be
\sigma{_{ab}}{_{A_1}}{^{B_1}}\dots {_{A'_m}}{^{B'_m}}
&=&
i
\frac{\partial}{\partial{}y^{ab}}
\Lambda{_{A_1}}{^{B_1}}\dots \Lambda{_{A'_m}}{^{B'_m}}\Big|_{{}y^{ab}=0}.
\ee
The explicit forms of the orbital parts,
\be
L_{ab}\phi{_{B_1\dots B_nB'_1\dots B'_m}}(x)
&=&
\sigma{_{ab}}{^r}{^s}
x_r\partial_s
\phi{_{B_1\dots B_nB'_1\dots B'_m}}(x)\\
L_{ab}\phi{_{B_1\dots B_nB'_1\dots B'_m}}(\bm p)
&=&
\sum_{{\bf j}=1}^3\sigma{_{ab}}{^{r}}{^{\bf j}}
p{_{r}}\frac{\partial}{\partial p^{\bf j}}
\phi{_{B_1\dots B_nB'_1\dots B'_m}}({\bm p}),
\ee
in general, will not be very important if we define spins by means of Pauli-Lubanski vectors.

\subsection{Bispinors and 2-spinors}

Simplest unitary representations of inhomogeneous SL(2,C) are most naturally introduced at the level of the Dirac equation, i.e. by means of bispinors. A bispinor is a direct sum of primed and unprimed 2-spinors. We will use the following abstract-index convention:
$
\psi^{\tt A}
=
\left(
\begin{array}{c}
\psi{^{A}}\\
\psi{^{A'}}
\end{array}
\right)
$ is a bispinor, Dirac's gamma matrices are represented by
\be
\gamma{_q}{_{\tt A}}{^{\tt B}}
=
\left(
\begin{array}{cc}
\gamma{_q}{_{A}}{^{B}} & \gamma{_q}{_{A}}{^{B'}}\\
\gamma{_q}{_{A'}}{^{B}} & \gamma{_q}{_{A'}}{^{B'}}
\end{array}
\right)
=
\sqrt{2}\left(
\begin{array}{cc}
0 & g{_{qA}}{^{B'}}\\
-g{_q}{^B}{_{A'}} & 0
\end{array}
\right),
\ee
and contractions of the bispinor indices are defined by $\alpha{^{\dots{\tt A}\dots}}\beta{_{\dots{\tt A}\dots}}=\alpha{^{\dots{A}\dots}}\beta{_{\dots{A}\dots}}+
\alpha{^{\dots{A'}\dots}}\beta{_{\dots{A'}\dots}}$. Generators of the bispinor $(\frac{1}{2},0)\oplus (0,\frac{1}{2})$ representation,
\be
\Lambda{_{\tt A}}{^{\tt B}}
&=&
\left(
\begin{array}{cc}
\Lambda{_{A}}{^{B}} & \Lambda{_{A}}{^{B'}}\\
\Lambda{_{A'}}{^{B}} & \Lambda{_{A'}}{^{B'}}
\end{array}
\right)
=
\left(
\begin{array}{cc}
\Lambda{_{A}}{^{B}} & 0\\
0 & \Lambda{_{A'}}{^{B'}}
\end{array}
\right),
\ee
satisfy
\be
\sigma{_{rs}}{_{\tt A}}{^{\tt B}}
=
\left(
\begin{array}{cc}
\sigma{_{rs}}{_{A}}{^{B}} & \sigma{_{rs}}{_{A}}{^{B'}}\\
\sigma{_{rs}}{_{A'}}{^{B}} & \sigma{_{rs}}{_{A'}}{^{B'}}
\end{array}
\right)
=
\left(
\begin{array}{cc}
\sigma{_{rs}}{_{A}}{^{B}} & 0\\
0 & \sigma{_{rs}}{_{A'}}{^{B'}}
\end{array}
\right).
\ee
Product of two gamma matrices,
\be
\gamma{_q}{_{\tt A}}{^{\tt B}}\gamma{_r}{_{\tt B}}{^{\tt C}}&=&
-2
\left(
\begin{array}{cc}
g{_{qA}}{^{B'}}g{_r}{^C}{_{B'}} & 0\\
0 & g{_q}{^B}{_{A'}} g{_{rB}}{^{C'}}
\end{array}
\right)\nonumber\\
&=&
\left(
\begin{array}{cc}
g{_{qr}}\varepsilon{_A}{^C}-2i\sigma{_{qr}} {_A}{^C} & 0\\
0 & g{_{qr}}\varepsilon{_{A'}}{^{C'}}-2i\sigma{_{qr}}
{_{A'}}{^{C'}}
\end{array}
\right)=
g{_{qr}}\ve{_{\tt A}}{^{\tt C}}-2i\sigma{_{qr}} {_{\tt A}}{^{\tt C}},\nonumber
\ee
implies
\be
\gamma{_q}{_{\tt A}}{^{\tt B}}\gamma{_r}{_{\tt B}}{^{\tt C}}
+
\gamma{_r}{_{\tt A}}{^{\tt B}}\gamma{_q}{_{\tt B}}{^{\tt C}}
&=&
2g{_{qr}}\ve{_{\tt A}}{^{\tt C}},\\
\gamma{_q}{_{\tt A}}{^{\tt B}}\gamma{_r}{_{\tt B}}{^{\tt C}}
-
\gamma{_r}{_{\tt A}}{^{\tt B}}\gamma{_q}{_{\tt B}}{^{\tt C}}
&=&
-4i\sigma{_{qr}} {_{\tt A}}{^{\tt C}}.
\ee
It is interesting that only at the abstract-index level one can realize that, as opposed to what one usually reads in relativistic quantum mechanics textbooks, $\gamma{_0}{_{\tt A}}{^{\tt B}}$ {\it cannot\/} be identified with the matrix $\beta$ occurring in the standard-notation formulas $\bar\psi\psi=\psi^{\dag}\beta\psi$ and $j_a=\bar\psi\gamma_a\psi=\psi^{\dag}\beta\gamma_a\psi$. Indeed, in the abstract-index notation the Dirac current reads \cite{PR}
\be
j_a
&=&
\sqrt{2}g_a{^{AA'}}\bigl(\psi_A\bar \psi_{A'} +
\psi_{A'}\bar \psi_{A}\bigr)
=\sqrt{2}\bigl(\psi_A\bar \psi_{A'} +
\psi_{A'}\bar \psi_{A}\bigr)
\label{spin-cur}\\
&=&
\sqrt{2}(\bar \psi^{A'},\bar \psi^A)
\left(
\begin{array}{cc}
0 & \varepsilon{_{A'}}{^{B'}}\\
-\varepsilon{_A}{^{B}} & 0
\end{array}
\right)
\left(
\begin{array}{cc}
0 & g{_{aB}}{^{C'}}\\
-g{_a}{^C}{_{B'}} & 0
\end{array}
\right)
\left(
\begin{array}{c}
\psi_C\\
\psi_{C'}
\end{array}
\right).
\ee
In particular,
\be
j_0&=&
\underbrace{
(\bar \psi^{A'},\bar \psi^A)}_{\bar\psi{^{\tt A'}}}
\underbrace{
\left(
\begin{array}{cc}
0 & \varepsilon{_{A'}}{^{B'}}\\
-\varepsilon{_A}{^{B}} & 0
\end{array}
\right)}_{\beta{_{\tt A'}}{^{\tt B}}}
\underbrace{\sqrt{2}
\left(
\begin{array}{cc}
0 & g{_{0B}}{^{C'}}\\
-g{_0}{^C}{_{B'}} & 0
\end{array}
\right)}_{\gamma_0{_{\tt B}}{^{\tt C}}}
\underbrace{\left(
\begin{array}{c}
\psi_C\\
\psi_{C'}
\end{array}
\right)}_{\psi{_{\tt C}}}\nonumber\\
&=&
\sqrt{2}
(\overline{\psi_{A}},\overline{\psi_{A'}})
\left(
\begin{array}{cc}
g{_0}{^A}{^{A'}} & 0\\
0 & g{_0}{^A}{^{A'}}
\end{array}
\right)
\left(
\begin{array}{c}
\psi_A\\
\psi_{A'}
\end{array}
\right).
\ee
Replacing Infeld-van der Waerden {\it tensors\/} by matrices of Infeld-van der Waerden {\it symbols\/} we find that the {\it matrices\/} corresponding to $\beta$ and $\gamma_0$ are indeed the same. However, a glimpse at the indices shows that these are matrices of maps that act in different linear spaces and in addition transform under SL(2,C) according to inequivalent representations (roughly speaking, $\beta{_{\tt A'}}{^{\tt B}}$ is a scalar, and not a timelike component of a world-vector). This is an example of a relativistic formal subtlety that can be appreciated only at the abstract-index 2-spinor level. In the following sections I will discuss more such subtleties, clarifying certain controversies about relativistic properties of qubits.

\section{Qubits associated with first-quantized Dirac equation}

Although 2-spinors and qubits are vectors from ${\bm C}^2$, not all qubits are 2-spinors, and not all 2-spinors are qubits. The reason is simple: 2-spinors correspond to finite-dimensional non-unitary representations of SL(2,C), but qubits --- by definition --- have to transform unitarily. Qubit representations of inhomogeneous SL(2,C) are infinite dimensional. Physically, the infinity of dimension means that relativistic qubits are {\it spinor fields\/} of some sort. These new types of spinors are directly related to 2-spinors by a kind of duality. The duality will be discussed at the end of this section, but before we do that we have to analyze some important preliminaries.

\subsection{Relativistic generalization of spin}

One expects that relativistic qubits are related to spin.
A conceptual difficulty one immediately encounters in this context is that there is no generally accepted definition of relativistic spin operator. It is not {\it a priori\/} evident that the discussed splitting of generators into `orbital' and `spin' parts corresponds, physically, to orbital angular momentum and spin. A simple illustration of the difficulty is the following. Let us take the Dirac Hamiltonian (units with $c=1$, $\hbar=1$)
\be
H &=& \bm\alpha\cdot\bm P+m\beta
\ee
in the non-manifestly-covariant formulation. The generator of rotations reads
\be
\bm J=\bm x\times \bm P+\bm s,
\ee
where $\bm s=\frac{1}{2}\left(\begin{array}{cc}\bm \sigma & 0\\0 & \bm \sigma\end{array}\right)$. It seems natural to identify spin with $\bm s$. However, since $[H,\bm s]\neq 0$ the projection of $\bm s$ on a unit vector $\bm a$ does not lead to a well defined quantum number, unless $\bm a$ is an operator parallel to momentum ${\bm P}=-i\bm \nabla$. The projection of spin on momentum $\bm P\cdot\bm s$ is a differential operator of first order, commuting with $H$. Knowledge of its eigenvectors is sufficient for vast majority of applications, but there are exceptions. Formulation of a relativistic generalization of the Bell inequality for electrons requires projections of spin on four different directions, so the knowledge of helicity states is not enough.

A possible solution is suggested by analogous problems with velocity of a free Dirac particle: $d\bm x/dt=\bm v=-i[\bm x,H]=\bm\alpha$, $[H,\bm v]\neq 0$. A way out is to start with the position operator
\be
\bm Q &=&\Pi_+ \bm x \Pi_+ +\Pi_- \bm x \Pi_-
\ee
where $\Pi_\pm$ project on states of positive or negative energy. Then, $d\bm Q/dt=\bm V=-i[\bm Q,H]=\bm P/H$, $[H,\bm V]=0$. $\bm Q$ has properties of relativistic center of mass, and $\bm V$ is the center-of-mass velocity. Applying this reasoning to spin, we arrive at
\be
\bm S &=&\Pi_+ \bm s \Pi_+ +\Pi_- \bm s \Pi_-=\bm J-\bm Q\times \bm P,\\
\bm J &=& \bm x\times \bm P+\bm s = \bm Q\times \bm P+\bm S.
\ee
Now $[\bm S,H]=0$, and $\bm a\cdot \bm S$ is a natural projection of spin with no restriction on the direction of $\bm a$. Apparently, the first application of the above idea to relativistic qubits (relativistic Einstein-Podolsky-Rosen-Bohm experiment) can be found in the unpublished preprint of mine from 1984 \cite{MC84}. The resulting spin operator depends on momentum in a very complicated way and does not satisfy the rotation Lie algebra typical of angular momentum. If $\bm a=\bm P/|\bm P|$ then $\bm a\cdot \bm S=\bm a\cdot \bm s=\bm a\cdot \bm J$ is the helicity operator.
If $\bm a$ is a unit vector perpendicular to $\bm P$ then the eigenvalues of $\bm a\cdot \bm S$ tend to 0 with $\bm p\to\infty$ (in momentum space), or with $m\to 0$. The latter property can be understood as a combined effect of two classical relativistic phenomena: the Lorentz flattening of the particle, and the M{\o}ller shift of the center of mass \cite{Moller,Fleming1}.

Another, manifestly covariant approach is to start with first-order differential operator known as the Pauli-Lubanski vector
\be
W^a{_{\tt X}}{^{\tt Y}} &=& P_b {}^*M^{ab}{_{\tt X}}{^{\tt Y}}=P_b {}^*\sigma^{ab}{_{\tt X}}{^{\tt Y}}
=\frac{1}{2}e^{abcd}P_b\sigma_{cd}{_{\tt X}}{^{\tt Y}}.
\ee
Its projection $W^0=\bm P\cdot \bm s$ on the `time' direction turns out to be the helicity times $|\bm P|$. Moreover, $W^a$ commutes with 4-momentum and $W_aW^a$ is a Casimir operator of the Poincar\'e group. In irreducible representations $W_aW^a=-m^2 j(j+1)$ where $j$ is the dimension of the associated representation of su(2). This is why another popular definition of spin is $w^a=W^a/m$. Its spacelike component $\bm w$ equals $\bm s$ for $\bm p=0$ (again, in momentum space). If $t^a$ is a constant world-vector then eigenvalues of $t^aW_a$ tend to infinity with $\bm p\to \infty$ (this is clear since in momentum space $W^0=\bm p\cdot\bm s$). It is interesting that $\bm S$ and $\bm W$ are proportional to each other, $\bm S=\bm W/H$, and thus possess the same eigenvectors. From the point of view of the Bell inequality, say, they are equivalent (cf. the analysis of this point given in \cite{MC-PRA,SPIE}). The components of $W^a$ do not satisfy the angular momentum Lie algebra.

The simplest way of deriving an explicit form of $W^a{_{\tt X}}{^{\tt Y}}$ is based on the following abstract-index identity:
If
$F_{ab}=-F_{ba}$
then (cf. Eq.~(3.4.21) in \cite{PR})
\be
F_{ab}
&=&
\phi_{AB}\ve_{A'B'}+\ve_{AB}\psi_{A'B'},\\
\phi_{AB}
&=&
\frac{1}{2}F{_{AX'}}{_{B}}{^{X'}},\\
\psi_{A'B'}
&=&
\frac{1}{2}F{_{XA'}}{^{X}}{_{B'}},\\
^*F_{ab}
&=&
-i\phi_{AB}\ve_{A'B'}+i\ve_{AB}\psi_{A'B'}.
\ee
Comparison with (\ref{sigma RS}), (\ref{sigma R'S'}) implies
\be
^*\sigma_{ab}{_R}{^S}
&=&
\frac{1}{2}\ve_{A'B'}(\ve_{AR}\ve{_{B}}{^{S}} + \ve{_{B}}{_{R}}\ve{_{A}}{^{S}})
\\
&=&
\frac{1}{2}(g_{aRX'}g_b{^{SX'}} - g_{bRX'}g_a{^{SX'}})
\label{*sigma RS},\\
^*\sigma_{ab}{_{R'}}{^{S'}}
&=&
-\frac{1}{2}\ve_{AB}(\ve_{A'R'}\ve{_{B'}}{^{S'}} + \ve_{B'R'}\ve{_{A'}}{^{S'}})\\
&=&
-\frac{1}{2}(g_{aXR'}g_b{^{XS'}} - g_{bXR'}g_a{^{XS'}})
\label{*sigma R'S'}.
\ee
It follows that
\be
W_{a}{_R}{^S}
&=&
\frac{1}{2}(g_{aRX'}P{^{SX'}} - P_{RX'}g_a{^{SX'}}),\label{W_a_R^S}\\
W_{a}{_{R'}}{^{S'}}
&=&
-\frac{1}{2}(g_{aXR'}P{^{XS'}} - P_{XR'}g_a{^{XS'}}),\\
W_{a}{_{\tt R}}{^{\tt S}}
&=&
\left(
\begin{array}{cc}
W_{a}{_R}{^S} & 0\\
0 & W_{a}{_{R'}}{^{S'}}
\end{array}
\right).
\ee
Projections of the Pauli-Lubanski vector on various directions will be discussed in a separate section.

Yet another possibility is to begin with the Newton-Wigner position operator $\bm Q_{NW}$, which differs from $\bm Q$ by a term commuting with $H$ \cite{NW}. The components of $\bm Q_{NW}$ commute, as opposed to those of $\bm Q$. One obtains $\bm S_{NW}=\bm J-\bm Q_{NW}\times \bm P$ whose components satisfy the angular momentum Lie algebra. It can be shown \cite{BLT} that this is the only axial vector operator linear in the Pauli-Lubanski vector, satisfying su(2), and transforming under rotations as a 3-dimensional vector. The above properties look natural. Spin based on $\bm S_{NW}$ has been extensively studied in the context of relativistic qubits and Bell inequalities by Rembieli\'nski and his group \cite{CRW1,CRW2,CR,RS}.

One can argue, however, that in classical mechanics of spinning particles one does {\it not\/} arrive at {\it commuting\/} center-of-mass coordinates. This happens whenever one imposes mass or spin constraints, which is the case here, and replaces Poisson brackets by Dirac brackets according to the rules of constrained dynamics. The position variable that commutes with respect to the unconstrained bracket becomes noncommutative with respect to the constrained one \cite{Mukunda,Zakrzewski}, and this is the bracket that should be employed if spin is 1/2 or mass is $m_{\rm electron}$. What is interesting, the resulting algebra is analogous to this of $\bm Q$ and not to $\bm Q_{NW}$. This is not so surprising if one realizes that position generates shifts of momentum. Mass hyperboloid in not a flat manifold so translations do not commute. If we relax the mass constraint, the momentum space becomes the Minkowski space, which is flat, so positions should commute, as it indeed happens in off-shell formalisms in quantum mechanics.

Classification of all spin operators linear in momentum can be found in \cite{Bagrov,IBB}, and different spins are used in practical applications to exact solutions of the Dirac equation coupled to electromagnetic fields.
From a formal point of view all the possible spin-like observables are possible candidates for quantum yes-no observables. Which of them are actually measured in experiments depends on experimental procedures.

\subsection{Beyond helicity: Two-spinor approach to polarization operators based on the Pauli-Lubanski vector}

Generators of four-translations are defined by
\be
e^{iy^a P_a}\psi{_{\tt A}}(x)=\psi{_{\tt A}}(x-y)=e^{-y^a \partial_a}\psi{_{\tt A}}(x),
\ee
so $P_a=i\partial_a$. The Pauli-Lubanski vector thus satisfies
\be
W_{a}{_{\tt R}}{^{\tt S}}e^{\mp i p\cdot x}
&=&
\underbrace{\pm\frac{1}{2}
\left(
\begin{array}{cc}
g_{aRX'}p{^{SX'}} - p_{RX'}g_a{^{SX'}} & 0\\
0 & -g_{aXR'}p{^{XS'}} + p_{XR'}g_a{^{XS'}}
\end{array}
\right)}_{W_{a}{_{\tt R}}{^{\tt S}}(\pm p)=\pm W_{a}{_{\tt R}}{^{\tt S}}(\bm p)}e^{\mp i p\cdot x}.\nonumber\\
\ee
In all the formulas we assume that $p^a$ is future-pointing, i.e. $p_0>0$.
Now, let us consider any symmetric spinor $\phi_{AB}$. Using $\phi{_A}{^A}=\phi{_0}{^0}+\phi{_1}{^1}=0$ we find
\be
\left|
\begin{array}{cc}
\phi{_0}{^0}-\lambda & \phi{_0}{^1}\\
\phi{_1}{^0} & \phi{_1}{^1}-\lambda
\end{array}
\right|
&=&
\lambda^2
+
\frac{1}{2}\phi{_A}{_B}\phi{^A}{^B},
\ee
so that the eigenvalues $\lambda^{(\pm)}$ of $\phi{_A}{^B}$ are given by
\be
\lambda^{(\pm)}
&=&
\pm\sqrt{-\phi{_X}{_Y}\phi{^X}{^Y}/2}.
\ee
Applying this formula to
$W{_{\tt R}}{^{\tt S}}(t,\bm p)=t^a(\bm p)W_{a}{_{\tt R}}{^{\tt S}}(\bm p)$, where $t^a(\bm p)$ is an arbitrary field of (real) world-vectors, we conclude that
the eigenvalues of such a general projection of the Pauli-Lubanski vector are given by
\be
\lambda^{(\pm)}(t,\bm p)
&=&
\pm \sqrt{-W{_{X}}{_{Y}}(t,\bm p)W{^{X}}{^{Y}}(t,\bm p)/2}
=
\pm \sqrt{-W{_{X'}}{_{Y'}}(t,\bm p)W{^{X'}}{^{Y'}}(t,\bm p)/2}
\nonumber\\
&=&
\pm \frac{1}{2}\sqrt{(t\cdot p)^2-t^2p^2}.\label{W(p,t)}
\ee
Here $t\cdot p=p_at^a(\bm p)$, $t^2=t_a(\bm p)t^a(\bm p)$, $p^2=p_ap^a=m^2$. Helicity, corresponding to
$t^{\bm a}(\bm p)=(1/|\bm p|,0,0,0)$, has eigenvalues
\be
\lambda^{(\pm)}(t,\bm p)
&=&
\pm \frac{1}{2}\sqrt{p_0^2/\bm p^2-m^2/\bm p^2}=\pm \frac{1}{2}.
\ee
If $t^a(\bm p)=p^a$ then $\lambda^{(\pm)}(t,\bm p)=0$, since $P^aW_a=0$ in any representation. The latter incidentally shows that projections of the Pauli-Lubanski vector possess a gauge freedom: $t^aW_a=(t^a+\theta P^a)W_a$, for any $\theta$. The same concerns the eigenvalues: $\lambda^{(\pm)}(t,\bm p)=\lambda^{(\pm)}(t+\theta p,\bm p)$.

For a particle at rest, $p^{\bf a}=(m,\bm 0)$, the Pauli-Lubanski vector reduces to non-relativistic spin multiplied by mass:
$W^{\bf a}=m(0,\bm s)$. The non-relativistic observable $\bm a\cdot\bm s=\frac{1}{m}\bm a\cdot\bm W$ can be covariantly written as $t^aW_a$, where
\be
t^{\bm a}=(t^0,\bm a/m)=(0,\bm a/m)+(t^0,\bm 0),\label{t gauge}
\ee
the last term being proportional to $p^{\bf a}$. This is the simplest explanation of the gauge freedom inherently present in the Pauli-Lubanski vector.
The fact that it is only the spacelike part of $t^a$ that counts in the definition of $t^aW_a$ does not mean that $t^a$ itself should be spacelike. A spacelike $t^a$ when shifted by a timelike $\theta p^a$ can become timelike or null (or remain spacelike).

Of particular interest turns out to be the null case: $t^2=0$, $t^ap_a=1$. Indeed, first of all the corresponding eigenvalues are identical to those of the helicity, $\lambda^{(\pm)}(t,\bm p)=\pm \frac{1}{2}$. Let $t^a=\omega^a(\bm p)=\omega^A(\bm p)\bar\omega^{A'}(\bm p)$, and define
\be
\pi{^{A}}(\bm p)
&=&
p{^{A}}{^{B'}}\bar\omega{_{B'}}(\bm p).\label{pi-omega}
\ee
The trace-reversal formula (Eq.~(3.4.13) in \cite{PR})
\be
p^{AB'}p^{BA'}
&=&
p^{AA'}p^{BB'}-\frac{m^2}{2}\ve^{AB}\ve^{A'B'}
=
p^ap^b-\frac{m^2}{2}g^{ab},
\ee
implies
\be
\pi^A(\bm p)\bar\pi^{A'}(\bm p)
&=&
p{^{A}}{^{B'}}p{^{B}}{^{A'}}\omega{_{B}}(\bm p)\bar\omega{_{B'}}(\bm p)
\nonumber\\
&=&
p{^{a}}
\underbrace{
p{^{b}}\omega{_{B}}(\bm p)\bar\omega{_{B'}}(\bm p)}_{p^bt_b=1}-\frac{m^2}{2}\omega{^{A}}(\bm p)\bar\omega{^{A'}}(\bm p),
\ee
and we arrive at the following important decomposition of $p^a$ into a combination of two null directions:
\be
p{^{a}}
&=&
\pi^A(\bm p)\bar\pi^{A'}(\bm p)
+
\frac{m^2}{2}\omega{^{A}}(\bm p)\bar\omega{^{A'}}(\bm p).\label{p}
\ee
Formula (\ref{p}) implies (\ref{pi-omega}). Indeed,
\be
p{^{AB'}}\bar\omega{_{B'}}(\bm p)
&=&
\Big(\pi^A(\bm p)\bar\pi^{B'}(\bm p)
+
\frac{m^2}{2}\omega{^{A}}(\bm p)\bar\omega{^{B'}}(\bm p)
\Big)\bar\omega{_{B'}}(\bm p)=\pi^A(\bm p).\nonumber
\ee
It follows that $\omega_A(\bm p)$ supplemented by (\ref{p}) determines its spin-partner $\pi^A(\bm p)$ uniquely. Moreover, since
\be
p_{AC'}\pi{^{A}}(\bm p)
&=&
p_{AC'}p{^{A}}{^{B'}}\bar\omega{_{B'}}(\bm p)=\frac{m^2}{2}\bar\omega{_{C'}}(\bm p),
\ee
the knowledge of $\pi{^{A}}(\bm p)$ and (\ref{p}) determines $\omega{_{A}}(\bm p)$ uniquely, unless $m=0$.

As opposed to the special case $m=0$, where $\pi^A(\bm p)$ is defined by its flagpole $p^a$ up to a phase (U(1) internal symmetry), the internal symmetry group is here U(2). The pair $(\pi^A(\bm p),\frac{m}{\sqrt{2}}\omega{^{A}}(\bm p))$ can be replaced by
$(\tilde\pi^A(\bm p),\frac{m}{\sqrt{2}}\tilde\omega{^{A}}(\bm p))$, where
\be
\left(
\begin{array}{c}
\tilde\pi^A(\bm p)\\
\frac{m}{\sqrt{2}}\tilde\omega{^{A}}(\bm p)
\end{array}
\right)
&=&
\left(
\begin{array}{cc}
\alpha(\bm p) & \beta(\bm p)\\
\gamma(\bm p) & \delta(\bm p)
\end{array}
\right)
\left(
\begin{array}{c}
\pi^A(\bm p)\\
\frac{m}{\sqrt{2}}\omega{^{A}}(\bm p)
\end{array}
\right),
\ee
and the matrix is unitary. Hughston showed \cite{H} that the existence of such splitting-of-$p^a$ ambiguities may be the actual geometric reason for internal symmetries occurring in gauge theories. From our point of view it is more important that a similar mechanism is responsible for the structure of unitary representations of the Poincar\'e group  \cite{MC-BW}.

\subsection{The simplest case: Projection on the null direction $t^a=\omega^a(\bm p)$}

Since $\omega^ap_a=\omega^A\bar \omega^{A'}\pi_A\bar \pi_{A'}=1$ independently of the value of $m$, we can assume --- without loss of generality --- spin-frame normalization $\omega_A\pi^A=1$.
Projections on the null direction,
\be
W(\omega,\bm p){_{A}}{^{B}}&=&
\frac{1}{2}\Bigl(
\pi_{A}\omega^{B}+ \omega_{A}\pi^{B}\Bigr),\label{W W omega}\\
W(\omega,\bm p){_{A'}}{^{B'}}&=&-
\frac{1}{2}\Bigl(
\bar \pi_{A'}\bar \omega^{B'}+
\bar \omega_{A'}\bar \pi^{B'}\Bigr),
\ee
look the same for both $m=0$ and $m\neq 0$. Also spin eigenvalue problems become in the null formalism particularly simple:
\be
W(\omega,\bm p){_{A}}{^{B}}\omega_{B}&=&
\frac{1}{2}\omega_{A},\label{pl000}\\
W(\omega,\bm p){_{A'}}{^{B'}}\bar \pi_{B'}&=&
\frac{1}{2}\bar \pi_{A'},\label{pl00'}\\
W(\omega,\bm p){_{A}}{^{B}}\pi_{B}&=&
-\frac{1}{2}\pi_{A},\label{pl00}\\
W(\omega,\bm p){_{A'}}{^{B'}}\bar \omega_{B'}&=&
-\frac{1}{2}\bar \omega_{A'}.\label{pl000'}
\ee
Spectral projectors associated with $\lambda^{(\pm)}(\omega,\bm p)=\pm \frac{1}{2}$,
\be
\Pi^{(+)}(\omega,\bm p){_{A}}{^{B}} &=& \omega{_{A}}\pi{^{B}},
\\
\Pi^{(-)}(\omega,\bm p){_{A}}{^{B}} &=& -\pi_{A}\omega^{B},
\\
\Pi^{(+)}(\omega,\bm p){_{A'}}{^{B'}}
&=&
-\bar\pi{_{A'}}\bar\omega{^{B'}},
\\
\Pi^{(-)}(\omega,\bm p){_{A'}}{^{B'}}
&=&
\bar\omega{_{A'}}\bar\pi{^{B'}},
\ee
trivially satisfy 2-spinor resolutions of unity (\ref{ve_A^B}),
\be
\Pi^{(+)}(\omega,\bm p){_{A}}{^{B}}+\Pi^{(-)}(\omega,\bm p){_{A}}{^{B}}
&=&
\omega{_{A}}\pi{^{B}}-\pi_{A}\omega^{B}=\ve{_{A}}{^{B}},\\
\Pi^{(-)}(\omega,\bm p){_{A'}}{^{B'}}
+
\Pi^{(+)}(\omega,\bm p){_{A'}}{^{B'}}
&=&
\bar\omega{_{A'}}\bar\pi{^{B'}}-\bar\pi{_{A'}}\bar\omega{^{B'}}=\ve{_{A'}}{^{B'}}.
\ee
At the bispinor level the projectors read
\be
\Pi^{(+)}(\omega,\bm p){_{\tt A}}{^{\tt B}}
&=&
\left(
\begin{array}{cc}
\Pi^{(+)}(\omega,\bm p){_{A}}{^{B}} & 0\\
0 & \Pi^{(+)}(\omega,\bm p){_{A'}}{^{B'}}
\end{array}
\right)
=
\left(
\begin{array}{cc}
\omega{_{A}}\pi{^{B}} & 0\\
0 & -\bar\pi{_{A'}}\bar\omega{^{B'}}
\end{array}
\right),\label{Pi^+}\\
\Pi^{(-)}(\omega,\bm p){_{\tt A}}{^{\tt B}}
&=&
\left(
\begin{array}{cc}
\Pi^{(-)}(\omega,\bm p){_{A}}{^{B}} & 0\\
0 & \Pi^{(-)}(\omega,\bm p){_{A'}}{^{B'}}
\end{array}
\right)
=
\left(
\begin{array}{cc}
-\pi_{A}\omega^{B} & 0\\
0 & \bar\omega{_{A'}}\bar\pi{^{B'}}
\end{array}
\right).\label{Pi^-}
\ee
\subsection{Wave functions associated with $W(\omega,\bm p){_{\tt A}}{^{\tt B}}$}

In order to introduce wave functions corresponding to $W(\omega,\bm p){_{\tt A}}{^{\tt B}}$ we first have to expand solutions of Dirac equation in terms of appropriate eigen-bispinors. A systematic procedure of deriving the bispinors can be based on `spin-energy' projectors.

The free Dirac equation
\be
D{_{\tt A}}{^{\tt B}}\psi{_{\tt B}}=0, \quad
D{_{\tt A}}{^{\tt B}}=i\partial{^r}\gamma{_r}{_{\tt A}}{^{\tt B}}-m\ve {_{\tt A}}{^{\tt B}}\label{Dirac}
\ee
is implicitly an orthogonality condition for `sign-of-energy' projectors $\Pi_\pm{_{\tt A}}{^{\tt B}}(\bm p)$ constructed as follows.
One begins with
\be
D{_{\tt A}}{^{\tt B}}e^{\mp i p\cdot x}
&=&
D_\pm{_{\tt A}}{^{\tt B}}(\bm p)e^{\mp i p\cdot x},
\ee
where
\be
D_\pm{_{\tt A}}{^{\tt B}}(\bm p)
&=&
\pm p{^r}\gamma{_r}{_{\tt A}}{^{\tt B}}-m\ve {_{\tt A}}{^{\tt B}}
=
\left(
\begin{array}{ccc}
-m\ve {_{A}}{^{B}} &, &\pm\sqrt{2} p{_{A}}{^{B'}}\\
\mp\sqrt{2}p{^B}{_{A'}} &, & -m\ve {_{A'}}{^{B'}}
\end{array}
\right).
\ee
The orthogonality conditions
\be
D_\pm{_{\tt A}}{^{\tt B}}(\bm p)
D_\mp{_{\tt B}}{^{\tt C}}(\bm p)
&=&
0,\\
D_\pm{_{\tt A}}{^{\tt B}}(\bm p)
D_\pm{_{\tt B}}{^{\tt C}}(\bm p)
&=&
-2m D_\pm{_{\tt A}}{^{\tt C}}(\bm p),
\ee
imply that the possible eigenvalues of $D_\pm{_{\tt A}}{^{\tt B}}(\bm p)$ are 0 and $-2m$.
Accordingly, the projectors we are interested in read
\be
\Pi_\pm{_{\tt A}}{^{\tt B}}(\bm p)=D_\pm{_{\tt A}}{^{\tt B}}(\bm p)/(-2m)
=
\frac{1}{2}
\left(
\begin{array}{cc}
\ve {_{A}}{^{B}}  &\mp\frac{\sqrt{2}}{m} p{_{A}}{^{B'}}\\
\pm\frac{\sqrt{2}}{m}p{^B}{_{A'}}  & \ve {_{A'}}{^{B'}}
\end{array}
\right).\label{Pi pm}
\ee
Multiplying (\ref{Pi pm}) by (\ref{Pi^+}) or (\ref{Pi^-}) we obtain `spin-energy' projectors,
\be
\Pi_\pm^{(+)}{_{\tt A}}{^{\tt C}}(\omega,\bm p)
&=&
\Pi_\pm{_{\tt A}}{^{\tt B}}(\bm p)
\Pi^{(+)}{_{\tt B}}{^{\tt C}}(\omega,\bm p)
=
\frac{1}{2}
\left(
\begin{array}{cc}
\omega{_{A}}\pi{^{C}}  &\mp\frac{m}{\sqrt{2}}\omega{_{A}}\bar\omega{^{C'}}\\
\pm\frac{\sqrt{2}}{m}\bar\pi{_{A'}}\pi{^{C}}  & -\bar\pi{_{A'}}\bar\omega{^{C'}}
\end{array}
\right),\label{Pi pm +}\\
\Pi_\pm^{(-)}{_{\tt A}}{^{\tt C}}(\omega,\bm p)
&=&
\Pi_\pm{_{\tt A}}{^{\tt B}}(\bm p)
\Pi^{(-)}{_{\tt B}}{^{\tt C}}(\omega,\bm p)
=
\frac{1}{2}
\left(
\begin{array}{cc}
-\pi{_{A}}\omega^{C}  &\mp\frac{\sqrt{2}}{m} \pi{_{A}}\bar\pi{^{C'}}\\
\pm\frac{m}{\sqrt{2}}\bar\omega{_{A'}}\omega^{C}  & \bar\omega{_{A'}}\bar\pi{^{C'}}
\end{array}
\right).\label{Pi pm -}
\ee
The completeness relation is
\be
\sum_{s,s'=\pm}\Pi_s^{(s')}{_{\tt A}}{^{\tt B}}(\omega,\bm p)=\ve{_{\tt A}}{^{\tt B}}
=\left(
\begin{array}{cc}
\ve {_{A}}{^{B}}  & 0\\
0  & \ve {_{A'}}{^{B'}}
\end{array}
\right).
\ee
It is interesting and important for later applications that (\ref{Pi pm +}) and (\ref{Pi pm -}) allow us to define basis bispinors that can be used also in the massless case. The trick is the following. Let
$
\phi{_{\tt A}}(\bm p)
=
2\left(
\begin{array}{c}
\pi{_{A}}(\bm p)\\
\bar\pi{_{A'}}(\bm p)
\end{array}
\right).
$
The eigen-bispinors
\be
\phi_\pm^{(+)}{_{\tt A}}(\omega,\bm p)
&=&
\Pi_\pm^{(+)}{_{\tt A}}{^{\tt B}}(\omega,\bm p)
\phi{_{\tt B}}(\bm p)
=
\left(
\begin{array}{c}
\pm\frac{m}{\sqrt{2}}\omega{_{A}}(\bm p)\\
\bar\pi{_{A'}}(\bm p)
\end{array}
\right),\\
\phi_\pm^{(-)}{_{\tt A}}(\omega,\bm p)
&=&
\Pi_\pm^{(-)}{_{\tt A}}{^{\tt B}}(\omega,\bm p)
\phi{_{\tt B}}(\bm p)
=
\left(
\begin{array}{c}
\pi{_{A}}(\bm p) \\
\mp\frac{m}{\sqrt{2}}\bar\omega{_{A'}}(\bm p)
\end{array}
\right),
\ee
are well defined also for $m=0$, and satisfy
\be
W(\omega,\bm p){_{\tt A}}{^{\tt B}}\phi_s^{(\pm)}{_{\tt B}}(\omega,\bm p)
&=&
\pm \frac{1}{2}\phi_s^{(\pm)}{_{\tt A}}(\omega,\bm p),\\
D{_{\tt A}}{^{\tt B}}
\phi_\pm^{(s')}{_{\tt B}}(\omega,\bm p)
e^{\mp ip\cdot x}
&=&
-2m\,
\phi_\pm^{(s')}{_{\tt A}}(\omega,\bm p)
e^{\mp ip\cdot x},\\
D{_{\tt A}}{^{\tt B}}
\phi_\mp^{(s')}{_{\tt B}}(\omega,\bm p)
e^{\mp ip\cdot x}
&=&
0.
\ee
A general solution of free Dirac equation can be finally written as
\be
\psi_{\tt A}(x)
&=&
\int dp \Big(\psi_{-\tt A}(\bm p)e^{-i p\cdot x}
+
\psi_{+\tt A}(\bm p)e^{+i p\cdot x}\Big)\nonumber\\
&=&
\sum_{{s}=\pm}\int dp \Big(\phi_-^{({s})}{_{\tt A}}(\omega,\bm p)f(s,\bm p)e^{-i p\cdot x}
+
\phi_+^{({s})}{_{\tt A}}(\omega,\bm p)\overline{g(-s,\bm p)}e^{+i p\cdot x}\Big).
\ee
Here $dp=d^3p/[(2\pi)^3 2\sqrt{\bm p^2+m^2}]$ is the invariant measure on mass-$m$ hyperboloid, and complex conjugation and the minus sign in $\overline{g(-s,\bm p)}$ are convenient for later applications. By definition,
\be
\psi_{+\tt A}(\bm p)
&=&
\left(
\begin{array}{c}
\frac{m}{\sqrt{2}}\omega{_{A}}(\bm p)\\
\bar\pi{_{A'}}(\bm p)
\end{array}
\right)
f(+,\bm p)
+
\left(
\begin{array}{c}
\pi{_{A}}(\bm p) \\
-\frac{m}{\sqrt{2}}\bar\omega{_{A'}}(\bm p)
\end{array}
\right)
f(-,\bm p),\\
\psi_{-\tt A}(\bm p)
&=&
\left(
\begin{array}{c}
-\frac{m}{\sqrt{2}}\omega{_{A}}(\bm p)\\
\bar\pi{_{A'}}(\bm p)
\end{array}
\right)
\overline{g(-,\bm p)}
+
\left(
\begin{array}{c}
\pi{_{A}}(\bm p) \\
\frac{m}{\sqrt{2}}\bar\omega{_{A'}}(\bm p)
\end{array}
\right)
\overline{g(+,\bm p)}.
\ee
Wave functions $f(s,\bm p)$ and $g(s,\bm p)$ are simultaneously scalar fields and spinors of a new type, as we will see shortly. They can be extracted from $\psi_{\pm\tt A}(\bm p)$ in a simple way:
\be
f(+,\bm p) &=& \bar\omega{_{A'}}(\bm p)\psi{_{+}}{^{A'}}(\bm p),\\
f(-,\bm p) &=& \omega{_{A}}(\bm p)\psi{_{+}}{^{A}}(\bm p),\\
\overline{g(-,\bm p)} &=& \bar\omega{_{A'}}(\bm p)\psi{_{-}}{^{A'}}(\bm p),\\
\overline{g(+,\bm p)} &=& \omega{_{A}}(\bm p)\psi{_{-}}{^{A}}(\bm p).
\ee
For future reference let us explicitly write the two components of the $m=0$ case:
\be
\psi_{A}(x)
&=&
\int dk\,
\pi{_{A}}(\bm k)\Big(f(-,\bm k)e^{-ik\cdot x}+\overline{g(+,\bm k)}e^{ik\cdot x}\Big),\label{psi m=01}\\
\psi_{A'}(x)
&=&
\int dk\,
\bar\pi{_{A'}}(\bm k)\Big(f(+,\bm k)e^{-ik\cdot x}+\overline{g(-,\bm k)}e^{ik\cdot x}\Big).
\label{psi m=02}
\ee
Massless 4-momenta are denoted here by $k^a$ and $dk=d^3k/[(2\pi)^3 2 |\bm k|]$ is the invariant measure on the light cone.

This is the right moment to explain the issue of unitary representations of the (covering space of the) Poincar\'e group. I will not follow the usual Wigner-Mackey procedure \cite{Wigner,Mackey} of induction from little groups. Instead, I will show by means of 2-spinor techniques that unitary representations of the group are encoded in shapes of solutions of relativistic wave equations. The latter statement is in itself not very original since links between unitary representations and relativistic wave equations were discussed already in \cite{BW} (for a modern analysis cf.  \cite{BR}). Nevertheless, the 2-spinor tricks I will use do not seem to be widely known and, apparently, were introduced for the first time in \cite{MC-BW}.

\subsection{Duality between active and passive SL(2,C) transformations: $\omega$-spinors}

Let us now consider the transformed solution
\be
\Lambda\psi_{\tt A}(x)
&=&
\Lambda{_{\tt A}}{^{\tt B}}\psi_{\tt B}(\Lambda^{-1}x),\nonumber\\
&=&
\sum_{{s}}\int dp \,\Lambda{_{\tt A}}{^{\tt B}}\Big(\phi_-^{({s})}{_{\tt B}}(\omega,\bm p)f(s,\bm p)e^{-i p\cdot \Lambda^{-1}x}
+
\phi_+^{({s})}{_{\tt B}}(\omega,\bm p)\overline{g(-s,\bm p)}e^{+i p\cdot \Lambda^{-1}x}\Big)\nonumber
\ee
Employing $p\cdot \Lambda^{-1}x=\Lambda p\cdot x$, changing variables in integrals, keeping in mind that $dp=d(\Lambda p)$, and finally expressing the transformed solution again in the basis $\phi_\pm^{({\pm})}{_{\tt A}}(\omega,\bm p)$, we get
\be
\Lambda\psi_{\tt A}(x)
&=&
\sum_{{s}}\int dp \Big(\Lambda{_{\tt A}}{^{\tt B}}\phi_-^{({s})}{_{\tt B}}(\omega,\bm{\Lambda^{-1}p})f(s,\bm{\Lambda^{-1}p})e^{-i p\cdot x}
\nonumber\\
&\pp=&
\pp{\sum_{{s}}\int dp \Big(}
+
\Lambda{_{\tt A}}{^{\tt B}}\phi_+^{({s})}{_{\tt B}}(\omega,\bm{\Lambda^{-1}p})\overline{g(-s,\bm{\Lambda^{-1}p})}e^{+i p\cdot x}\Big)\nonumber\\
&=&
\sum_{{s}}\int dp \Big(\phi_-^{({s})}{_{\tt A}}(\omega,\bm p)\Lambda f(s,\bm{p})e^{-i p\cdot x}
+
\phi_+^{({s})}{_{\tt A}}(\omega,\bm p)\overline{\Lambda g(-s,\bm{p})}e^{+i p\cdot x}\Big)\nonumber\\
\ee
Comparing appropriate terms we find
\be
&{}&
\left(
\begin{array}{c}
\frac{m}{\sqrt{2}}\Lambda \omega{^{A}}(\bm{p})\\
\overline{\Lambda\pi}{^{A'}}(\bm{p})
\end{array}
\right)
f(+,\bm{\Lambda^{-1}p})
+
\left(
\begin{array}{c}
\Lambda \pi{^{A}}(\bm{p}) \\
-\frac{m}{\sqrt{2}}\overline{\Lambda\omega}{^{A'}}(\bm{p})
\end{array}
\right)
f(-,\bm{\Lambda^{-1}p})
\nonumber\\
&{}&\pp=
=
\left(
\begin{array}{c}
\frac{m}{\sqrt{2}}\omega{^{A}}(\bm p)\\
\bar\pi{^{A'}}(\bm p)
\end{array}
\right)
\Lambda f(+,\bm p)
+
\left(
\begin{array}{c}
\pi{^{A}}(\bm p) \\
-\frac{m}{\sqrt{2}}\bar\omega{^{A'}}(\bm p)
\end{array}
\right)
\Lambda f(-,\bm p),
\ee
where $\Lambda\pi{_{A}}(\bm p)=\Lambda{_{A}}{^{B}}\pi{_{B}}(\bm{\Lambda^{-1}p})$, $\Lambda\omega{_{A}}(\bm p)=\Lambda{_{A}}{^{B}}\omega{_{B}}(\bm{\Lambda^{-1}p})$, and analogously with the primed spin-frames and complex-conjugated anti-particle wave functions
$\overline{\Lambda g(-s,\bm{p})}$. Denoting
\be
\left(
\begin{array}{c}
\psi_{-\it 0}(\bm p)\\
\psi_{-\it 1}(\bm p)\\
\psi_{+\it 0}(\bm p)\\
\psi_{+\it 1}(\bm p)
\end{array}
\right)
&=&
\left(
\begin{array}{c}
f(+,\bm p)\\
f(-,\bm p)\\
g(+,\bm p)\\
g(-,\bm p)\\
\end{array}
\right),\label{t-bispinor}
\ee
we obtain
\be
\Lambda \psi_{\pm\cal A}(\bm p) &=& U(\Lambda,\bm p){_{\cal A}}{^{\cal B}}\psi_{\pm\cal B}(\bm{\Lambda^{-1}p}),\label{unitary0}
\ee
\be
\left(
\begin{array}{c}
\Lambda \psi_{\pm\it 0}(\bm p)\\
\Lambda \psi_{\pm\it 1}(\bm p)
\end{array}
\right)
&=&
\underbrace{
\left(
\begin{array}{cc}
{\bar\omega}{_{A'}}(\bm{p})\overline{\Lambda\pi}{^{A'}}(\bm{p})  & -\frac{m}{\sqrt{2}}{\bar\omega}{_{A'}}(\bm{p})\overline{\Lambda\omega}{^{A'}}(\bm{p})
\\
\frac{m}{\sqrt{2}}\omega{_{A}}(\bm{p})\Lambda \omega{^{A}}(\bm{p})  &
{\omega}{_{A}}(\bm{p}){\Lambda\pi}{^{A}}(\bm{p})
\end{array}
\right)}_{U(\Lambda,\bm p){_{\cal A}}{^{\cal B}}}
\left(
\begin{array}{c}
\psi_{\pm\it 0}(\bm{\Lambda^{-1}p})\\
\psi_{\pm\it 1}(\bm{\Lambda^{-1}p})
\end{array}
\right).\nonumber\\
\label{unitary1}
\ee
Notice that new a type of binary indices has been introduced: ${\cal A},{\cal B}={\it 0}, {\it 1}$. They correspond to local SU(2) spinors. (Local since spinor fields taken at different points of the mass-$m$ hyperboloid transform by different SU(2) transformations.)

It is obvious from the construction that the latter {\it passive\/} \cite{PR} transformation of wave functions is equivalent
to the bispinor-field {\it acive\/} transformation $\Lambda\psi_{\tt A}(x)
=
\Lambda{_{\tt A}}{^{\tt B}}\psi_{\tt B}(\Lambda^{-1}x)$ of {\it solutions of the Dirac equation\/}. One can explicitly check \cite{MC-BW}
that the matrix in (\ref{unitary1}) is unitary and has unit determinant, and that the map $f(s,\bm p)\to \Lambda f(s,\bm p)$ is a representation of SL(2,C). The choice of $f(s,\bm p)$ is completely arbitrary and unrelated to the Dirac equation itself, but the equation is encoded in the form of the matrix occurring in (\ref{unitary1}) ({\it via\/} its implicit dependence on `sign-of-energy' projectors). Translations $x^a\to x^a+y^a$ are represented at the level of wave functions by $f(s,\bm p)\to e^{ip\cdot y}f(s,\bm p)$. The map $f(s,\bm p)\to e^{ip\cdot y}\Lambda f(s,\bm p)$ is thus nothing else but the {\it unitary\/} spin-1/2, mass-$m$ {\it representation\/} of inhomogeneous SL(2,C). The massless limit $m=0$ is trivially obtained from (\ref{unitary1}) and shows that the representation splits for $m=0$ into direct sum of two representations.

The above properties are well known in the context of induced representations of inhomogeneous SL(2,C) \cite{Wigner}: The case $m>0$ corresponds to the little group SU(2) of $p^{\bm a}=(m,\bm 0)$; for $m=0$ the little group is E(2) and its irreducible discrete-spin representations are 1-dimensional. The formulation I have presented in this section depends on two crucial technical elements, the decomposition (\ref{p}) and the choice of $t^a=\omega^a$, but does not use the idea of induction from little groups.

Local SU(2) spinors associated with projections of $W^a$ on $t^a$ are termed the $t$-spinors. A link between general $t$-spinors and the $\omega$-spinors is given by an SU(2) transformation whose explicit form can be found in \cite{MC-BW}. For a non-null $t^a$ the transformation is much more cumbersome than the case of null $\omega^a$, but choosing $t^{\bm a}=(1/|\bm p|,\bm 0)$ we reconstruct the standard helicity formalism \cite{Weinberg}.
The direct sum (\ref{t-bispinor}) of particle and anti-particle representations can be called an $\omega$-bispinor.

\subsection{Pauli-Lubanski vector associated with (\ref{unitary1})}

It pays to compute explicitly the Pauli-Lubanski vector corresponding to (\ref{unitary1}),
\be
W_a(\bm p){_{\cal A}}{^{\cal B}}
&=&
p^b {^*}i\frac{\partial}{\partial{}y^{ab}}
\left(
\begin{array}{cc}
{\bar\omega}{_{A'}}(\bm{p})\overline{\Lambda\pi}{^{A'}}(\bm{p})  & -\frac{m}{\sqrt{2}}{\bar\omega}{_{A'}}(\bm{p})\overline{\Lambda\omega}{^{A'}}(\bm{p})
\\
\frac{m}{\sqrt{2}}\omega{_{A}}(\bm{p})\Lambda \omega{^{A}}(\bm{p})  &
{\omega}{_{A}}(\bm{p}){\Lambda\pi}{^{A}}(\bm{p})
\end{array}
\right)\Bigg|_{{}y^{ab}=0}\nonumber\\
&=&
-\left(
\begin{array}{cc}
{\bar\omega}{^{R'}}(\bm{p})W_a(\bm p){_{R'}}{^{S'}}\overline{\pi}{_{S'}}(\bm{p})
&
-\frac{m}{\sqrt{2}}{\bar\omega}{^{R'}}(\bm{p})W_a(\bm p){_{R'}}{^{S'}}\overline{\omega}{_{S'}}(\bm{p})
\\
\frac{m}{\sqrt{2}}\omega{^{R}}(\bm{p})W_a(\bm p){_{R}}{^{S}}\omega{_{S}}(\bm{p})
&
{\omega}{^{R}}(\bm{p})W_a(\bm p){_{R}}{^{S}}{\pi}{_{S}}(\bm{p})
\end{array}
\right)\nonumber\\
&=&
\frac{1}{2}\left(
\begin{array}{cc}
\pi{_{A}}(\bm p)\bar\pi{_{A'}}(\bm p)-\frac{m^2}{2}\omega{_{A}}(\bm p)\bar\omega{_{A'}}(\bm p)
&
-m\sqrt{2}\pi_{A}(\bm p)\bar\omega{_{A'}}(\bm p)
\\
-m\sqrt{2}\omega_{A}(\bm p)\bar\pi{_{A'}}(\bm p)
&
\frac{m^2}{2}\omega{_{A}}(\bm p)\bar\omega{_{A'}}(\bm p)
-\pi{_{A}}(\bm p)\bar\pi{_{A'}}(\bm p)
\end{array}
\right)\nonumber\\
\label{W_ARS}
\ee
The projection
\be
\omega{^{A}}(\bm p)\bar\omega{^{A'}}(\bm p)W_a(\bm p){_{\cal A}}{^{\cal B}}
&=&
W(\omega,\bm p){_{\cal A}}{^{\cal B}}
=
\frac{1}{2}
\left(
\begin{array}{cc}
1 & 0
\\
0  & -1
\end{array}
\right),
\ee
is the same for both $m=0$ and $m>0$. For $m\neq 0$, one can employ
\be
\pi{^{A}}(\bm p)\bar\omega{^{A'}}(\bm p)W_a(\bm p){_{\cal A}}{^{\cal B}}
&=&
\frac{m}{\sqrt{2}}
\left(
\begin{array}{cc}
0 & 0
\\
1  & 0
\end{array}
\right),\\
\omega{^{A}}(\bm p)\bar\pi{^{A'}}(\bm p)W_a(\bm p){_{\cal A}}{^{\cal B}}
&=&
\frac{m}{\sqrt{2}}
\left(
\begin{array}{cc}
0 & 1
\\
0  & 0
\end{array}
\right),
\ee
to define relativistically invariant analogs of the remaining two Pauli matrices,

The massless case,
\be
W_a(\bm k){_{\cal A}}{^{\cal B}}
&=&
\frac{1}{2}\left(
\begin{array}{cc}
1
&
0
\\
0
&
-1
\end{array}
\right)k_a
\label{W_ARSm=0}
\ee
explains why projections of the Pauli-Lubanski vectors cannot violate the Bell inequality if $m=0$: All such observables commute with one another.
The ultrarelativistic limit of massive particles is more complicated \cite{MC-PRA,CRW1}.

\subsection{Peres-Scudo-Terno problem and an $\omega$-spinor way to circumvent it}

True relativistic qubits are, by definition, fields that transform unitarily under inhomogeneous SL(2,C), i.e. the $t$-spinors, or the $t$-bispinors. Let us concentrate on the $t$-spinor case. The corresponding Hilbert space is equipped with the scalar product
\be
\langle f_1|f_2\rangle
&=&
\sum_{s=\pm}\int dp\,
\overline{f_1(s,\bm p)}f_2(s,\bm p).
\ee
The expression
$
\rho(s,\bm p;s',\bm p')=f(s,\bm p)\overline{f(s',\bm p')}
$
is an example of a pure-state one-particle density matrix of a relativistic qubit. The qubit is characterized by two quantum numbers --- spin $s$ and momentum $\bm p$.
Peres, Scudo and Terno  posed the following problem \cite{PST}: Assume that our detectors do not distinguish between different $\bm p$, but detect the sign $s$ in $f(s,\bm p)$. Can we use this $s$ as a quantum bit? The probability of finding a given $s$ is given by $\int dp\,\rho(s,\bm p;s,\bm p)$. It is therefore justified to introduce the reduced spin density matrix $\rho(s,s')=\int dp\,\rho(s,\bm p;s',\bm p)$. The problem is that the entropy of $\rho(s,s')$ is not relativistically invariant. Indeed, let $f(s,\bm p)=F(s)G(\bm p)$, where $\int dp\,|G(\bm p)|^2=1$, $\sum_s|F(s)|^2=1$, and $\rho(s,s')=F(s)\overline{F(s')}$. The wave function $F(s)$ is a pure state so the entropy of $\rho(s,s')$ is zero. Now, the matrix $U(\Lambda,\bm p){_{\cal A}}{^{\cal B}}$ depends on momenta in a highly nontrivial way, and its form is even more complicated if one works with helicity or other $t$-spinors. In consequence, $\Lambda f(s,\bm p)$ does not separate into a product $\Lambda F(s)\Lambda G(\bm p)$, and thus involves a nontrivial entanglement between $s$ and $\bm p$. But then it is well known that the reduced density matrix $\Lambda\rho(s,s')$ is mixed: Its entropy is nonzero. The result has attracted attention of many authors and various variations on the theme can be found in the literature. With the exception of \cite{MCMW,MC05} apparently all the authors work in the helicity formalism.

But we know that qubits can be defined in terms of all $t$-spinors --- there is nothing fundamental about helicity, a $t$-spinor associated with direction parallel to some $t^{\bm a}=(1,0,0,0)$. In particular, let us take an arbitrary null direction $t^a=\tau^A\bar\tau^{A'}$. If $p^2=m^2>0$ then $p\cdot t$ is nonzero for all $p^a$. The null vector $\omega^a=\tau^A\bar\tau^{A'}/(p\cdot t)$ satisfies $p\cdot \omega=1$, so $\omega^A(\bm p)
=\tau^A/\sqrt{p\cdot t}$ can be used to split 4-momentum $p^a$ according to (\ref{p}), and the whole construction of $\omega$-spinors can be repeated.
Now, let $\Lambda{_A}{^B}$ be an arbitrary SL(2,C) transformation. It is known (Sec.~3.6 in \cite{PR}) that any SL(2,C) transformation possess at least one eigenvector, $\Lambda{_A}{^B}\tau_B=\lambda \tau_A$, $\lambda=|\lambda|e^{i\varphi}$, so let us take this $\tau_A$ in our definition of $t^a$. Then,
\be
\Lambda{_A}{^B}\omega_B(\bm{\Lambda^{-1}p})
&=&
\frac{\Lambda{_A}{^B}\tau_B}{\sqrt{\Lambda^{-1}p\cdot t}}=
\frac{\lambda\tau_A}{\sqrt{p\cdot \Lambda t}}=
e^{i\varphi}\omega_A(\bm p),\\
\Lambda{_A}{^B}\pi_B(\bm{\Lambda^{-1}p})
&=&
\frac{\Lambda{_A}{^B}(\Lambda^{-1}p){_B}{^{B'}}\bar\tau{_{B'}}}{{\sqrt{\Lambda^{-1}p\cdot t}}}
=
\frac{\Lambda{_A}{^B}\Lambda^{-1}{_B}{^C}p{_C}{^{B'}}\Lambda{_{B'}}{^{C'}}\bar\tau{_{C'}}}{{\sqrt{p\cdot \Lambda t}}}\nonumber\\
&=&
\frac{p{_A}{^{B'}}\bar\lambda\bar\tau{_{B'}}}{{|\lambda|\sqrt{p\cdot t}}}=e^{-i\varphi}\pi_A(\bm{p}).
\ee
The transformation matrix becomes
\be
&{}&
\left(
\begin{array}{cc}
{\bar\omega}{_{A'}}(\bm{p})\overline{\Lambda\pi}{^{A'}}(\bm{p})  & -\frac{m}{\sqrt{2}}{\bar\omega}{_{A'}}(\bm{p})\overline{\Lambda\omega}{^{A'}}(\bm{p})
\\
\frac{m}{\sqrt{2}}\omega{_{A}}(\bm{p})\Lambda \omega{^{A}}(\bm{p})  &
{\omega}{_{A}}(\bm{p}){\Lambda\pi}{^{A}}(\bm{p})
\end{array}
\right)
\nonumber\\
&{}&\pp==
\left(
\begin{array}{cc}
e^{i\varphi}{\bar\omega}{_{A'}}(\bm{p})\overline{\pi}{^{A'}}(\bm{p})  & -e^{-i\varphi}\frac{m}{\sqrt{2}}{\bar\omega}{_{A'}}(\bm{p})\overline{\omega}{^{A'}}(\bm{p})
\\
e^{i\varphi}\frac{m}{\sqrt{2}}\omega{_{A}}(\bm{p})\omega{^{A}}(\bm{p})  &
e^{-i\varphi}{\omega}{_{A}}(\bm{p}){\pi}{^{A}}(\bm{p})
\end{array}
\right)
=
\left(
\begin{array}{cc}
e^{i\varphi}  & 0
\\
0  &
e^{-i\varphi}
\end{array}
\right),
\ee
which is independent of $\bm p$. Actually, the transformation matrix is simply an ordinary rotation by $2\varphi$ which, of course, cannot change any entanglement. The reduced density matrix of Peres, Scudo and Terno has not changed its entropy, even though the SL(2,C) transformation was arbitrary.

The construction I have just presented was introduced in \cite{MCMW}, and can be regarded as a relativistic error correction algorithm. What it means is that it is always possible to adjust $t^a$ in $t^a W_a$ in a way that ensures that qubits defined by this concrete observable do not change their Peres-Scudo-Terno entropy. This trick can be generalized to any number of particles that propagate in arbitrary ways, not necessarily  inertially, and described by any states --- entangled or not. Another consequence is that entanglement between Peres-Scudo-Terno qubits will not change, no matter what kind of motion is considered, if one defines qubits by means of $t^a$ which are principal null directions \cite{PR,PR2} of Lorentz transformations that define the motion of observers.

\section{Transformation properties of fields of spin frames for massless particles}

Let us consider a massless particle whose 4-momentum $k^a$ is a future-pointing null but nonzero world-vector, $k^ak_a=0$, $k^a\neq 0$. We know that such a $k^a$ determines, up to a phase, a 2-spinor $\pi^A$ satisfying $k^a=\pi^A\bar\pi^{A'}$. The latter equation can be understood in several ways. The approach I prefer takes the future light cone as a given 3-dimensional manifold on which a spinor field $\pi^A=\pi^A(\bm k)$ is defined, but treats this spinor field as a fundamental physical object related to 4-momentum by $k^a=\pi^A(\bm k)\bar\pi^{A'}(\bm k)$. A consequence is apparently paradoxical. The spinor field transforms under SL(2,C) transformations as
$\pi_A(\bm k)\mapsto\Lambda\pi_A(\bm k)=\Lambda{_A}{^B}\pi_B(\bm{\Lambda}^{-1}\bm k)$. But then $\Lambda\pi_A(\bm k)\overline{\Lambda\pi}_{A'}(\bm k)=\Lambda{_a}{^b}\Lambda{^{-1}}{_b}{^c}k_c=k_a$. Accordingly, the 4-momentum does not change. This type of approach is consistent with the idea that the fundamental fields are those of 2-spinors and not the 4-vectors. An important consequence is that both $\Lambda\pi_A(\bm k)$ and $\pi_A(\bm k)$ possess the same flagpole and, thus, differ at most by a phase factor: $\Lambda\pi_A(\bm k)=e^{-i\Theta(\Lambda,\bm k)}\pi_A(\bm k)$. For any spin-frame partner $\omega_A(\bm k)$ of $\pi_A(\bm k)$, we find $\omega_A(\bm k)\Lambda\pi^A(\bm k)=e^{-i\Theta(\Lambda,\bm k)}$ and
$\bar\omega{_{A'}}(\bm k)\overline{\Lambda\pi}{^{A'}}(\bm k)=e^{i\Theta(\Lambda,\bm k)}$. The latter two expressions have been encountered before in the diagonal elements of the matrix in (\ref{unitary1}).

Now let us consider the `orbital' part of the 6-angular momentum antisymmetric tensor of the massless particle: $M_{ab}=x_a k_b-x_bk_a$. Any antisymmetric (real) second-rank tensor can be written as (Eq. (3.4.20) in \cite{PR})
\be
M_{ab} &=& \mu_{AB}\ve_{A'B'}+\ve_{AB}\bar\mu_{A'B'},\\
\mu_{AB} &=& \frac{1}{2}M_{AA'BB'} \ve^{A'B'}=\mu_{BA},\\
\bar\mu_{A'B'} &=& \overline{\mu_{AB}},
\ee
the relation between $M_{ab}$ and $\mu_{AB}$ being expressible also by means of the generator of $\Lambda{_A}{^B}$:
\be
M{^{ab}}\sigma_{ab}{_R}{^S}
&=&
M{^{ab}}
\frac{i}{2}(g_{aRX'}g_b{^{SX'}} - g_{bRX'}g_a{^{SX'}})
=i\, M{^{ab}}g_{aRX'}g_b{^{SX'}}\nonumber\\
&=&
i\, M_{RX'}{^{SX'}}=2i\,\mu_{R}{^{S}}.
\ee
Let us apply this procedure to $M_{ab}$:
\be
\mu_{AB}
&=&
\frac{1}{2}(x_a k_b-x_bk_a)\ve^{A'B'}\nonumber\\
&=&
\frac{1}{2}(x_{AA'} \pi_B\bar\pi_{B'}-x_{BB'}\pi_A\bar\pi_{A'})\ve^{A'B'}
\nonumber\\
&=&
\frac{1}{2}(\underbrace{x_{AA'}\bar\pi^{A'}}_{\omega_A} \pi_B\underbrace{x_{BB'}\bar\pi^{B'}}_{\omega_B}\pi_A)
\nonumber\\
&=&
\frac{1}{2}(\omega_A\pi_B+\omega_B\pi_A).\label{W W W omega}
\ee
The orbital part thus turns out to be determined by two spinors: $\pi_A$ whose flagpole is the 4-momentum, and $\omega_A=\omega_A(x,\pi)=x_{AA'}\bar\pi^{A'}$.

The bispinor $(\omega_A,\bar\pi_{A'})$, with the characteristic bilinear dependence of $\omega_A$ on $x$ and $\bar\pi$, is known as a {\it twistor} \cite{PR2}.
Treating $\pi_A$ as a spinor field $\pi_A(\bm k)$ we obtain $\omega_A=\omega_A(x,\bm k)=x_{AA'}\bar\pi^{A'}(\bm k)$ which is a field defined on the Cartesian product of the Minkowski space with the momentum-space light cone. The field transforms under SL(2,C) transformations as follows
\be
\omega_{A}(x,\bm k)
&\to&
\Lambda{_A}{^B}\omega_B(\Lambda^{-1}x,{\bm \Lambda}^{-1}{\bm k})
\nonumber\\
&=&
\Lambda{_A}{^B}\Lambda^{-1}{_B}{^C}\Lambda^{-1}{_{B'}}{^{C'}}x_{CC'}\bar\pi^{B'}({\bm \Lambda}^{-1}{\bm k})
\nonumber\\
&=&
\ve{_A}{^C}x_{CC'}\bar\pi^{B'}({\bm \Lambda}^{-1}{\bm k})\Lambda^{-1}{_{B'}}{^{C'}}
\nonumber\\
&=&
x_{AC'}\overline{\Lambda\pi}{^{C'}}({\bm k})
\nonumber\\
&=&
x_{AA'}e^{i\Theta(\Lambda,\bm k)}\bar\pi{^{A'}}({\bm k})
\nonumber\\
&=&
e^{i\Theta(\Lambda,\bm k)}\omega_A(x,\bm k).
\ee
The contraction
\be
\omega_A(x,\bm k)\pi^A(\bm k)
&=&
x_{AA'}\pi^A(\bm k)\bar\pi^{A'}(\bm k)=x_ak^a=x\cdot k
\ee
shows that for $x_ak^a\neq 0$ we can define a spin-frame field
\be
\ii_A(\bm k) &=& \pi_A(\bm k),\label{!1}\\
o_A(x,\bm k) &=& \frac{\omega_A(x,\bm k)}{x\cdot k}=\frac{x_{AA'}\bar\pi^{A'}(\bm k)}{x\cdot k},
\ee
satisfying simultaneously
\be
o_A(x,\bm k)\ii^A(\bm k) &=& 1,
\\
\ii_A(\bm k)\bar\ii_{A'}(\bm k) &=& k_a,\\
\Lambda{_A}{^B}\ii_B(\bm{\Lambda}^{-1}\bm k)
&=&
e^{-i\Theta(\Lambda,\bm k)}\ii_A(\bm k),\\
\Lambda{_A}{^B}o_B(\Lambda^{-1}x,{\bm \Lambda}^{-1}{\bm k})
&=&
e^{i\Theta(\Lambda,\bm k)}o_A(x,\bm k).\label{!2}
\ee
It is an instructive exercise to check what would have been the consequences of assuming that
$x_{AA'}\bar\pi^{A'}(\bm k)/(x\cdot k)$ is a spinor field defined on the light cone only, and not on the `phase space' $(x,\bm k)$. Then we would have to assume the transformation rule
\be
\Lambda o_A({\bm k})
&=&
\Lambda{_A}{^B}o_B({\bm \Lambda}^{-1}{\bm k})
=
e^{i\Theta(\Lambda,\bm k)}\Lambda x_{AA'}\bar\pi^{A'}(\bm k)/(\Lambda x\cdot k).
\label{wrong}
\ee
The right side of (\ref{wrong}) is not proportional to $o_A({\bm k})$ but also contains a component parallel to $\ii_A({\bm k})$.
In effect, we find
\be
\Lambda o_A({\bm k})
&=&
\Lambda{_A}{^B}o_B({\bm \Lambda}^{-1}{\bm k})
=
e^{i\Theta(\Lambda,\bm k)}\tilde o_A({\bm k})\label{tilde o}
\ee
where, in spite of the fact that $\tilde o_A({\bm k})\ii^A({\bm k})=1$, $\tilde o_A({\bm k})=o_A({\bm k})+\phi(\Lambda,\bm k)\ii^A({\bm k})$.
The appearance of the function $\phi(\Lambda,\bm k)=o_A({\bm k})\tilde o^A({\bm k})$, which is different for different explicit forms of the spin frame, is responsible, as we shall see later, for a gauge transformation of the electromagnetic four-potential and has consequences for field quantization.
This change of gauge seems to be a generic property of spin frames defined on mass-$m$ hyperboloids.

However, we will see that there are reasons to believe that Eq.~(\ref{!2}) is of fundamental importance for relativistic transformations of electromagnetic qubits, but one should not identify the point $x^a$ with an {\it event\/} in Minkowski space. A more natural interpretation arises if one treats the twistor as defined with respect to some internal coordinate, say $R^a$, which is timelike (and thus $R\cdot k$ is nowhere vanishing if $k^a\neq 0$). At the level of quantized electromagnetic field the approach will lead to an intriguing splitting of the field in momentum space into three massless fields: one field transforming according to the unitary spin-1 representation of the Poincar\'e group, and two scalars. Photon degrees of freedom will be defined with respect to the spin-1 part.

\section{Qubits associated with electromagnetic field}

Qubits associated with electromagnetic field can be discussed at both first- and second-quantized levels. In this section we concentrate on the first-quantized description, which is important since transformation properties of polarization degrees of freedom will remain unchanged after field quantization.

Electromagnetic field tensor is antisymmetric, so can be decomposed in the usual way: $F_{ab}(x)=F_{AB}(x)\ve_{A'B'}+\ve_{AB}\bar F_{A'B'}(x)$, $F_{A'B'}(x)=\overline{F_{AB}(x)}$, $F_{AB}(x)=F_{BA}(x)$. Source-free Maxwell equations, $\partial^a F_{ab}(x)=\partial^a {^*}F_{ab}(x)=0$, are equivalent to the spinor equation
\be
\partial^{AA'}F_{AB}(x)=0=\partial^{AA'}F_{A'B'}(x).
\ee
In Fourier space
\be
F_{AB}(x)
=
\int dk\,\Big(F_{-AB}(\bm k)e^{-ik\cdot x}+F_{+AB}(\bm k)e^{ik\cdot x}\Big),\quad
F_{\pm AB}(\bm k)\pi{^{A}}(\bm k)\bar\pi{^{A'}}(\bm k)
=
0.\nonumber\\
\label{F-AB}
\ee
Expanding $F_{\pm AB}(\bm k)$ in terms of spin frames we find that the only symmetric solution of (\ref{F-AB}) is $F_{\pm AB}(\bm k)\sim
\pi{_{A}}(\bm k)\pi{_{B}}(\bm k)$. The field spinor can be thus written as
\be
F_{AB}(x)
&=&
-\int dk\,
\pi_A(\bm k)\pi_B(\bm k)
\Big(\alpha(-,\bm k)e^{-ik\cdot x}
+\overline{\alpha(+,\bm k)}e^{ik\cdot x}\Big),\label{F AB1}\\
F_{A'B'}(x)
&=&
-\int dk\,
\bar\pi_{A'}(\bm k)\bar\pi_{B'}(\bm k)
\Big(\alpha(+,\bm k)e^{-ik\cdot x}
+\overline{\alpha(-,\bm k)}e^{ik\cdot x}\Big).\label{F AB2}
\ee
The signs and conjugations are a matter of convention.
Similarity of (\ref{F AB1})-(\ref{F AB2}) to the zero-mass solution of Dirac's equation (\ref{psi m=01})-(\ref{psi m=02}) is not accidental and is typical of massless Bargmann-Wigner equations of any spin \cite{MC-BW,Woodhouse}. SL(2,C) transformations of (\ref{F AB1})-(\ref{F AB2}) can be deduced from analogous formulas found for the Dirac field,
\be
&{}&
\Lambda{_{A}}{^{C}}\Lambda{_{B}}{^{D}}F_{CD}(\Lambda^{-1}x)
\nonumber\\
&{}&\pp==
-\int dk\,
\pi_A(\bm k)\pi_B(\bm k)
e^{-2i\Theta(\Lambda,\bm k)}\Big(\alpha(-,\bm{\Lambda^{-1}k})e^{-ik\cdot x}
+\overline{\alpha(+,\bm{\Lambda^{-1}k})}e^{ik\cdot x}\Big),\nonumber\\
\label{F AB1'}\\
&{}&
\Lambda{_{A'}}{^{C'}}\Lambda{_{B'}}{^{D'}}F_{C'D'}(\Lambda^{-1}x)
\nonumber\\
&{}&\pp==
-\int dk\,
\bar\pi_{A'}(\bm k)\bar\pi_{B'}(\bm k)
e^{2i\Theta(\Lambda,\bm k)}\Big(\alpha(+,\bm{\Lambda^{-1}k})e^{-ik\cdot x}
+\overline{\alpha(-,\bm{\Lambda^{-1}k})}e^{ik\cdot x}\Big).\nonumber\\
\label{F AB2'}
\ee
\subsection{Four-potential $A_a(x)$ based on spin frames defined on the light cone}

Four-potential is defined by
\be
{F}_{ab}(x)
=
\partial_a A_b(x)-\partial_b A_a(x)
\label{F}.
\ee
We know that given $\pi^A(\bm k)$ we cannot find a unique spin-partner $\omega^A(\bm k)$, since
for any function $\phi(\bm k)$ the new field
\be
\tilde \omega^A(\bm k)=\omega^A(\bm k)+\phi(\bm k)\pi^A(\bm k)
\label{sim}
\ee
also satisfies
$\tilde \omega_A(\bm k)\pi^A(\bm k)=1$. This leads to the equivalence relation:
$\tilde \omega^A(\bm k)\sim \omega^A(\bm k)$ iff
$\tilde \omega^A(\bm k)-\omega^A(\bm k)$ is proportional to
$\pi^A(\bm k)$.

The 4-potential in a Lorenz gauge\footnote{Note: `Lorenz gauge' (L. Lorenz) and  `Lorentz group' (H. A. Lorentz).}, can be taken in the form
(cf.~\cite{Ashtekar,MCJN})
\be
A_a(x) &=& i\int dk\,e^{-ik\cdot x}
\Big(
\omega_A(\bm k)\bar\pi_{A'}(\bm k)
\alpha(+,\bm k)
+
\pi_{A}(\bm k)\bar\omega_{A'}(\bm k)
\alpha(-,\bm k)
\Big)
+ {\,\rm c.c.}\label{A}
\ee
Now, if we replace $\omega^A(\bm k)$ by $\tilde \omega^A(\bm k)$
satisfying (\ref{sim}),
then
\be
{A}_a(x) &\mapsto&
\tilde {A}_a(x)
=
{A}_a(x)
-
\partial_a
\Phi(x)\label{tilde omega}
\ee
where
\be
\Phi(x)
&=&
\int dk\,
\phi(\bm k)
\Big(\alpha(+,\bm k)e^{-ik\cdot x}
+
\overline{\alpha(-,\bm k)}e^{ik\cdot x}
\Big)
+\,{\rm c.c.}
\ee
is a solution of $\partial^a\partial_a \Phi(x)=0$. The equivalence
class of spin-frames corresponds to an equivalence class of
Lorenz-gauge potentials.

(\ref{F AB1'})-(\ref{F AB2'}) show the duality between active spinor transformations and passive zero-mass, spin-1 unitary transformations of wave functions. An analogous construction does not exactly work for $A_a(x)$. Let us have a closer look at this important subtlety since it has nontrivial implications for field quantization.

The problem is with changes of gauge of the form (\ref{tilde omega}) and (\ref{tilde o}):
\be
\Lambda{_a}{^b}A_b(\Lambda^{-1}x)
&=& i\int dk\,e^{-ik\cdot x}
e^{2i\Theta(\Lambda,\bm k)}
\tilde\omega_A(\bm k)\bar\pi_{A'}(\bm k)
\alpha(+,\bm{\Lambda^{-1} k})
+\dots\nonumber
\\
&=&i\int dk\,e^{-ik\cdot x}
e^{2i\Theta(\Lambda,\bm k)}
\omega_A(\bm k)\bar\pi_{A'}(\bm k)
\alpha(+,\bm{\Lambda^{-1} k})
+
\partial_a\Phi(x)+\dots\nonumber
\\
\ee
In consequence, if wave functions transform according to massless spin-1 unitary representations of inhomogeneous SL(2,C), then
\be
A_a(x)\to \Lambda{_a}{^b}A_b(\Lambda^{-1}x)-\partial_a\Phi(x),
\ee
meaning that $A_a(x)$ is not a four-vector field. And the other way around, if we assume that the four-potential is a four-vector field, its corresponding wave functions do not carry massless spin-1 unitary representations.

\subsection{Null vs. Minkowski tetrads, or circular vs. linear polarizations}

The polarization world vectors occurring in (\ref{A}) can be regarded as elements of a null tetrad in Minkowski space (we skip the arguments $\bm k$):
\be
\left(
\begin{array}{c}
g^{a}_{~~00'}\\
g^{a}_{~~01'}\\
g^{a}_{~~10'}\\
g^{a}_{~~11'}
\end{array}
\right) =
\left(
\begin{array}{c}
\varepsilon^{A}_{~~0}\varepsilon^{A'}_{~~0'}\\
\varepsilon^{A}_{~~0}\varepsilon^{A'}_{~~1'}\\
\varepsilon^{A}_{~~1}\varepsilon^{A'}_{~~0'}\\
\varepsilon^{A}_{~~1}\varepsilon^{A'}_{~~1'}
\end{array}
\right)=
\left(
\begin{array}{c}
\omega^{A}\bar\omega^{A'}\\
\omega^{A}\bar\pi^{A'}\\
\pi^{A}\bar\omega^{A'}\\
\pi^{A}\bar\pi^{A'}
\end{array}
\right)
=
\left(
\begin{array}{c}
\omega^{a}\\
m^{a}\\
\bar{m}^{a}\\
k^{a}
\end{array}
\right)\label{4}
\end{eqnarray}
and dually
\begin{eqnarray}
 \left(
\begin{array}{c}
g_{a}^{~~00'}\\
g_{a}^{~~01'}\\
g_{a}^{~~10'}\\
g_{a}^{~~11'}
\end{array}
\right) =
\left(
\begin{array}{c}
\varepsilon_{A}^{~~0}\varepsilon_{A'}^{~~0'}\\
\varepsilon_{A}^{~~0}\varepsilon_{A'}^{~~1'}\\
\varepsilon_{A}^{~~1}\varepsilon_{A'}^{~~0'}\\
\varepsilon_{A}^{~~1}\varepsilon_{A'}^{~~1'}
\end{array}
\right)
=
\left(
\begin{array}{c}
\pi_{A}\bar\pi_{A'}\\
-\pi_{A}\bar\omega_{A'}\\
-\omega_{A}\bar\pi_{A'}\\
\omega_{A}\bar\omega_{A'}
\end{array}
\right)
=
\left(
\begin{array}{c}
k_{a}\\
-\bar{m}_{a}\\
-m_{a}\\
\omega_{a}
\end{array}
\right)\label{5}
\end{eqnarray}
The associated Minkowski tetrad
can be expressed by means of Infeld-van der Waerden symbols,
$g^{a}{_{\bf{a}}}
=
g_{\bf{a}}{^{\bf{BB}'}}g^{a}{_{\bf{BB}'}}
$,
\begin{eqnarray}
\left(
\begin{array}{c}
t^{a}\\
x^{a}\\
y^{a}\\
z^{a}
\end{array}
\right) = \left(
\begin{array}{c}
g^{a}_{~~0}\\
g^{a}_{~~1}\\
g^{a}_{~~2}\\
g^{a}_{~~3}
\end{array}
\right) = \frac{1}{\sqrt{2}} \left(
\begin{array}{cccc}
1&0&0&1\\
0&1&1&0\\
0&i&-i&0\\
1&0&0&-1
\end{array}
\right) \left(
\begin{array}{c}
\omega^{A}\bar\omega^{A'}\\
\omega^{A}\bar\pi^{A'}\\
\pi^{A}\bar\omega^{A'}\\
\pi^{A}\bar\pi^{A'}
\end{array}
\right)
=
 \frac{1}{\sqrt{2}}
 \left(
\begin{array}{c}
\omega^{a}+k^{a}\\
m^{a}+\bar{m}^{a}\\
i m^{a} - i\bar{m}^{a}\\
\omega^{a}-k^{a}
\end{array}
\right)
\end{eqnarray}
and dually,
$g_{a}{^{\bf{a}}}
=
g^{\bf{a}}{_{\bf{BB}'}}g_{a}{^{\bf{BB}'}}
$,
\begin{eqnarray}
\left(
\begin{array}{c}
t_{a}\\
-x_{a}\\
-y_{a}\\
-z_{a}
\end{array}
\right) = \left(
\begin{array}{c}
g_{a}^{~~0}\\
g_{a}^{~~1}\\
g_{a}^{~~2}\\
g_{a}^{~~3}
\end{array}
\right) = \frac{1}{\sqrt{2}} \left(
\begin{array}{cccc}
1&0&0&1\\
0&1&1&0\\
0&-i&i&0\\
1&0&0&-1
\end{array}
\right) \left(
\begin{array}{c}
\pi_{A}\bar\pi_{A'}\\
-\pi_{A}\bar\omega_{A'}\\
-\omega_{A}\bar\pi_{A'}\\
\omega_{A}\bar\omega_{A'}
\end{array}
\right)=
 \frac{1}{\sqrt{2}}
 \left(
\begin{array}{c}
k_{a}+\omega_{a}\\
-\bar{m}_{a}-m_{a}\\
i \bar{m}_{a} -i m_{a}\\
k_{a}-\omega_{a}
\end{array}
\right)\nonumber
\end{eqnarray}
The meaning of $x_a$ and $y_a$ can be deduced from
\be
x_a(\bm k)
\alpha_1(\bm k)
+
y_a(\bm k)
\alpha_2(\bm k)
&=&
m_a(\bm k)
\alpha(+,\bm k)
+
\bar m_a(\bm k)
\alpha(-,\bm k),\\
\alpha(\pm,\bm k)
&=&
\frac{1}{\sqrt{2}}\Big(\alpha_1(\bm k)\pm i\,\alpha_2(\bm k)\Big).
\ee
Elements $m_a(\bm k)$, $\bar m_a(\bm k)$ of the null tetrad correspond to two circular polarizations (since after SL(2,C) transformations they get multiplied by appropriate Wigner phase factors). The two elements of the Minkowski tetrad, $x_a(\bm k)$, $y_a(\bm k)$, define linear polarizations,
\be
A_a(x) &=& i\int dk\,e^{-ik\cdot x}
\Big(
g{_{a}}{^{1}}(\bm k)
\alpha_1(\bm k)
+
g{_{a}}{^{2}}(\bm k)
\alpha_2(\bm k)
\Big)
+ {\,\rm c.c.}\label{A L}
\ee

\subsection{$A_a(x)$ based on twistor-like spin-frames}

Let $R^2=1$, $R_0>0$, $dR=d^3R/[(2\pi)^3 2\sqrt{1+\bm R^2}]$, and consider spin-frames analogous to (\ref{!1})-(\ref{!2}):
\be
\omega_A(R,\bm k)\pi^A(\bm k) &=& 1,\label{!1'}
\\
\pi_A(\bm k)\bar\pi_{A'}(\bm k) &=& k_a,\\
\Lambda{_A}{^B}\pi_B(\bm{\Lambda}^{-1}\bm k)
&=&
e^{-i\Theta(\Lambda,\bm k)}\pi_A(\bm k),\\
\Lambda{_A}{^B}\omega_B(\bm{\Lambda^{-1}R},{\bm \Lambda}^{-1}{\bm k})
&=&
e^{i\Theta(\Lambda,\bm k)}\omega_A(\bm R,\bm k).\label{!2'}
\ee
The world-vector $R^a$ has properties of four-velocity. In this sense it is analogous to the four-velocity parameter occurring in theories with preferred reference frame or based on non-standard clock synchronization schemes \cite{RS,R,CR2}. Another interesting analogy is with fields defined by Kaiser on phase-space \cite{Kaiser,Kjmp} --- here $R^a$ would be an analogue of the temper vector.
The four-potential
\be
A_a(x) &=&
i\int dk\,dR\,e^{-ik\cdot x}
\nonumber\\
&{}&\times
\Big(
\omega_A(\bm R,\bm k)\bar\pi_{A'}(\bm k)
\alpha(+,\bm R,\bm k)
+
\pi_{A}(\bm k)\bar\omega_{A'}(\bm R,\bm k)
\alpha(-,\bm R,\bm k)
\Big)
+ {\,\rm c.c.}\nonumber\\
&=& i\int dk\,dR\,e^{-ik\cdot x}
\Big(
g{_{a}}{^{1}}(\bm R,\bm k)
\alpha_1(\bm R,\bm k)
+
g{_{a}}{^{2}}(\bm R,\bm k)
\alpha_2(\bm R,\bm k)
\Big)
+ {\,\rm c.c.}\label{A L R}
\ee
transforms as follows
\be
\Lambda{_a}{^b}A_b(\Lambda^{-1}x)
&=&
i\int dk\,dR\,e^{-ik\cdot \Lambda^{-1}x}
\Lambda{_a}{^b}
\omega_B(\bm R,\bm k)\bar\pi_{B'}(\bm k)
\alpha(+,\bm R,\bm k)
+\dots\nonumber\\
&=&
i\int dk\,dR\,e^{-ik\cdot x}
\Lambda{_A}{^B}
\omega_B(\bm R,\bm{\Lambda^{-1}k})\overline{\Lambda\pi}{_{B'}}(\bm k)
\alpha(+,\bm R,\bm{\Lambda^{-1}k})
+\dots\nonumber\\
&=&
i\int dk\,dR\,e^{-ik\cdot x}
e^{2i\Theta(\Lambda,\bm k)}
\omega_A(\bm{R},\bm{k})\bar\pi{_{A'}}(\bm k)
\alpha(+,\bm{\Lambda^{-1}R},\bm{\Lambda^{-1}k})
+\dots\nonumber\\
\ee
Note that linear polarization vectors are elements of the Minkowski tetrad transforming as follows:
\be
\left(
\begin{array}{c}
\Lambda t^{a}(\bm R,\bm k)\\
\Lambda x^{a}(\bm R,\bm k)\\
\Lambda y^{a}(\bm R,\bm k)\\
\Lambda z^{a}(\bm R,\bm k)
\end{array}
\right)
&=&
\frac{1}{\sqrt{2}} \left(
\begin{array}{cccc}
1&0&0&1\\
0&1&1&0\\
0&i&-i&0\\
1&0&0&-1
\end{array}
\right) \left(
\begin{array}{c}
\Lambda \omega^{A}(\bm R,\bm k)\overline{\Lambda \omega}{^{A'}}(\bm R,\bm k)\\
\Lambda \omega^{A}(\bm R,\bm k)\overline{\Lambda \pi}{^{A'}}(\bm k)\\
\Lambda \pi^{A}(\bm k)\overline{\Lambda \omega}{^{A'}}(\bm R,\bm k)\\
\Lambda \pi^{A}(\bm k)\overline{\Lambda \pi}{^{A'}}(\bm k)
\end{array}
\right)\label{LLL-R}\\
&=&
\left(
\begin{array}{c}
t^a(\bm R,\bm k)
\\
\cos 2\Theta(\Lambda,\bm k) x^a(\bm R,\bm k)
+\sin 2\Theta(\Lambda,\bm k) y^a(\bm R,\bm k)
\\
-\sin 2\Theta(\Lambda,\bm k) x^a(\bm R,\bm k)
+\cos 2\Theta(\Lambda,\bm k) y^a(\bm R,\bm k)
\\
z^a(\bm R,\bm k)
\end{array}
\right)\label{LLLL-R}
\ee
$t^a(\bm R,\bm k)$ and $z^a(\bm R,\bm k)$ do not appear in (\ref{A L R}), but the space-like plane spanned by $x^a(\bm R,\bm k)$ and $y^a(\bm R,\bm k)$ is invariant under SL(2,C). So the twistor-like spin-frames lead to Minkowski tetrads that span two orthogonal subspaces, both invariant under SL(2,C). This does not hold for tetrads corresponding to spin-frames that depend only on $\bm k$.

The passive unitary representation of the amplitudes reads
\be
\Lambda\alpha(\pm,\bm{R},\bm{k})
&=&
e^{\pm 2i\Theta(\Lambda,\bm k)}
\alpha(\pm,\bm{\Lambda^{-1}R},\bm{\Lambda^{-1}k}).
\ee
There is no change of gauge. (Change of gauge means that the plane spanned by $m_a$ and $\bar m_a$ is not invariant, since a contribution parallel to $k_a$ occurs. In quantum-field-theory terms this would mean that SL(2,C) transformations mix transverse photons with `longitudinal or timelike' spin-0 `photons'; see the discussion in the next Section.)

The field spinors
\be
F_{AB}(x)
&=&
-\int dk\,dR\,
\pi_A(\bm k)\pi_B(\bm k)
\Big(\alpha(-,\bm R,\bm k)e^{-ik\cdot x}
+\overline{\alpha(+,\bm R,\bm k)}e^{ik\cdot x}\Big)\label{F AB1 R}\\
F_{A'B'}(x)
&=&
-\int dk\,dR\,
\bar\pi_{A'}(\bm k)\bar\pi_{B'}(\bm k)
\Big(\alpha(+,\bm R,\bm k)e^{-ik\cdot x}
+\overline{\alpha(-,\bm R,\bm k)}e^{ik\cdot x}\Big),\label{F AB2 R}
\ee
are identical to (\ref{F AB1})-(\ref{F AB2}) if $\int dR\,\alpha(\pm,\bm R,\bm k)=\alpha(\pm,\bm k)$. Moreover, relativistic invariance of $dR$ implies that
\be
\Lambda\alpha(\pm,\bm k)
&=&
e^{\pm 2i\Theta(\Lambda,\bm k)}
\int dR\,\alpha(\pm,\bm{\Lambda^{-1}R},\bm{\Lambda^{-1}k})
=
e^{\pm 2i\Theta(\Lambda,\bm k)}\alpha(\pm,\bm{\Lambda^{-1}k}).
\label{int dR a}
\ee
Let $\int dR\,\alpha_{j}(\bm R,\bm k)=\alpha_{j}(\bm k)$, $j=1,2$.
SL(2,C) simply rotates the amplitude 2-dimensional vector $(\alpha_{1}(\bm k), \alpha_{2}(\bm k))$ by spin-1 Wigner angle $2\Theta(\Lambda,\bm k)$.

The world-vector $R^a$ plays a role of an internal degree of freedom, invisible at the level of field tensors. Its role can be fully appreciated only at the level of quantized fields \cite{MCKW}. Let us stress that the choice of a time-like $R^a$ is perhaps not necessary --- in principle $R^a$ could be space-like or null, but only for a time-like $R^a$ the product $R\cdot k$ is, for $k^a\neq 0$, non-vanishing, which allows us to define spin-frames globally for all $k^a$ and $R^a$. The globality condition is convenient but not necessary.

\section{Qubits associated with quantized electromagnetic field}

An automatic identification of the Pauli-Lubanski vector with polarization of a quantum field leads to a problem: $W^a$ is {\it quadratic\/} in elements of Poincar\'e algebra, $W^a=P_b{^*}M^{ab}$, and thus --- in terms of number operators --- $W^a$ is quadratic as well. However, spin observables, similarly to 4-momentum and 6-angular momentum, are expected to be linear in numbers of particles. The subtlety is not so visible in first quantization since observables are there defined at 1-particle levels. A possible way out is to treat the {\it harmonic oscillator algebra\/} as the fundamental algebraic level, and treat inhomogeneous SL(2,C) as a structure derived from it. The choice of spin-fames --- possessing appropriate transformation properties --- becomes then essential for a correct relativistic formulation of the formalism.

\subsection{Manifestly covariant quantization of $A_a(x)$ based on twistor-like spin-frames}

One typically quantizes $A_a(x)$ in a way that does not guarantee the four-vector behavior of the field. An approach to relativistic EPR correlations of photons based on such a non-manifestly covariant formalism was recently discussed in \cite{Caban,Caban2}. In what follows I will outline an alternative but not widely known formulation based on tetrads built from twistor-like spin-frames \cite{MCKW}.

The potential is defined by
\be
A_a(x)
&=& i\int dk\,dR\,e^{-ik\cdot x}
\nonumber\\
&\pp=&
\times
\Big(
g{_{a}}{^{1}}(\bm R,\bm k)
a_1(\bm R,\bm k)
+
g{_{a}}{^{2}}(\bm R,\bm k)
a_2(\bm R,\bm k)
\nonumber\\
&\pp=&
\pp{\times\Big(}
+
g{_{a}}{^{3}}(\bm R,\bm k)
a_3(\bm R,\bm k)
+
g{_{a}}{^{0}}(\bm R,\bm k)
a_0(\bm R,\bm k)^{\dag}
\Big)
+ {\,\rm c.c.}
\ee
The field $g{_{a}}{^{\bf a}}(\bm R,\bm k)$, by definition of a Minkowski tetrad, satisfies $g{_{a}}{^{\bf a}}(\bm R,\bm k)g{^{a}}{^{\bf b}}(\bm R,\bm k)
=g^{\bf ab}$, $g{_{a}}{_{\bf a}}(\bm R,\bm k)g{_{b}}{^{\bf a}}(\bm R,\bm k)
=g_{ab}$.
The amplitude operators are assumed to satisfy the ordinary harmonic-oscillator Lie algebra,
\be
{[a_{\bf a}(\bm R,\bm k),a_{\bf b}(\bm R',\bm k')^{\dag}]}
&=&
\delta_{\bf ab}\delta(\bm R,\bm k,\bm R',\bm k')I(\bm R,\bm k),\label{ccr1}\\
{[a_{\bf a}(\bm R,\bm k),n_{\bf b}(\bm R',\bm k')]}
&=&
\delta_{\bf ab}\delta(\bm R,\bm k,\bm R',\bm k')a_{\bf a}(\bm R,\bm k),\label{nccr1}\\
{[a_{\bf a}(\bm R,\bm k)^{\dag},n_{\bf b}(\bm R',\bm k')]}
&=&
-\delta_{\bf ab}\delta(\bm R,\bm k,\bm R',\bm k')a_{\bf a}(\bm R,\bm k)^{\dag},\label{nccr2}\\
{[a_{\bf a}(\bm R,\bm k),I(\bm R',\bm k')]}
&=&
{[a_{\bf a}(\bm R,\bm k)^{\dag},I(\bm R',\bm k')]}
={[n_{\bf a}(\bm R,\bm k)^{\dag},I(\bm R',\bm k')]}=0,\label{ccr2}\\
{[a_{\bf a}(\bm R,\bm k),a_{\bf b}(\bm R',\bm k')]}
&=&
{[a_{\bf a}(\bm R,\bm k)^{\dag},a_{\bf b}(\bm R',\bm k')^{\dag}]}=0.
\ee
Here $\delta_{\bf ab}$, ${\bf a}, {\bf b}=0,1,2,3$, is the Kronecker delta; $\delta(\bm R,\bm k,\bm R',\bm k')=\delta_1(\bm R,\bm R')
\delta_0(\bm k,\bm k')$, where the latter two distributions are SL(2,C) invariant Dirac deltas on $R^2=1$ and $k^2=0$ hyperboloids, respectively. Central element
$I(\bm R,\bm k)$ is a multiple of identity in irreducible representations of the algebra. However, I believe that a class of reducible representations discussed below is more physical \cite{MCJN,MCKW,MWMC} so I keep $I(\bm R,\bm k)$ at this stage unspecified. How to choose the representation of the algebra becomes particularly important if one considers the so called {\it entanglement with vacuum\/}, which turns out to be a notion representation dependent \cite{MPMC}. Number operator $n_{\bf b}(\bm R',\bm k')$ is added explicitly as an element of the Lie algebra as there exist representations where $n_{\bf b}(\bm R,\bm k)\neq a_{\bf b}(\bm R,\bm k)^{\dag}a_{\bf b}(\bm R,\bm k)$.

Let us note that $e^{-ik\cdot x}g{_{a}}{^{0}}(\bm R,\bm k)$ is accompanied by the {\it creation\/} operator $a_0(\bm R,\bm k)^{\dag}$. Had we employed in this place an annihilation operator, as it is typically done in the literature, we would have to resign either from hermiticity of the 4-potential or positivity of the scalar product in the space of states. Both cases lead to nonunitary evolutions. Our formulation leads to a manifestly covariant quantization procedure with Hermitian 4-potential and does not involve unphysical indefinite-metric `Hilbert space' \cite{MCJN,MCKW}.
The field commutator
\be
{[A_a(x),A_b(y)]}
&=&
ig_{ab} D(x-y)\label{[A,A]}
\ee
involves a generalized Jordan-Pauli `function'
\be
D(x)
&=&
i\int d k\,dR\,I(\bm R,\bm k)
\big(
e^{-ik\cdot x}
-
e^{ik\cdot x}
\big).
\label{J-P0}
\ee
The fact that we work with twistor-like spin-frames guarantees that the four operators occurring in (\ref{ccr1})--(\ref{ccr2}) split into two different massless representations of inhomogeneous SL(2,C):
$a_{1}(\bm R,\bm k)$, $a_{2}(\bm R,\bm k)$ (spin 1), and $a_{0}(\bm R,\bm k)$, $a_{3}(\bm R,\bm k)$ (spin 0). Transformation
$x_a\mapsto \Lambda{_a}{^b}x_b+y_a$ has to be represented unitarily at the operator level, i.e. there must exist $U(\Lambda,y)$ such that
\be
U(\Lambda,y)^{\dag}a(\pm,\bm{R},\bm{k})U(\Lambda,y)
&=&
e^{ip\cdot y}e^{\pm 2i\Theta(\Lambda,\bm k)}
a(\pm,\bm{\Lambda^{-1}R},\bm{\Lambda^{-1}k}),\label{spin1}\\
U(\Lambda,y)^{\dag}a_3(\bm{R},\bm{k})U(\Lambda,y)
&=&
e^{ip\cdot y}
a_3(\bm{\Lambda^{-1}R},\bm{\Lambda^{-1}k}),\\
U(\Lambda,y)^{\dag}a_0(\bm{R},\bm{k})^{\dag}U(\Lambda,y)
&=&
e^{ip\cdot y}
a_0(\bm{\Lambda^{-1}R},\bm{\Lambda^{-1}k})^{\dag}.
\ee
Alternatively,
\be
U(\Lambda,y)^{\dag}A_a(x)U(\Lambda,y)
&=&
\Lambda{_a}{^b}A_b\big(\Lambda^{-1}(x-y)\big).
\ee
For twistor-like spin-frames the above two sets of transformations are {\it not\/} inconsistent.

Electromagnetic qubits are associated only with the spin-1 part (\ref{spin1}).
How to explicitly construct $U(\Lambda,y)$ depends on the choice of representation of the oscillator algebra. In general, formulas involving irreducible representations with $I(\bm R,\bm k)=Z=$~const lead to mathematical inconsistencies typical of standard quantum field theory.

In order to define a mathematically precise representation one has to carefully define Dirac delta, and take a nontrivial $I(\bm R,\bm k)$.
This type of quantization is nonstandard but I believe it is exactly what should be done (to understand how it works for fields interacting with sources, see \cite{MCKW}).

\subsection{Reducible representations of harmonic-oscillator Lie algebra and transformation properties of field operators}

One begins with four operators, $a_0$, $a_1$, $a_2$, $a_3$,
satisfying commutation relations typical of an
{\it irreducible\/} representation of CCR:
$[a_{\bf a},a_{\bf b}^{\dag}]=\delta_{\bf ab}1$.
Let $|0\rangle$ denote their common vacuum,
i.e.~$a_{\bf a}|0\rangle=0$.
Now take kets $|\bm k\rangle$ and $|\bm R\rangle$, normalized with respect
to SL(2,C)-invariant M-shaped delta functions (see Appendix),
$
\langle\bm k|\bm k'\rangle
=
\delta_0(\bm k,\bm k')
$,
$\langle\bm R|\bm R'\rangle
=
\delta_1(\bm R,\bm R')$.
The reducible representation is parametrized by a natural number $N$. For $N=1$ the Hilbert space, denoted by ${\cal H}(1)$, is spanned by
kets of the form
\be
|\bm R,\bm k,n_0,n_1,n_2,n_3\rangle
=
|\bm R,\bm k\rangle\otimes
\frac{(a_0^{\dag})^{n_0}(a_1^{\dag})^{n_1}(a_2^{\dag})^{n_2}(a_3^{\dag})^{n_3}}
{\sqrt{n_0!n_1!n_2!n_3!}}|0\rangle,\nonumber
\ee
$|\bm R,\bm k\rangle=|\bm R\rangle\otimes|\bm k\rangle$.
The 1-oscillator representation is defined by
\be
a_{\bf a}(\bm R,\bm k,1) &=&|\bm R,\bm k\rangle\langle\bm R,\bm k|\otimes a_{\bf a},\\
n_{\bf a}(\bm R,\bm k,1) &=&|\bm R,\bm k\rangle\langle\bm R,\bm k|\otimes a_{\bf a}^{\dag}a_{\bf a},\label{n_a}\\
I(\bm R,\bm k,1) &=&|\bm R,\bm k\rangle\langle\bm R,\bm k|\otimes 1.
\ee
Operators
$I(\bm R,\bm k,1)$ form a resolution of unity
\be
\int d k{\,}d R{\,} I(\bm R,\bm k,1)
&=&
I\otimes I\otimes 1
=I(1),\label{I(1)-R}
\ee
Let us note that (\ref{n_a}) is constructed independently of $a_{\bf a}(\bm R,\bm k,1)$. The form $n_{\bf a}(\bm R,\bm k,1)=a_{\bf a}(\bm R,\bm k,1)^{\dag}a_{\bf a}(\bm R,\bm k,1)$ is possible  if one works with M-shaped Dirac deltas `regular at zero', i.e. $\delta_0(\bm k,\bm k)=\delta_1(\bm R,\bm R)=1$ (see Appendix). Quantized analogs of the wave functions occurring in (\ref{int dR a}),
\be
a_{\bf a}(\bm k,1) &=&\int dR\,a_{\bf a}(\bm R,\bm k,1)=I\otimes|\bm k\rangle\langle\bm k|\otimes a_{\bf a},
\ee
are in this representation well defined.
Now, let
\be
P_a(1)
&=&
-I\otimes {\int} d k{\,}k_a|\bm k\rangle\langle \bm k|\otimes
a^{\dag}_{\bf a}a^{\bf a}
\label{P-R}\\
&=&
{\int} dk{\,}k_a
\big(
n_1(\bm k,1)
+
n_2(\bm k,1)
+
n_3(\bm k,1)
-
n_0(\bm k,1)
\big)
\label{Pn-R},\\
&=&
{\int} dk{\,}k_a
\big(
n_+(\bm k,1)
+
n_-(\bm k,1)
+
n_3(\bm k,1)
-
n_0(\bm k,1)
\big)
\label{Pn-R+-},\\
n_{\bf a}(\bm k,1) &=&
\int dR\,n_{\bf a}(\bm R,\bm k,1)=I\otimes|\bm k\rangle\langle\bm k|\otimes a_{\bf a}^{\dag}a_{\bf a},\quad {\bf a}=0,1,2,3,+,-,\\
a_\pm
&=&
\frac{1}{\sqrt{2}}(a_1\pm i a_2),\\
J_3
&=&
i(a_{1}^{\dagger}a_{2}-a_{2}^{\dagger}a_{1})=
a_{+}^{\dagger}a_{+}-a_{-}^{\dagger}a_{-}\label{L=J}.
\ee
The representation of inhomogeneous SL(2,C) then reads
\be
U(\Lambda,0,1) &=&\int dk{\,}d R{\,}|\bm R,\bm k\rangle\langle
\bm{\Lambda^{-1}R},\bm{\Lambda^{-1}k}|\otimes
e^{2i \Theta(\Lambda,\bm k) J_3},\label{U-R}\\
U(\bm 1,y,1) &=&e^{iy^aP_a(1)}.
\ee
One immediately checks that
\be
U(\Lambda,y,1)^{\dag}a(\pm,\bm{R},\bm{k},1)U(\Lambda,y,1)
&=&
e^{ip\cdot y}e^{\pm 2i\Theta(\Lambda,\bm k)}
a(\pm,\bm{\Lambda^{-1}R},\bm{\Lambda^{-1}k},1),\\
U(\Lambda,y,1)^{\dag}a_3(\bm{R},\bm{k},1)U(\Lambda,y,1)
&=&
e^{ip\cdot y}
a_3(\bm{\Lambda^{-1}R},\bm{\Lambda^{-1}k},1),\\
U(\Lambda,y,1)^{\dag}a_0(\bm{R},\bm{k},1)^{\dag}U(\Lambda,y,1)
&=&
e^{ip\cdot y}
a_0(\bm{\Lambda^{-1}R},\bm{\Lambda^{-1}k},1)^{\dag},\\
U(\Lambda,y,1)^{\dag}A_a(x,1)U(\Lambda,y,1)
&=&
\Lambda{_a}{^b}A_b\big(\Lambda^{-1}(x-y),1\big).
\ee
Moreover, splitting the 4-momentum into electromagnetic and scalar parts,
\be
P_a(1)
&=&
P^{\{12\}}_a(1)+P^{\{3\}}_a(1)-P^{\{0\}}_a(1),\\
P^{\{12\}}_a(1)
&=&
{\int} dk{\,}k_a
\big(
n_+(\bm k,1)
+
n_-(\bm k,1)
\big)
\label{P 12},\\
P^{\{3\}}_a(1)
&=&
{\int} dk{\,}k_a
n_3(\bm k,1)
\label{P 3},\\
P^{\{0\}}_a(1)
&=&
{\int} dk{\,}k_a
n_0(\bm k,1)
\label{P 0},
\ee
we note that the three types of 4-momenta are not mixed with one another by SL(2,C) transformations:
\be
U(\Lambda,y,1)^{\dag}P^{\{12\}}_a(1)U(\Lambda,y,1)
&=&
{\int} dk{\,}k_a
\big(
n_+(\bm{\Lambda^{-1}k},1)
+
n_-(\bm{\Lambda^{-1}k},1)
\big)\nonumber\\
&=&
\Lambda{_a}{^b}P^{\{12\}}_b(1),\\
U(\Lambda,y,1)^{\dag}P^{\{3\}}_a(1)U(\Lambda,y,1)
&=&
{\int} dk{\,}k_a
n_3(\bm{\Lambda^{-1}k},1)
=
\Lambda{_a}{^b}P^{\{3\}}_b(1),\\
U(\Lambda,y,1)^{\dag}P^{\{0\}}_a(1)U(\Lambda,y,1)
&=&
{\int} dk{\,}k_a
n_0(\bm{\Lambda^{-1}k},1)
=
\Lambda{_a}{^b}P^{\{0\}}_b(1).
\ee
For arbitrary $N$ the representation is constructed as follows.
Define
\be
{\cal H}(N)=\underbrace{{\cal H}(1)\otimes\dots\otimes {\cal H}(1)}_N={\cal H}(1)^{\otimes N}
\ee
and let $A$ be an arbitrary operator defined for $N=1$. Let
\be
A^{(n)}=\underbrace{I(1)\otimes\dots\otimes I(1)}_{n-1}\otimes A
\otimes \underbrace{I(1)\otimes\dots\otimes I(1)}_{N-n}.
\ee
The $N$ oscillator extension of $a_{\bf a}(\bm R,\bm k,1)$ is defined by
\be
a_{\bf a}(\bm R,\bm k,N)
&=&
\frac{1}{\sqrt{N}}\sum_{n=1}^N a_{\bf a}(\bm R,\bm k,1)^{(n)},\\
n_{\bf a}(\bm R,\bm k,N)
&=&
\sum_{n=1}^N n_{\bf a}(\bm R,\bm k,1)^{(n)},\\
I(\bm R,\bm k,N)
&=&
\frac{1}{N}\sum_{n=1}^N I(\bm R,\bm k,1)^{(n)},\\
U(\Lambda,y,N) &=& U(\Lambda,y,1)^{\otimes N}.
\ee
Note that for $N>1$ $n_{\bf a}(\bm R,\bm k,N)\neq a_{\bf a}(\bm R,\bm k,N)^{\dag}a_{\bf a}(\bm R,\bm k,N)$.
The 4-momentum describes $N$ noninteracting 4-dimensional oscillators of indefinite frequency
\be
P_a(N)
&=&
\sum_{n=1}^NP_a(1)^{(n)}\\
&=&{\int} dk{\,}k_a
\big(
n_+(\bm k,N)
+
n_-(\bm k,N)
+
n_3(\bm k,N)
-
n_0(\bm k,N)
\big).
\ee
The latter formula justifies the definition of $n_{\bf a}(\bm R,\bm k,N)$.
$I(\bm R,\bm k,N)$ again form a resolution of unity
\be
\int d k{\,}d R{\,} I(\bm R,\bm k,N)
&=&
I(N)=I(1)^{\otimes N}.\label{I(N)-R}
\ee
Of course,
\be
U(\Lambda,y,N)^{\dag}a(\pm,\bm{R},\bm{k},N)U(\Lambda,y,N)
&=&
e^{ip\cdot y}e^{\pm 2i\Theta(\Lambda,\bm k)}
a(\pm,\bm{\Lambda^{-1}R},\bm{\Lambda^{-1}k},N),\\
U(\Lambda,y,N)^{\dag}a_3(\bm{R},\bm{k},N)U(\Lambda,y,N)
&=&
e^{ip\cdot y}
a_3(\bm{\Lambda^{-1}R},\bm{\Lambda^{-1}k},N),\\
U(\Lambda,y,N)^{\dag}a_0(\bm{R},\bm{k},N)^{\dag}U(\Lambda,y,N)
&=&
e^{ip\cdot y}
a_0(\bm{\Lambda^{-1}R},\bm{\Lambda^{-1}k},N)^{\dag},
\ee
and
\be
U(\Lambda,y,N)^{\dag}A_a(x,N)U(\Lambda,y,N)
&=&
\Lambda{_a}{^b}A_b\big(\Lambda^{-1}(x-y),N\big),
\ee
so the representation of inhomogeneous SL(2,C) again splits into spin-1 and spin-0 parts in the Fourier space, but at the level of Minkowski space
$A_a(x,N)$ is an ordinary world-vector field.

\subsection{Polarization operators for fields quantized in reducible representations of harmonic-oscillator Lie algebras}

Our goal is to define polarization operators that have transformation properties analogous to those based on Pauli-Lubanski vectors, but that are nevertheless linear in number-of-particles operators. In order to do so, let us note that for $N=1$ there exists an analog of first-quantized 4-momentum,
\be
K_a &=& I\otimes\int dk\,k_a |\bm k\rangle\langle \bm k|\otimes 1,\\
U(\Lambda,y,1)^{\dag}K_aU(\Lambda,y,1)
&=&
\Lambda{_a}{^b}K_b.
\ee
This operator naturally occurs if one starts with 4-momentum that includes vacuum energy (see Appendix). Indeed,
\be
P_a(1)+K_a
&=&
P_a(1)
+
I\otimes {\int} d k{\,}k_a|\bm k\rangle\langle \bm k|\otimes
\frac{1}{2}\Big(1+1+1-1\Big)\nonumber\\
&=&
-I\otimes {\int} d k{\,}k_a|\bm k\rangle\langle \bm k|\otimes
\frac{1}{2}\Big(a^{\dag}_{\bf a}a^{\bf a}+a^{\bf a}a^{\dag}_{\bf a}\Big)
.\nonumber
\ee
$K_a$ is in this representation well defined and commutes with elements of the oscillator Lie algebra. Generator
\be
M_{ab}(1)
&=&
i\frac{\partial U(\Lambda,0,1)}{\partial {}y^{ab}}\Big|_{{}y^{ab}=0}=L_{ab}(1)+S_{ab}(1)
\ee
splits into orbital and spin parts, where
\be
S_{ab}(1)
&=&
I\otimes
\int dk\,|\bm k\rangle\langle\bm{k}|\otimes
i\frac{\partial }{\partial {}y^{ab}}
e^{2i \Theta(\Lambda,\bm k) J_3}
\Big|_{{}y^{ab}=0}\nonumber\\
&=&
2I\otimes
\int dk\,|\bm k\rangle\langle\bm{k}|i\frac{\partial }{\partial {}y^{ab}}
{\bar\omega}{_{A'}}(\bm{k})\overline{\Lambda\pi}{^{A'}}(\bm{k})\Big|_{{}y^{ab}=0}\otimes J_3
\ee
and $\omega_A(\bm k)$ is any spin-partner of $\pi^A(\bm k)$.
Now consider
\be
W_a(1)
&=&
K^b(1){^*}S_{ab}(1)=
I\otimes
\int dk\,k_a|\bm k\rangle\langle\bm{k}|\otimes
J_3\\
&=&
\int dk\,k_a \Big(n_+(\bm k,1)-n_-(\bm k,1)\Big).
\ee
An extension to $N>1$, which is linear in $n_{\bf a}(\bm k,N)$, satisfies
\be
W_a(N)
&=&\sum_{n=1}^N W_a(1)^{(n)}=
\int dk\,k_a \Big(n_+(\bm k,N)-n_-(\bm k,N)\Big).
\ee
This type of extension is analogous to the relation between $P_a(1)$ and $P_a(N)$.

\subsection{Linear polarizations}

For $N=1$ let us define
\be
V_{\theta}(1)=I\otimes \int dk|\bm k\rangle\langle\bm k|\otimes e^{\theta(\bm k) (a_1^{\dag}a_2-a_2^{\dag}a_1)}
\ee
satisfying
\be
V_{\theta}(1)a_1(\bm k,1)V_{\theta}(1)^{\dag}
&=&
a_1(\bm k,1)\cos\theta(\bm k) - a_2(\bm k,1)\sin\theta(\bm k)=a_{\theta}(\bm k,1),\\
V_{\theta}(1)a_2(\bm k,1)V_{\theta}(1)^{\dag}
&=&
a_2(\bm k,1)\cos\theta(\bm k) + a_1(\bm k,1)\sin\theta(\bm k)=a_{\theta'}(\bm k,1).
\ee
For arbitrary $N\geq 1$ we define
\be
V_{\theta}(N)
&=&
V_{\theta}(1)^{\otimes N},\\
V_{\theta}(N)a_1(\bm k,N)V_{\theta}(N)^{\dag}
&=&
a_1(\bm k,N)\cos\theta(\bm k) - a_2(\bm k,N)\sin\theta(\bm k)=a_{\theta}(\bm k,N),\\
V_{\theta}(N)a_2(\bm k,N)V_{\theta}(N)^{\dag}
&=&
a_2(\bm k,N)\cos\theta(\bm k) + a_1(\bm k,N)\sin\theta(\bm k)=a_{\theta'}(\bm k,N),\\
V_{\theta}(N)a(\pm,\bm k,N)V_{\theta}(N)^{\dag}
&=&
e^{\pm i\theta(\bm k)}a(\pm,\bm k,N)=\frac{1}{\sqrt{2}}\Big(a_{\theta}(\bm k,N)\pm i\,a_{\theta'}(\bm k,N)\Big).
\ee
Number operators corresponding to linear polarizations are defined analogously
\be
V_{\theta}(N)n_1(\bm k,N)V_{\theta}(N)^{\dag}
&=&
n_{\theta}(\bm k,N),\\
V_{\theta}(N)n_2(\bm k,N)V_{\theta}(N)^{\dag}
&=&
n_{\theta'}(\bm k,N).
\ee
A yes-no observable associated with linear polarization can be defined as
\be
Y_{\theta}(\bm k,N)
&=&
n_{\theta}(\bm k,N)-n_{\theta'}(\bm k,N).\label{Y-N}
\ee

\section{Problem of relativistic analogues of EPR states of photons}

EPR states involve maximal entanglement in at least two different polarization bases (say, circular and linear). Such states can be easily invented also for photons, but the problem is with relativistic invariance of the EPR condition. At first glance the difficulty should not be a serious one since we know that the form (\ref{ve_A_B}) is SL(2,C) invariant: $\ve_{AB}$ is a scalar representation of SL(2,C). However, for relativistic qubits the scalar-field condition has to be formulated at the level of unitary representation of SL(2,C), and it seems the problem is far from being fully systematized as yet. The basic question is if we can find scalar fields that involve maximal entanglement in two different polarization degrees of freedom, and in all reference frames.

\subsection{Vacuum and multiphoton states}

To begin with, let us note that vacuum is any pure state annihilated by all annihilation and number operators.
In reducible representations
\be
|0,N\rangle
&=&
|0,1\rangle^{\otimes N},\\
|0,1\rangle
&=&
\int dk\,dR\, O(\bm R,\bm k)|\bm R,\bm k\rangle \otimes|0\rangle,\quad \int dk\,dR\, |O(\bm R,\bm k)|^2=1.
\ee
The analysis given in \cite{MCKW} suggests that the most natural choice of vacuum wave function is the separable state
$O(\bm R,\bm k)=O_0(\bm k)O_1(\bm R)$. Vacuum states are in this representation Bose-Einstein condensates at zero temperature. As such they are not unique, but the whole subspace of vacuum states is a relativistically invariant subspace of the Hilbert space of states. In field quantization based on twistor-like spin-frames an inhomogeneous SL(2,C) transformation cannot produce particles, a fact which is not so obvious in general since Lorentz transformations in more standard approaches squeeze vacuum (a detailed discussion of this problem can be found in \cite{MCKW}).

It is important to understand that the parameter $N$ that characterizes a given reducible representation is unrelated to the number of photons. For example,
\be
n_{\bf a}(\bm R,\bm k,N)
\Big(a_{\bf a}(\bm R,\bm k,N)^{\dag}\Big)^j|0,N\rangle
&=&
j\Big(a_{\bf a}(\bm R,\bm k,N)^{\dag}\Big)^j|0,N\rangle.
\ee
Weak (i.e. performed at the level of averages) limits  $N\to\infty$ reconstruct standard {\it regularized\/} formulas known from irreducible representations (for details see \cite{MCJN,MCKW,MWMC}).

\subsection{Scalar states}

The scalar-field condition means that
\be
\Psi(N)=\sum_{s,s'=\pm}\int dR\,dR'\,dk \,dk'\,
\Psi(s,\bm R,\bm k;s',\bm R',\bm k')
a(s,\bm R,\bm k,N)^{\dag}
a(s',\bm R',\bm k',N)^{\dag}\nonumber\\
\ee
satisfies
\be
U(\Lambda,0,N)^{\dag}\Psi(N)U(\Lambda,0,N)
=
\Psi(N).
\ee
Gauge independence implies that $\Psi(s,\bm R,\bm k;s',\bm R',\bm k')$ can be constructed  only by means of spinor fields $\pi_A$ (and thus does not depend on $\bm R$ and $\bm R'$) so that
\be
\Psi(N) &=&
\sum_{s,s'=\pm}\int dR\,dR'\,dk \,dk'\,
\Psi(s,\bm k;s',\bm k')
a(s,\bm R,\bm k,N)^{\dag}
a(s',\bm R',\bm k',N)^{\dag}\nonumber\\
&=&
\sum_{s,s'=\pm}\int dk \,dk'\,
\Psi(s,\bm k;s',\bm k')
a(s,\bm k,N)^{\dag}
a(s',\bm k',N)^{\dag}\label{Psi bez R}.
\ee
Since
\be
&{}&
U(\Lambda,0,N)^{\dag}a(s,\bm k,N)^{\dag}a(s',\bm k',N)^{\dag}U(\Lambda,0,N)
\nonumber\\
&{}&\pp=
=
e^{-2is\Theta(\Lambda,\bm k)}e^{-2is'\Theta(\Lambda,\bm k')}
a(s,\bm{\Lambda^{-1}k},N)^{\dag}a(s',\bm{\Lambda^{-1}k'},N)^{\dag},
\nonumber
\ee
the problem reduces to finding $\Psi$ satisfying
\be
\Psi(s,\bm k;s',\bm k')
e^{-2is\Theta(\Lambda,\bm k)}e^{-2is'\Theta(\Lambda,\bm k')}
=
\Psi(s,\bm{\Lambda^{-1}k};s',\bm{\Lambda^{-1}k'}).
\label{Psi}
\ee
Recalling that
\be
e^{-i\Theta(\Lambda,\bm k)}\pi_A(\bm k)
&=&
\Lambda{_A}{^B}\pi_B(\bm{\Lambda}^{-1}\bm k)
,\\
e^{i\Theta(\Lambda,\bm k)}\bar\pi_{A'}(\bm k)
&=&
\Lambda{_{A'}}{^{B'}}\bar\pi_{B'}(\bm{\Lambda}^{-1}\bm k)
\ee
we observe that only two types of contractions have the required form (\ref{Psi}):
\be
\Big(\pi_A(\bm k)\pi^A(\bm k')\Big)^2e^{-2i\Theta(\Lambda,\bm k)}e^{-2i\Theta(\Lambda,\bm k')}
&=&
\Big(\pi_A(\bm{\Lambda}^{-1}\bm k)\pi^A(\bm{\Lambda}^{-1}\bm k')\Big)^2
,\\
\Big(\bar\pi_{A'}(\bm k)\bar\pi^{A'}(\bm k')\Big)^2e^{2i\Theta(\Lambda,\bm k)}e^{2i\Theta(\Lambda,\bm k')}
&=&
\Big(\bar\pi_{A'}(\bm{\Lambda}^{-1}\bm k)\bar\pi^{A'}(\bm{\Lambda}^{-1}\bm k')\Big)^2.
\ee
Since for two null vectors, $k_a=\pi_A(\bm k)\bar\pi_{A'}(\bm k)$,
$l_a=\pi_A(\bm l)\bar\pi_{A'}(\bm l)$, we find
\be
k\cdot l=k_al^a = \pi_A(\bm k)\pi^A(\bm l)\bar\pi_{A'}(\bm k)\bar\pi^{A'}(\bm l)=\big|\pi_A(\bm k)\pi^A(\bm l)\big|^2,
\ee
the expressions $\big(\bar\pi_{A'}(\bm k)\bar\pi^{A'}(\bm k')\big)^{-1}$ and $\pi_A(\bm k)\pi^A(\bm k')$ are proportional to each other, the proportionality factor $k\cdot k'$ being SL(2,C) invariant.

So the corresponding solution reads
\be
\Psi(N)
&=&
\int dk\,dk'\,
F_+\big(\pi_A(\bm k)\pi^A(\bm k')\big)
a(+,\bm k,N)^{\dag}a(+,\bm k',N)^{\dag}\nonumber\\
&+&
\int dk\, dk'\,
F_-\big(\bar\pi_{A'}(\bm k)\bar\pi^{A'}(\bm k')\big)
a(-,\bm k,N)^{\dag}a(-,\bm k',N)^{\dag},
\label{Psi1}
\ee
where $F_\pm$ are any functions satisfying $F_\pm(e^{i\phi} z)=e^{2i\phi} F_\pm(z)$.
In reducible representations the integrals in (\ref{Psi1}) are well defined.

Another solution of (\ref{Psi}) is $\Psi(\pm,\bm k;\pm,\bm k')=0$,
\be
\Psi(\pm,\bm k;\mp,\bm k')
&=&
c_\pm \delta_0(\bm k,\bm k'),\\
\Psi(N)
&=&
c\int dk \,a(+,\bm k,N)^{\dag}a(-,\bm k,N)^{\dag}.\label{Psi2}
\ee
States generated by both types of $\Psi(N)$ are maximally entangled in circular polarizations. Now take a closer look at linear-polarization entanglement of (\ref{Psi1}). Inserting
\be
a(\pm,\bm k,N)^{\dag}=\frac{1}{\sqrt{2}}e^{\pm i\theta(\bm k)}\Big(a_{\theta}(\bm k,N)^{\dag}\mp i\,a_{\theta'}(\bm k,N)^{\dag}\Big),
\ee
into (\ref{Psi1}),
\be
\Psi(N)
&=&
\int dk\,dk'\,
\frac{1}{2}
\Big(
F_+(\dots)
e^{i[\theta(\bm k)+\theta(\bm k')]}
+
F_-(\dots)
e^{-i[\theta(\bm k)+\theta(\bm k')]}
\Big)
\nonumber\\
&\pp=&\times
\Big(
a_{\theta}(\bm k,N)^{\dag}a_{\theta}(\bm k',N)^{\dag}
-
a_{\theta'}(\bm k,N)^{\dag}a_{\theta'}(\bm k',N)^{\dag}
\Big)
\nonumber\\
&+&
\int dk\,dk'\,
\frac{1}{2i}
\Big(
F_+(\dots)
e^{i[\theta(\bm k)+\theta(\bm k')]}
-
F_-(\dots)
e^{-i[\theta(\bm k)+\theta(\bm k')]}
\Big)
\nonumber\\
&\pp=&\times
\Big(
a_{\theta'}(\bm k,N)^{\dag}a_{\theta}(\bm k',N)^{\dag}
+
a_{\theta}(\bm k,N)^{\dag}a_{\theta'}(\bm k',N)^{\dag}
\Big),
\ee
we obtain a condition for maximally entangled linear polarizations,
\be
F_-\big(\bar\pi_{A'}(\bm k)\bar\pi^{A'}(\bm k')\big)
=
\pm
F_+\big(\pi_A(\bm k)\pi^A(\bm k')\big)e^{2i[\theta(\bm k)+\theta(\bm k')]}
.
\ee
Combining this with (\ref{Psi1}), we arrive at
\be
\Psi(N)
&=&
\int dk\,dk'\,
F_+\big(\pi_A(\bm k)\pi^A(\bm k')\big)
\nonumber\\
&\pp=&\times
\Big(a(+,\bm k,N)^{\dag}a(+,\bm k',N)^{\dag}\pm e^{2i[\theta(\bm k)+\theta(\bm k')]}a(-,\bm k,N)^{\dag}a(-,\bm k',N)^{\dag}\Big),
\nonumber\\
\label{Psi5}
\ee
which {\it does not\/} satisfy (\ref{Psi}). State $\Psi(N)|0,N\rangle$ is scalar but does not satisfy linear-polarization EPR condition. It seems there is no difficulty with extending the above argument to general elliptic polarizations.

The second solution (\ref{Psi2}) is more interesting. Indeed, (\ref{Psi2}) can be written as
\be
\Psi(N)
&=&
\frac{c}{2}
\int dk
\Big(
a_{\theta}(\bm k,N)^{\dag}a_{\theta}(\bm k,N)^{\dag}
+
a_{\theta'}(\bm k,N)^{\dag}a_{\theta'}(\bm k,N)^{\dag}
\Big),\label{Psi4}
\ee
which is simultaneously maximally entangled in all linear polarizations. Let us note that although (\ref{Psi4}) involves photons of the same momenta, this does not exclude applications to experiments with pairs of detectors localized in space in arbitrary locations.

\subsection{EPR states involving different momenta}

Let us again consider the operator (\ref{Psi bez R}) but with the kernel satisfying
\be
\Psi(s,\bm k;s',\bm k')=-\Psi(s',\bm k;s,\bm k')=-\Psi(s,\bm k';s',\bm k).\label{anti-s}
\ee
Denoting $\Psi(+,\bm k;-,\bm k')=\psi(\bm k,\bm k')/2=-\psi(\bm k',\bm k)/2$, we find
\be
\Psi(N) &=&
\int dk \,dk'\,
\psi(\bm k,\bm k')
a(+,\bm k,N)^{\dag}a(-,\bm k',N)^{\dag}
\label{Psi6}\\
&=&
(\alpha\delta-\beta\gamma) \int dk \,dk'\,
\psi(\bm k,\bm k')
b_1(\bm k,N)^{\dag}b_2(\bm k',N)^{\dag}
\nonumber
\ee
where
\be
\left(
\begin{array}{c}
b_1(\bm k,N)^{\dag}\\
b_2(\bm k,N)^{\dag}
\end{array}
\right)
&=&
\left(
\begin{array}{cc}
\alpha & \beta\\
\gamma & \delta
\end{array}
\right)^{-1}
\left(
\begin{array}{c}
a(+,\bm k,N)^{\dag}\\
a(-,\bm k,N)^{\dag}
\end{array}
\right)
,
\ee
and the matrix is unitary and independent of $\bm k$. In particular, if
\be
\left(
\begin{array}{cc}
\alpha & \beta\\
\gamma & \delta
\end{array}
\right)^{-1}
=
\frac{1}{\sqrt{2}}
\left(
\begin{array}{cc}
e^{-i\theta} & e^{i\theta}\\
ie^{-i\theta} & -ie^{i\theta}
\end{array}
\right)
\ee
the state $\Psi(N)|0,N\rangle$ is maximally entangled in both linear and circular polarizations.
One can replace the antisymmetric condition (\ref{anti-s}) by a symmetric one, but then we arrive at the state of the form (\ref{Psi5}).

Let us now consider the issue of SL(2,C) invariance of linear-polarization entanglement of (\ref{Psi6}). The transformed operator
\be
&{}&
U(\Lambda,0,N)^{\dag}\Psi(N)U(\Lambda,0,N)
\nonumber\\
&{}&\pp==
\int dk \,dk'\,
\psi(\bm{\Lambda k},\bm{\Lambda k'})
e^{-2i\Theta(\Lambda,\bm{\Lambda k})}a(+,\bm{k},N)^{\dag}
e^{2i\Theta(\Lambda,\bm{\Lambda k'})}a(-,\bm{k'},N)^{\dag}
\label{Psi8}
\ee
is maximally entangled in all linear polarizations defined by operators
\be
\frac{1}{\sqrt{2}}
\Big(e^{-i\theta(\bm k)}a(+,\bm{k},N)^{\dag}+e^{i\theta(\bm k)}a(-,\bm{k},N)^{\dag}\Big)
,\\
\frac{i}{\sqrt{2}}
\Big(e^{-i\theta(\bm k)}a(+,\bm{k},N)^{\dag}-e^{i\theta(\bm k)}a(-,\bm{k},N)^{\dag}\Big),
\ee
where
$\theta(\bm k)=\theta+2\Theta(\Lambda,\bm{\Lambda k})$. In general, components corresponding to different wave vectors get rotated by different angles.

Putting it differently, the antisymmetry of $\psi(\bm{k},\bm{k'})$ is not conserved by SL(2,C) transformations since
\be
\psi(\bm{\Lambda k},\bm{\Lambda k'})
e^{-2i\Theta(\Lambda,\bm{\Lambda k})}
e^{2i\Theta(\Lambda,\bm{\Lambda k'})}
\ee
is not antisymmetric in $\bm k$ and $\bm k'$, so that the state is no longer maximally entangled in the original linear polarizations.

The latter effect is essentially the massless version of the Peres-Scudo-Terno phenomenon. In the massless case the polarization observables are not defined with respect to projections of the Pauli-Lubanski vector, and thus the argument based on $\omega$-spinors cannot be directly employed. On the other hand, Wigner phases depend only on directions of $\bm k$, so all parallel wave vectors correspond to the same rotation angle. In order to maintain maximal entanglement in EPR-type experiments, one has to employ momentum-dependent polarization operators that compensate the presence of the Wigner angle $2\Theta(\Lambda,\bm{\Lambda k})$.

\subsection{Normalization of 2-photon states in reducible representation}

Let us pause here for a moment to discuss the issue of normalization of 2-photon states.
As an exercise consider the simple scalar case $F_+(z)=z^2$, $F_-(z)=0$:
\be
&{}&\int dk\,dk'\,dl\,dl' \Big(\bar\pi_{A'}(\bm l)\bar\pi^{A'}(\bm l')\Big)^2 \Big(\pi_A(\bm k)\pi^A(\bm k')\Big)^2
\nonumber\\
&{}&\pp{\int dk\,dk'\,dl\,dl'}\times
\langle 0,N|a(+,\bm l,N)a(+,\bm l',N)a(+,\bm k,N)^{\dag}a(+,\bm k',N)^{\dag}|0,N\rangle\nonumber\\
&{}&\pp=
=
\int dk\,dk'\,dl\,dl' \Big(\bar\pi_{A'}(\bm l)\bar\pi^{A'}(\bm l')\Big)^2 \Big(\pi_A(\bm k)\pi^A(\bm k')\Big)^2
\nonumber\\
&{}&\pp{\int dk\,dk'\,dl\,dl'}\times
\Big(\delta_0(\bm k,\bm l')\delta_0(\bm k',\bm l)+\delta_0(\bm k',\bm l')\delta_0(\bm k,\bm l)\Big)
\langle 0,N|I(\bm k,N)I(\bm k',N)|0,N\rangle\nonumber\\
&{}&\pp=
=
2\int dk\,dk'\,(k\cdot k')^2 \langle 0,N|I(\bm k,N)I(\bm k',N)|0,N\rangle\nonumber\\
&{}&\pp=
=
2\Big(1-\frac{1}{N}\Big)\int dk\,dk'\,(k\cdot k')^2 |O_0(\bm k)|^2|O_0(\bm k')|^2. \nonumber
\ee
Let us note that in irreducible representations, where $I(\bm k)=Z\bm 1$, $Z=$~const, an analogous calculation would involve the divergent integral
\be
2\int dk\,dk'\,(k\cdot k')^2 \langle 0|I(\bm k)I(\bm k')|0\rangle
=
2Z^2\int dk\,dk'\,(k\cdot k')^2.
\ee
Invariance of the norm under inhomogeneous SL(2,C) is a consequence of the fact that the vacuum wave function is a scalar field:
$O_0(\bm k)\to O_0(\bm{\Lambda}^{-1}\bm k)$ is equivalent to $|0,N\rangle\to U(\Lambda,y,N)|0,N\rangle$. (Inclusion of vacuum 4-momentum $K_a$ would imply $O_0(\bm k)\to e^{ik\cdot y}O_0(\bm{\Lambda}^{-1}\bm k)$.)

Performing an analogous calculation for (\ref{Psi6}), we obtain
\be
\langle 0,N|\Psi(N)^{\dag}\Psi(N)|0,N\rangle
=
\Big(1-\frac{1}{N}\Big)
\int dk\,dk'\,
|\psi(\bm k,\bm k')|^2
|O(\bm k)|^2 |O(\bm k')|^2.
\ee
In reducible representations the state $\Psi(N)|0,N\rangle$, generated by (\ref{Psi2}), is normalizable as well. Since this exercise is also instructive, let us explicitly compute the norm for $N=1$ in two alternative ways (put $c=1$ for simplicity). The first approach explicitly employs the form of the vacuum state and normalization of kets to M-shaped deltas,
\be
\Psi(1)|0,1\rangle
&=&
\int dk \,a(+,\bm k,1)^{\dag}a(-,\bm k,1)^{\dag}\int dk'\, dR'\, O_1(\bm R')O_0(\bm k')|\bm R',\bm k',0,0,0,0\rangle\nonumber\\
&=&
\int dk \,dR\,|\bm R,\bm k\rangle\langle\bm R,\bm k|\otimes a_+^{\dag}a_-^{\dag}\int dk'\, dR'\, O_1(\bm R')O_0(\bm k')|\bm R',\bm k',0,0,0,0\rangle\nonumber\\
&=&
\int dk\, dR\, O_1(\bm R)O_0(\bm k)|\bm R,\bm k\rangle\otimes a_+^{\dag}a_-^{\dag}|0\rangle\nonumber,
\ee
so that $\langle 0,1|\Psi(1)^{\dag}\Psi(1)|0,1\rangle=\int dR\, |O_1(\bm R)|^2\int dk\,|O_0(\bm k)|^2= 1$. The second way is based on properties of the M-shaped deltas (see Appendix \ref{subsec 75}), and the fact that annihilation operators annihilate vacuum:
\be
{}&{}&
\langle 0,1|\Psi(1)^{\dag}\Psi(1)|0,1\rangle
\nonumber\\
&{}&\pp==
\int dk \int dk'
\langle 0,1|a(+,\bm k,1)a(-,\bm k,1)a(+,\bm k',1)^{\dag}a(-,\bm k',1)^{\dag}|0,1\rangle\nonumber\\
&{}&\pp=
=
\lim_{n_1\to\infty}\lim_{n_2\to\infty}\lim_{n_3\to\infty} \lim_{n_4\to\infty}
\int dk \int dk'
\nonumber\\
&{}&\pp{\lim_{n_1\to\infty}\lim_{n_2\to\infty}}
\times
\langle 0,1|a(-,\bm k,1,\textstyle{\frac{1}{n_1}})a(+,\bm k,1,\textstyle{\frac{1}{n_2}})a(+,\bm k',1,\textstyle{\frac{1}{n_3}})^{\dag}a(-,\bm k',1,\textstyle{\frac{1}{n_4}})^{\dag}|0,1\rangle\nonumber\\
&{}&\pp=
=
\lim_{n_1\to\infty}\lim_{n_4\to\infty}
\int dk \int dk'
\langle 0,1|a(-,\bm k,1,\textstyle{\frac{1}{n_1}})\delta_0(\bm k,\bm k')I(\bm k,1)a(-,\bm k',1,\textstyle{\frac{1}{n_4}})^{\dag}|0,1\rangle\nonumber\\
&{}&\pp=
=
\int dk
\langle 0,1|I(\bm k,1)^2|0,1\rangle
=
\langle 0,1|\int dk\,I(\bm k,1)|0,1\rangle=\langle 0,1|0,1\rangle=1.
\nonumber
\ee
One analogously checks that, for arbitrary $N\geq 1$ and $c=1$,  (\ref{Psi2}) implies
\be
\langle 0,N|\Psi(N)^{\dag}\Psi(N)|0,N\rangle
&=&
\frac{1}{N^2}+\Big(1-\frac{1}{N}\Big)\int dk\,|O_0(\bm k)|^4.
\ee
\section{EPR averages for linearly polarized photons --- the reducible representation approach}

Let us turn to the question of EPR averages for quantum electromagnetic fields quantized in reducible representations of harmonic oscillator Lie algebras. The issue is important since all the experiments performed so far were analyzed in terms of {\it irreducible\/} representations, so one might be tempted to conclude that the standard approach to quantization is supported by EPR-type predictions.

Let us consider the EPR state (\ref{Psi6}). Yes-no observables are defined by (\ref{Y-N}). The EPR average
\be
&{}&
\frac{\langle 0,N|\Psi(N)^{\dag}Y_{\beta}(\bm l',N)Y_{\alpha}(\bm l,N)\Psi(N)|0,N\rangle}
{\langle 0,N|\Psi(N)^{\dag}\Psi(N)|0,N\rangle}
\nonumber\\
&{}&\pp=
=
2\cos 2(\alpha-\beta)\frac{\delta_0(\bm l',\bm l)
|O_0(\bm l)|^2
\int  dk\,
|\psi(\bm l,\bm k)|^2
|O_0(\bm k)|^2
-
|\psi(\bm l,\bm l')|^2
|O_0(\bm l)|^2 |O_0(\bm l')|^2}
{\int dk\,dk'\,
|\psi(\bm k,\bm k')|^2
|O_0(\bm k)|^2 |O_0(\bm k')|^2}
\nonumber\\
\ee
involves sharp wave vectors $\bm l$, $\bm l'$, and  is independent of $N$ if $N>1$ ($\Psi(N)|0,N\rangle=0$ for $N=1$). Since localization of detectors leads to momentum spread, $\bm l\in \Omega$, $\bm l'\in \Omega'$, say, yes-no operators integrated over both sets
\be
Y_{\alpha}(N)
&=&
\int_\Omega dl\,Y_{\alpha}(\bm l,N),\\
Y'_{\beta}(N)
&=&
\int_{\Omega'} dl'\,Y_{\beta}(\bm l',N),
\ee
can be used to compute more realistic cases. For disjoint detectors, $\Omega \cap \Omega'=\phi$,
\be
\frac{\langle 0,N|\Psi(N)^{\dag}Y'_{\beta}(N)Y_{\alpha}(N)\Psi(N)|0,N\rangle}
{\langle 0,N|\Psi(N)^{\dag}\Psi(N)|0,N\rangle}
&=&
-
\cos 2(\alpha-\beta)p_{\Omega\times\Omega'},
\label{Y'Y}
\ee
where
\be
p_{\Omega\times\Omega'}
&=&
p_{\Omega'\times\Omega}\nonumber\\
&=&
2\frac{
\int_\Omega dl\,\int_{\Omega'} dl'\,|\psi(\bm l,\bm l')|^2
|O_0(\bm l)|^2 |O_0(\bm l')|^2
}
{\int_{\bm R^3} dk \int_{\bm R^3}dk'\,
|\psi(\bm k,\bm k')|^2
|O_0(\bm k)|^2 |O_0(\bm k')|^2}
\nonumber\\
&=&
2\frac{
\int_\Omega dl\,\int_{\Omega'} dl'\,|\psi(\bm l,\bm l')|^2
\chi(\bm l)\chi(\bm l')
}
{\int_{\bm R^3} dk \int_{\bm R^3}dk'\,
|\psi(\bm k,\bm k')|^2
\chi(\bm k)\chi(\bm k')}.\label{p red}
\ee
The form (\ref{p red}) employs the cutoff function $\chi(\bm k)=|O(\bm k)|^2/Z$,  $Z=\max_{\bm k}|O(\bm k)|^2$, $0\leq \chi(\bm k)\leq 1$, whose appearance is typical of predictions based on the reducible representation.

For identical detectors, $\Omega =\Omega'$,
\be
\frac{\langle 0,N|\Psi(N)^{\dag}Y'_{\beta}(N)Y_{\alpha}(N)\Psi(N)|0,N\rangle}
{\langle 0,N|\Psi(N)^{\dag}\Psi(N)|0,N\rangle}
&=&
\cos 2(\alpha-\beta)p_{\Omega\times(\bm R^3-\Omega)}.
\label{YY}
\ee
The average (\ref{YY}) vanishes if $\Omega=\Omega'=\bm R^3$.
In the most general case, with arbitrary overlap $\Omega_0=\Omega\cap\Omega'$, $\Omega=\Omega_1\cup\Omega_0$, $\Omega'=\Omega_0\cup\Omega_1'$,  one finds
\be
-
\cos 2(\alpha-\beta)
\Big(
p_{\Omega_1\times\Omega'_1}
+
p_{\Omega_1\times\Omega_0}
+
p_{\Omega_0\times\Omega'_1}
-
p_{\Omega_0\times(\bm R^3-\Omega_0)}
\Big).\label{overlap}
\ee
Let us note that Bell's inequality can be violated only if
\be
p_{\Omega_1\times\Omega'_1}
+
p_{\Omega_1\times\Omega_0}
+
p_{\Omega_0\times\Omega'_1}
-
p_{\Omega_0\times(\bm R^3-\Omega_0)}>\frac{1}{\sqrt{2}},
\ee
For example, let
\be
\psi(\bm k,\bm k')=f(\bm k)g(\bm k')-f(\bm k')g(\bm k),
\ee
where supports $X_f={\rm supp}(f)$ and $X_g={\rm supp}(g)$ are disjoint (say, photons of two different colors are emitted, like in parametric down-conversion experiments). If $X_f\subset \Omega$, $X_g\subset \Omega'$,
$\Omega\cap\Omega'=\phi$, then
\be
p_{\Omega\times\Omega'}
&=&
\frac{
2\int_\Omega dl\,\int_{\Omega'} dl'\,|f(\bm l)g(\bm l')|^2\chi(\bm l) \chi(\bm l')
+
2\int_\Omega dl\,\int_{\Omega'} dl'\,|f(\bm l')g(\bm l)|^2\chi(\bm l) \chi(\bm l')
}
{
\int_\Omega dk\,\int_{\Omega'} dk'\,|f(\bm k)g(\bm k')|^2\chi(\bm k) \chi(\bm k')
+
\int_{\Omega'} dk\,\int_{\Omega} dk'\,|f(\bm k')g(\bm k)|^2\chi(\bm k') \chi(\bm k)}
\nonumber\\
&=&
p_{\Omega_1\times\Omega'_1}
+
p_{\Omega_1\times\Omega_0}
+
p_{\Omega_0\times\Omega'_1}
-
p_{\Omega_0\times(\bm R^3-\Omega_0)}
=
1,
\ee
so that the Bell inequality can be violated. If $X_f\cup X_g\subset \Omega=\Omega'$
\be
\int_\Omega dl\,\int_{\bm R^3-\Omega} dl'\,|f(\bm l)g(\bm l')|^2\chi(\bm l) \chi(\bm l')
+
\int_\Omega dl\,\int_{\bm R^3-\Omega} dl'\,|f(\bm l')g(\bm l)|^2\chi(\bm l) \chi(\bm l')
=0,\nonumber
\ee
and no violation is found.

In irreducible representations, where $I(\bm k)=Z\bm 1$, we get identical formulas but with
\be
p_{\Omega\times\Omega'}
&=&
2\frac{
\int_\Omega dl\,\int_{\Omega'} dl'\,|\psi(\bm l,\bm l')|^2
}
{\int_{\bm R^3} dk \int_{\bm R^3}dk'\,
|\psi(\bm k,\bm k')|^2 \label{p irred}
}.
\ee
In practice, probabilities (\ref{p red}) and (\ref{p irred}) are not easy to distinguish from one another if $\chi(\bm k)$ is sufficiently flat in the optical-range set of wave vectors. The EPR averages thus do not provide us with any practical clue about the choice of representation of harmonic oscillator Lie algebras appropriate for field quantization.

Relativistic properties of EPR averages cannot be directly inferred from (\ref{Y'Y}) but one has to take into account the remarks made after (\ref{Psi8}) and repeat the whole calculation. I will not pursue the matter further here.

\section{Remarks on two related issues}

One can find in the literature a discussion of `entanglement with vacuum' (where a single-photon state is entangled) and violation of Bell's inequality in vacuum (where the vacuum state is entangled). In light of what I have written above on EPR states an existence of such phenomena may seem weird, but the problem is purely semantic --- different authors have different things in mind when they speak of `vacuum'.

\subsection{Entanglement with vacuum}

The idea of entanglement with vacuum comes from a specific representation of the Lie algebra
\be
[a_m,a_n^{\dag}]=\delta_{nm}\bm 1, \label{stand}
\ee
where $m$, $n$ are integers, and
\be
a_n=\dots 1\otimes  1\otimes a\otimes  1\otimes  1\dots
\ee
with $a$, $[a,a^{\dag}]=1$, located on an ``$n$th position". $a_n$ acts in the non-separable Hilbert space spanned by infinite tensor products of number sates $|n\rangle$, $a^{\dag}a|n\rangle=n|n\rangle$. Infinite tensor products make the set of basis vectors uncountable (`Orlov states' \cite{Orlov}, such as $\dots |0\rangle\otimes  |3\rangle\otimes  |1\rangle\otimes  |4\rangle\otimes|1\rangle\otimes|5\rangle\otimes|9\rangle\otimes  \dots$, with numbers taken from the digits of $\pi$, are indexed by real numbers).

The identity at the right side of (\ref{stand}) is given by the infinite tensor product of identities
\be
\bm 1 &=& \dots 1\otimes  1\otimes  1\dots
\ee
while the vacuum is represented by
\be
|\bm 0\rangle &=& \dots |0\rangle\otimes  |0\rangle\otimes  |0\rangle\dots
\ee
A single photon state that exists in a superposition of two different modes, say,
\be
\big(a_n^{\dag}+a_m^{\dag}\big)|\bm 0\rangle
&=&
\dots |0\rangle\otimes  |1\rangle\otimes \dots \otimes  |0\rangle \otimes  |0\rangle\dots\nonumber\\
&\pp=&+
\dots |0\rangle\otimes  |0\rangle\otimes \dots \otimes  |0\rangle \otimes  |1\rangle\dots
\nonumber
\ee
resembles, up to all the problems with infinite tensor products, the 2-particle entangled state
\be
|1\rangle\otimes  |0\rangle
+
|0\rangle\otimes  |1\rangle.
\ee
This representation is often treated (especially in the quantum optics literature) as {\it the\/} representation of the harmonic oscillator Lie algebra typical of quantum fields.

We have seen, however, that EPR correlations of photons can be computed at a much more general level and do not rely on specific tensor product structures of Hilbert spaces in question. The reducible representations we have worked with allow us to speak of all the field modes, but involve only $N$th tensor powers. Attempts of interpreting, say,
\be
\Big(a(+,\bm k,N)^{\dag}+a(-,\bm k,N)^{\dag}\Big)|0,N\rangle
\ee
as an EPR state do not make much sense. Moreover, one does not have to resort to reducible representations to show that a single-photon 2-mode state may not always be interpretable in terms of entangled states. It is sufficient (see \cite{MPMC}) to take the original (1932) representation of the Fock space \cite{Fock,Berezin}.

\subsection{Nonlocal properties of vacuum states in algebraic quantum field theory}

Summers and Werner \cite{SW0} showed that vacuum states, equipped with all their properties assumed in axiomatic quantum field theory, are enough to violate the Bell inequality. The paper is simultaneously the first published account of relativistic Bell theorem (see, however, \cite{MC84}). The results from \cite{SW0} were subsequently generalized in a number of works (cf. \cite{Summers} for a recent review).

In order to grasp the main idea let us consider a simpler but physically similar problem suggested in \cite{Redhead}. Consider two systems, ${\cal O}_1$, ${\cal O}_2$, equipped with certain algebras ${\cal A}({\cal O}_1)$, ${\cal A}({\cal O}_2)$. We assume commutativity $[A_1,A_2]=0$ if $A_1\in {\cal A}({\cal O}_1)$ and $A_2\in {\cal A}({\cal O}_2)$.
The algebraic structure can be represented in a Hilbert space ${\cal H}={\cal H}_1\otimes {\cal H}_2$ where representatives of $A_1$ and $A_2$ are of the form $A\otimes 1$ and $1\otimes B$, respectively. The crucial assumption about a vacuum state $|\Omega\rangle$ is its cyclicity with respect to both ${\cal A}({\cal O}_1)$ and ${\cal A}({\cal O}_2)$. What it means is that acting on $|\Omega\rangle$ with operators of {\it either\/} ${\cal A}({\cal O}_1)$ or ${\cal A}({\cal O}_2)$ one can generate any vector in ${\cal H}$.

To make our analysis as concrete and simple as possible, assume that ${\cal H}_1$ and ${\cal H}_2$ are 2-dimensional. If vacuum $|\Omega\rangle$ is cyclic in the above sense, then {\it all\/} basis vectors of ${\cal H}$ can be written as $(A\otimes 1)|\Omega\rangle$. Let $|0_j\rangle$, $|1_j\rangle$, span ${\cal H}_j$. It is obvious that $|\Omega\rangle$ cannot be a product state. However, any state from the Bell basis
\be
|\Psi_\pm\rangle
&=&
\frac{1}{\sqrt{2}}\Big(|0_1\rangle\otimes |1_2\rangle\pm |1_1\rangle\otimes |0_2\rangle\Big),\\
|\Phi_\pm\rangle
&=&
\frac{1}{\sqrt{2}}\Big(|0_1\rangle\otimes |0_2\rangle\pm |1_1\rangle\otimes |1_2\rangle\Big),
\ee
can play the role of $|\Omega\rangle$. Indeed, take $A=|0_1\rangle\langle 0_1|-|1_1\rangle\langle 1_1|$,
$B=|0_1\rangle\langle 1_1|+|1_1\rangle\langle 0_1|$, and $|\Omega\rangle=|\Psi_+\rangle$. Then
\be
(1\otimes 1)|\Omega\rangle &=& |\Psi_+\rangle,\\
(A\otimes 1)|\Omega\rangle &=& |\Psi_-\rangle,\\
(B\otimes 1)|\Omega\rangle &=& |\Phi_+\rangle,\\
(AB\otimes 1)|\Omega\rangle &=& |\Phi_-\rangle.
\ee
The same effect is obtained if one acts on the second qubit.
Moreover, we can replace $|\Psi_+\rangle$ by any entangled state, say, a ground state of some 2-qubit Hamiltonian. $|\Omega\rangle$ is then the lowest energy state which is cyclic with respect to commuting von Neumann algebras ${\cal A}({\cal O}_j)$. So this is precisely the vacuum in the sense of algebraic quantum field theory, but for a trivial toy model. The fact that `vacuum' can maximally violate the Bell inequality is no longer weird.

Our vacuum state $|0,N\rangle$ is different. It belongs to a Poincar\'e invariant subspace which is uniquely defined, but a single $|0,N\rangle$ is neither unique nor relativistically invariant. $|0,N\rangle$ is not cyclic either --- it does not satisfy the axioms employed by Summers and Werner. In spite of that, the formalism leads to a well defined field theory, hopefully with no divergences, which is surprisingly close to the standard one in all the applications considered so far, simultaneously leading to new effects, testable at least in principle \cite{MCKW,MWMC}.

\section*{Appendices}

\section{Dirac delta regular at zero}

\subsection{M-shaped delta-sequences}

Let us consider the function shown in the upper part of Fig.~1. It is a particular example, for $a=1$ and $\epsilon=1/2$, of
\be
\delta(k,a,\epsilon)
&=&
\left\{
\begin{array}{crc}
0 & \textrm{for} & k < -\frac{\epsilon}{2} \\
\big(\frac{4k}{\epsilon} +2\big)\big(\frac{2}{\epsilon} - \frac{a}{2}\big) & \textrm{for} & -\frac{\epsilon}{2} \leq k < -\frac{\epsilon}{4}\\
-\frac{4k}{\epsilon}\big(\frac{2}{\epsilon} - \frac{3a}{2}\big) + a & \textrm{for} & -\frac{\epsilon}{4} \leq k < 0\\
\frac{4k}{\epsilon}\big(\frac{2}{\epsilon} - \frac{3a}{2}\big) + a & \textrm{for} & 0\leq k<\frac{\epsilon}{4}\\
\big(-\frac{4k}{\epsilon} +2\big)\big(\frac{2}{\epsilon} - \frac{a}{2}\big) & \textrm{for} & \frac{\epsilon}{4} \leq k < \frac{\epsilon}{2}\\
0 & \textrm{for} & \frac{\epsilon}{2}\leq k,
\end{array}\label{1}
\right.\nonumber
\\
\ee
($a>0$, $\epsilon>0$).
\begin{figure}
\includegraphics[width=8cm]{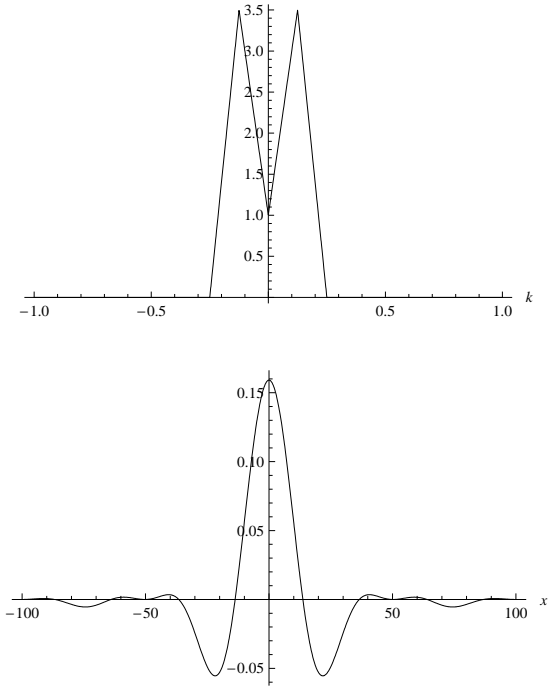}
\caption{The M-shaped function (\ref{1}) with $a=1$, $\epsilon=1/2$ (upper), and its Fourier transform (lower).}
\end{figure}
Its Fourier transform,
\be
\hat\delta(x,a,\epsilon)
&=&
\frac{1}{2\pi}
\int_{-\infty}^\infty \delta(k,a,\epsilon) e^{ikx}dk\nonumber\\
&=&
\frac{8}{\pi}
\frac{\epsilon a + (4 - \epsilon a) \cos\frac{\epsilon x}{4}}{\epsilon^2 x^2}\sin^2\frac{\epsilon x}{8},
\\
\lim_{\epsilon\to 0}\hat\delta(x,a,\epsilon) &=& \frac{1}{2\pi},
\ee
is a real function shown in the lower part of Fig.~1 . The sequence $\delta(k,a,\textstyle{\frac{1}{n}})$, with natural $n$ (i.e. $\epsilon=1/n$), is an example of what I call an M-shaped delta-sequence, and is in fact a particular example of a delta-sequence in the sense of \cite{Sikorski}.
Indeed,
\be
\int_{-\infty}^\infty \delta(k,a,\textstyle{\frac{1}{n}})dk &=& 1,
\ee
and for a function $f$ left- and right-continuous at 0
\be
\lim_{n\to\infty}\int_{-\infty}^\infty f(k)\delta(k,a,\textstyle{\frac{1}{n}})dk &=&\frac{f(0_-)+f(0_+)}{2}.
\ee
A peculiarity of M-shaped delta-sequences is their regularity at 0,
\be
\delta(0,a,\textstyle{\frac{1}{n}}) &=& a,
\ee
for all $n$, so that
\be
\lim_{n\to\infty}\delta(0,a,\textstyle{\frac{1}{n}}) &=& a.
\ee
The fact that delta-sequences do not have to be divergent at the origin is, perhaps, not widely known but in itself is not new (an example of an analogous  `filtering function', vanishing at the origin, can be found in \cite{Pol}).
In what follows we restrict our analysis to M-shaped delta-sequences normalized by $a=1/(2\pi)$,
\be
\delta_M(k,\textstyle{\frac{1}{n}})=\delta(k,\textstyle{\frac{1}{2\pi}},\textstyle{\frac{1}{n}}).
\ee
Let us note that, in spite of regularity at 0, one finds
\be
\lim_{n\to\infty}\int_{-\infty}^\infty \delta_M(k,\textstyle{\frac{1}{n}})\delta_M(k,\textstyle{\frac{1}{n}})dk
&=&\infty.
\ee
The square of Dirac's delta thus will not exist even if we define delta in terms of M-shaped delta-sequences. However,
\be
\lim_{m\to\infty}\lim_{n\to\infty}\int_{-\infty}^\infty \delta_M(k,\textstyle{\frac{1}{m}})\delta_M(k,\textstyle{\frac{1}{n}})dk
&=&
\lim_{m\to\infty}\delta_M(0,\textstyle{\frac{1}{m}})
=\frac{1}{2\pi}.
\ee

\subsection{Plane waves and M-shaped delta-sequences}

We are heading towards an analysis of plane waves in terms of deltas that are regular at 0. To do so, we need delta-sequences that can be represented as scalar products of square-integrable functions. The simplest strategy is to start with the convolution of two M-shaped delta-sequences,
\be
\delta_M^*(k,\textstyle{\frac{1}{n}},\textstyle{\frac{1}{m}})
&=&
\int_{-\infty}^\infty\delta_M(k-k',\textstyle{\frac{1}{n}})\delta_M(k',\textstyle{\frac{1}{m}})dk'
=
\delta_M^*(k,\textstyle{\frac{1}{m}},\textstyle{\frac{1}{n}}),\\
\lim_{m\to\infty}
\delta_M^*(k,\textstyle{\frac{1}{n},\frac{1}{m}})
&=&
\lim_{m\to\infty}
\int_{-\infty}^\infty\delta_M(k-k',\textstyle{\frac{1}{n}})\delta_M(k',\textstyle{\frac{1}{m}})dk'
=
\delta_M(k,\textstyle{\frac{1}{n}}).
\ee
The new sequence is again a delta-sequence,
\be
\int_{-\infty}^\infty\delta_M^*(k,\textstyle{\frac{1}{n}},\textstyle{\frac{1}{m}})dk
&=&
\int_{-\infty}^\infty\int_{-\infty}^\infty\delta_M(k-k',\textstyle{\frac{1}{n}})\delta_M(k',\textstyle{\frac{1}{m}})dk'dk
\nonumber\\
&=&\int_{-\infty}^\infty\delta_M(k',\textstyle{\frac{1}{m}})dk'
=1,
\ee
\be
{}&{}&
\lim_{n\to\infty}\lim_{m\to\infty}\int_{-\infty}^\infty f(k)\delta_M^*(k,\textstyle{\frac{1}{n}},\textstyle{\frac{1}{m}})dk
\nonumber\\
&{}&\pp{\lim_{n\to\infty}\lim_{m\to\infty}}=
\lim_{n\to\infty}\lim_{m\to\infty}\int_{-\infty}^\infty\int_{-\infty}^\infty f(k)\delta_M(k-k',\textstyle{\frac{1}{n}})\delta_M(k',\textstyle{\frac{1}{m}})dk'dk
\nonumber\\
&{}&\pp{\lim_{n\to\infty}\lim_{m\to\infty}}=
\lim_{n\to\infty}\int_{-\infty}^\infty f(k)\delta_M(k,\textstyle{\frac{1}{n}})dk
=
\displaystyle{\frac{f(0_-)+f(0_+)}{2},}
\ee
but is not exactly M-shaped in the sense of the previous subsection. Indeed,
\be
\delta_M^*(0,\textstyle{\frac{1}{n}},\textstyle{\frac{1}{m}})
&=&
\int_{-\infty}^\infty\delta_M(0-k',\textstyle{\frac{1}{n}})\delta_M(k',\textstyle{\frac{1}{m}})dk'
\nonumber\\
&=&
\int_{-\infty}^\infty\delta_M(k',\textstyle{\frac{1}{n}})\delta_M(k',\textstyle{\frac{1}{m}})dk'
\ee
in general depends on $n$ and $m$. The other properties are nevertheless analogous to M-shaped delta-sequences,
\be
\lim_{m\to\infty}
\delta_M^*(0,\textstyle{\frac{1}{n}},\textstyle{\frac{1}{m}})
&=&
\delta_M(0,\textstyle{\frac{1}{n}})=1/2\pi,\\
\lim_{n\to\infty}\lim_{m\to\infty}
\delta_M^*(0,\textstyle{\frac{1}{n}},\textstyle{\frac{1}{m}})
&=&
1/2\pi,
\ee
and
\be
\lim_{n\to\infty}
\delta_M^*(0,\textstyle{\frac{1}{n}},\textstyle{\frac{1}{n}})
&=&
\int_{-\infty}^\infty\delta_M(k',\textstyle{\frac{1}{n}})\delta_M(k',\textstyle{\frac{1}{n}})dk'=\infty
\nonumber.
\ee
Employing
\be
\hat\delta_M^*(x,\textstyle{\frac{1}{n}},\textstyle{\frac{1}{m}})
&=&
\frac{1}{2\pi}
\int_{-\infty}^\infty\delta_M^*(k,\textstyle{\frac{1}{n}},\textstyle{\frac{1}{m}})e^{ikx}dk
=
2\pi \hat\delta_M(x,\textstyle{\frac{1}{n}})\hat\delta_M(x,\textstyle{\frac{1}{m}})\label{FF}
\ee
we can write
\be
\delta_M^*(k-k',\textstyle{\frac{1}{n}},\textstyle{\frac{1}{m}})
&=&
\int_{-\infty}^\infty \hat\delta_M^*(x,\textstyle{\frac{1}{n}},\textstyle{\frac{1}{m}}) e^{-i(k-k')x}dx
\nonumber\\
&=&
2\pi\int_{-\infty}^\infty
\overline{\hat\delta_M(x,\textstyle{\frac{1}{n}})e^{ikx}}
\hat\delta_M(x,\textstyle{\frac{1}{m}})e^{ik'x}dx
\nonumber\\
&=&
\frac{1}{2\pi}\langle k,\textstyle{\frac{1}{n}}|k',\textstyle{\frac{1}{m}}\rangle=
\displaystyle{\frac{1}{2\pi}}\langle k,\textstyle{\frac{1}{m}}|k',\textstyle{\frac{1}{n}}\rangle,
\ee
since $\hat\delta_M(x,\textstyle{\frac{1}{n}})$ is real. Of course, $\langle k,\textstyle{\frac{1}{n}}|k',\textstyle{\frac{1}{m}}\rangle<\infty$ for all $k,k'$. Now let us recall that
\be
\lim_{n\to\infty}\hat\delta_M(x,\textstyle{\frac{1}{n}}) &=& \frac{1}{2\pi},
\ee
and thus, formally, it is justified to write
\be
\lim_{n\to\infty}\lim_{m\to\infty}\langle k,\textstyle{\frac{1}{n}}|k',\textstyle{\frac{1}{m}}\rangle
\textrm{ `$=$' }
\int_{-\infty}^\infty
e^{i(k'-k)x}dx,\label{'delta'}
\ee
while simultaneously we have shown that, for $k=k'$,
\be
\lim_{n\to\infty}\lim_{m\to\infty}\langle k,\textstyle{\frac{1}{n}}|k,\textstyle{\frac{1}{m}}\rangle
&=&
2\pi \lim_{n\to\infty}\lim_{m\to\infty}\delta_M^*(0,\textstyle{\frac{1}{n}},\textstyle{\frac{1}{m}})=1.
\ee
There is no contradiction between the above two formulas --- simply, integration does not commute with the limits $n,m\to\infty$ (integration must be performed first).

A similar situation occurs with derivatives. Denoting
\be
\langle x|k,\textstyle{\frac{1}{n}}\rangle
&=&
2\pi\hat\delta_M(x,\textstyle{\frac{1}{n}})e^{ikx}
\ee
and taking into account
\be
\lim_{n\to\infty}\frac{d^N}{dx^N}\hat\delta_M(x,\textstyle{\frac{1}{n}}) &=& 0, \quad N=1,2\dots,
\ee
we find
\be
\lim_{n\to\infty}\Big(\frac{1}{i}\frac{d}{dx}\Big)^N\langle x|k,\textstyle{\frac{1}{n}}\rangle
&=&
k^N\lim_{n\to\infty}\langle x|k,\textstyle{\frac{1}{n}}\rangle=
k^Ne^{ikx}.
\ee
Collecting all the formulas we have derived so far we are ready for generalization.

\subsection{Generalized function $\delta_M^*(k)$}

From the point of view of the sequential approach to distributions \cite{Sikorski} the sequences $\delta_\Lambda(k,\frac{1}{n})=\delta(k,4n,\frac{1}{n})$, $\delta_M(k,\frac{1}{n})=\delta(k,\frac{1}{2\pi},\frac{1}{n})$, and $\delta_M^*(k,\frac{1}{n},\frac{1}{m})$ belong to the same equivalence class and thus define the same distribution.
\begin{figure}
\includegraphics[width=8cm]{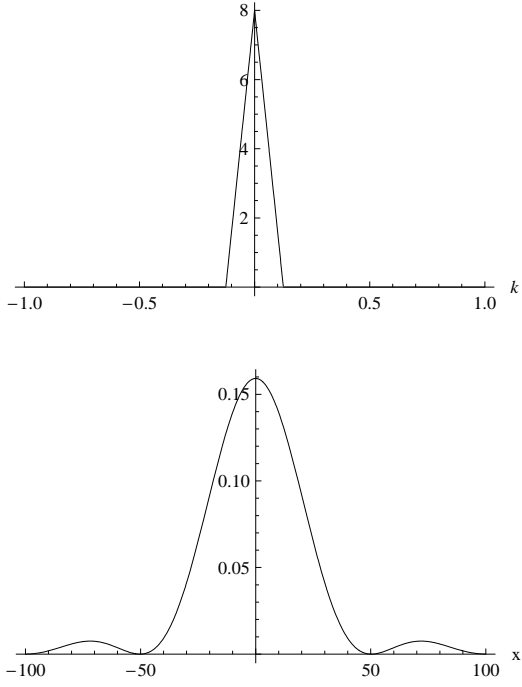}
\caption{The usual ($\Lambda$-shaped) delta-sequences are special cases of M-shaped delta-sequences --- to see this one puts $a=4/\epsilon$ and defines $\delta_\Lambda(k,\epsilon)=\delta(k,4/\epsilon,\epsilon)$. Here $\epsilon=1/2$, $a=8=4/\epsilon$ (upper), and its Fourier transform (lower).}
\end{figure}
In consequence, any of these delta-sequences can be treated as a representative of the equivalence class and employed in computations involving Dirac deltas.

However, we want to do more, and in particular want to include expressions such as Dirac delta at the origin. To do so,
we will treat the limits $\lim_{n\to\infty}\lim_{m\to\infty}\delta_M^*(k,\frac{1}{n},\frac{1}{m})$ as a new type of generalized function, $\delta_M^*(k)$.
The basic properties are as follows
\be
\textstyle{\int_{-\infty}^\infty} f(k)\delta_M^*(k)dk
&=&
\lim_{n\to\infty}\lim_{m\to\infty}\textstyle{\int_{-\infty}^\infty} f(k)\delta_M^*(k,\textstyle{\frac{1}{n}},\textstyle{\frac{1}{m}})dk
=
\displaystyle{\frac{f(0_-)+f(0_+)}{2}},\\
\textstyle{\int_{-\infty}^\infty} \delta_M^*(k)dk
&=&
1,\\
\textstyle{\int_{-\infty}^\infty} \delta_M^*(k)^2dk
&=&
\infty,\\
\delta_M^*(0)
&=&
\frac{1}{2\pi},\\
\hat\delta_M^*(x)
&=&
\frac{1}{2\pi}.
\ee
Plane waves defined by
\be
\langle x|k\rangle
&=&
\lim_{n\to\infty}\langle x|k,\textstyle{\frac{1}{n}}\rangle
=
e^{ikx}
\ee
satisfy
\be
\Big(\frac{1}{i}\frac{d}{dx}\Big)^N\langle x|k\rangle
&=&
\lim_{n\to\infty}\Big(\frac{1}{i}\frac{d}{dx}\Big)^N\langle x|k,\textstyle{\frac{1}{n}}\rangle\nonumber\\
&=&
k^N\langle x|k\rangle,\\
\langle k|k'\rangle
&=&
\lim_{n\to\infty}\lim_{m\to\infty}\langle k,\textstyle{\frac{1}{n}}|k',\textstyle{\frac{1}{m}}\rangle
\nonumber\\
&=& 2\pi \delta_M^*(k-k')\\
&=&
\lim_{n\to\infty}\lim_{m\to\infty}\textstyle{\int_{-\infty}^\infty}
\langle k,\textstyle{\frac{1}{n}}|x\rangle\langle x|k',\textstyle{\frac{1}{m}}\rangle dx,\nonumber\\
&\pp=&
\\
\langle k|k\rangle
&=&
\lim_{n\to\infty}\lim_{m\to\infty}\langle k,\textstyle{\frac{1}{n}}|k,\textstyle{\frac{1}{m}}\rangle
=
2\pi \delta_M^*(0)=1
\ee
These rules allow for all the standard computations involving the Dirac delta, but one can perform also certain new operations. For example, consider the expression
\be
\int\langle x|k\rangle\langle k|k'\rangle\langle k'|y\rangle f(k)dk,
\ee
where $f(k)$ is, say, square-integrable, and which should be understood in the following sense
\be
&{}&
\lim_{n_1\to\infty}\lim_{n_2\to\infty}\lim_{n_3\to\infty}\lim_{n_4\to\infty}
\int\langle x|k,\textstyle{\frac{1}{n_1}}\rangle\langle k,\textstyle{\frac{1}{n_2}}|k',\textstyle{\frac{1}{n_3}}\rangle\langle k',\textstyle{\frac{1}{n_4}}|y\rangle f(k)dk
\nonumber\\
&{}&\pp{\lim_{n_1\to\infty}\lim_{n_2\to\infty}\lim_{n_3\to\infty}\lim_{n_4\to\infty}}
=
\lim_{n_2\to\infty}\lim_{n_3\to\infty}
\int\langle x|k\rangle 2\pi\delta_M^*(k-k',\textstyle{\frac{1}{n_2},\frac{1}{n_3}})\langle k'|y\rangle f(k)dk
\nonumber\\
&{}&\pp{\lim_{n_1\to\infty}\lim_{n_2\to\infty}\lim_{n_3\to\infty}\lim_{n_4\to\infty}}
=\lim_{n_2\to\infty}
\int\langle x|k\rangle2\pi\delta_M(k-k',\textstyle{\frac{1}{n_2}})\langle k'|y\rangle f(k)dk
\nonumber\\
&{}&\pp{\lim_{n_1\to\infty}\lim_{n_2\to\infty}\lim_{n_3\to\infty}\lim_{n_4\to\infty}}
=2\pi\langle x|k'\rangle\langle k'|y\rangle f(k')
\nonumber\\
&{}&\pp{\lim_{n_1\to\infty}\lim_{n_2\to\infty}\lim_{n_3\to\infty}\lim_{n_4\to\infty}}
=2\pi\int\langle x|k\rangle\delta(k-k')\langle k'|y\rangle f(k)dk
\ee
where the last formula was written in terms of the `standard' Dirac delta, to make it more familiar, but we could replace $\delta(k)$ by $\delta_M^*(k)$.
Simultaneously,
\be
{}&{}&
\lim_{n_1\to\infty}\lim_{n_2\to\infty}\lim_{n_3\to\infty}\lim_{n_4\to\infty}
\int\langle x|k,\textstyle{\frac{1}{n_1}}\rangle\langle k,\textstyle{\frac{1}{n_2}}|k,\textstyle{\frac{1}{n_3}}\rangle\langle k,\textstyle{\frac{1}{n_4}}|y\rangle f(k)dk
\nonumber\\
&{}&\pp{\lim_{n_1\to\infty}\lim_{n_2\to\infty}\lim_{n_3\to\infty}\lim_{n_4\to\infty}}
=
\int\langle x|k\rangle\langle k|y\rangle f(k)dk
\ee
The formulas reduce to
\be
|k\rangle\langle k|k'\rangle\langle k'|
&=&
2\pi \delta(k-k')|k\rangle\langle k|,\label{11}
\ee
and
\be
|k\rangle\langle k|k\rangle\langle k|
=
|k\rangle\langle k|,\label{22}
\ee
that are mutually consistent if one treats each $|k\rangle$ as a limit of a separate sequence $|k,\frac{1}{n}\rangle$. Eqs. (\ref{11})--(\ref{22})
can be conveniently written as
\be
|k\rangle\langle k|k'\rangle\langle k'|
&=&
2\pi \delta_M^*(k-k')|k\rangle\langle k|.
\ee
\subsection{M-shaped deltas with respect to more general measures}

In this section we will concentrate on a generalization that is useful from the point of view of relativistic applications.
Let us treat explicitly only a one-dimensional case. Let us assume that instead of $dp$ we have to use a measure $d\mu(p)=\rho(p)dp$, and an appropriate delta is needed,
\be
\int d\mu(p')\delta_{\mu M}(p,p')f(p')=f(p),
\ee
with $\delta_{\mu M}(p,p)=a$, say, where $a$ is a constant.
The standard solution,
\be
\delta_\mu(p,p')=\rho(p')^{-1}\delta(p-p'),\label{mu}
\ee
if generalized to M-shaped deltas by
\be
\delta_{\mu M}(p,p')=\rho(p')^{-1}\delta_M(p-p'),\label{muM}
\ee
implies $\delta_{\mu M}(p,p)=\rho(p)^{-1}\delta_M(0)$ and will not lead to $a$ independent of $p$. It follows that we have to proceed in a way different from (\ref{muM}).
So, let $\delta(p,a,\frac{1}{n})$, $\delta(0,a,\frac{1}{n})=a$, be an arbitrary M-shaped delta-sequence discussed in the preceding sections. The sequence
\be
\delta_\mu(p,p',\textstyle{\frac{1}{n}})
&=&
\rho(p')^{-1}\delta\big(p-p',a\rho(p),\textstyle{\frac{1}{n}}\big),
\ee
$\rho(p)=d\mu(p)/dp$, has the following properties
\be
\lim_{n\to \infty}
\int d\mu(p')\delta_\mu(p,p',\textstyle{\frac{1}{n}})f(p')
&=&
\frac{f(p_-)+f(p_+)}{2},\\
\delta_\mu(p,p,\textstyle{\frac{1}{n}})
&=&
a
\ee
and defines distribution $\delta_{\mu M}(p,p')$ satisfying
\be
\int d\mu(p')\delta_{\mu M}(p,p')f(p')
&=&
\frac{f(p_-)+f(p_+)}{2},\\
\delta_{\mu M}(p,p)
&=&
a.
\ee
\subsection{The meaning of products of operators occurring in harmonic-oscillator Lie algebra}
\label{subsec 75}

Consider $N=1$ reducible representation of harmonic-oscillator Lie algebra.
Define
\be
a_{\bm b}(\bm R,\bm k,1,\textstyle{\frac{1}{n_1},\frac{1}{n_2}})=|\bm R,\textstyle{\frac{1}{n_1}}\rangle
\langle\bm R,\textstyle{\frac{1}{n_1}}|
\otimes
|\bm k,\textstyle{\frac{1}{n_2}}\rangle\langle\bm k,\textstyle{\frac{1}{n_2}}|\otimes a_{\bm b}.\label{XXXX}
\ee
The kets and bras are understood in the sense described in this Appendix. The products such as, say, $a_{\bm b}(\bm R,\bm k,1)a_{\bm c}(\bm R',\bm k',1)^{\dag}$ are understood in the sense of $M$-shaped Dirac deltas:
\be
&{}&a_{\bm b}(\bm R,\bm k,1)a_{\bm c}(\bm R',\bm k',1)^{\dag}
\nonumber\\
&{}&
\pp{XX}=
\lim_{n_1\to\infty}\lim_{n_2\to\infty}\lim_{n'_1\to\infty}\lim_{n'_2\to\infty}
a_{\bm b}(\bm R,\bm k,1,\textstyle{\frac{1}{n_1},\frac{1}{n_2}})a_{\bm c}(\bm R',\bm k',1,\textstyle{\frac{1}{n'_1},\frac{1}{n'_2}})^{\dag}.
\ee
The order of limits is irrelevant.

\section{Noetherian construction of generators of $N=1$ reducible representation --- example of 4-momentum}

Consider any two fields $A(x)$, $B(x)$ satisfying d'Alembert equation. For example, let $A(x)=A_b(x,1,\textstyle{\frac{1}{n_1},\frac{1}{n_2}})$, $B(x)=A_c(x,1,\textstyle{\frac{1}{n'_1},\frac{1}{n'_2}})$, where the dependence on sequences means that the corresponding amplitude operators are constructed according to the recipe (\ref{XXXX}). d'Alembert equation implies conservation of the energy-momentum tensor
\be
T_{ab}(x,\textstyle{\frac{1}{n_1},\frac{1}{n_2},\frac{1}{n'_1},\frac{1}{n'_2}})
&=&
-\frac{1}{2} \Big(\partial_a A(x)\partial_b B(x)+\partial_b A(x)\partial_a B(x)-g_{ab}\partial_c A(x)\partial^c B(x)\Big),\nonumber\\
\ee
i.e. $\partial^aT_{ab}(x,\textstyle{\frac{1}{n_1},\frac{1}{n_2},\frac{1}{n'_1},\frac{1}{n'_2}})=0$,  and
\be
P_{a}(\textstyle{\frac{1}{n_1},\frac{1}{n_2},\frac{1}{n'_1},\frac{1}{n'_2}})
=
\displaystyle{\int d^3x T_{a0}(x_0,\bm x,\textstyle{\frac{1}{n_1},\frac{1}{n_2},\frac{1}{n'_1},\frac{1}{n'_2}})}
\ee
is independent of $x_0$ for all $n_1,\dots,n'_2$. The limit
\be
P_a
&=&
\lim_{n_1\to\infty}\lim_{n_2\to\infty}\lim_{n'_1\to\infty}\lim_{n'_2\to\infty}
P_{a}(\textstyle{\frac{1}{n_1},\frac{1}{n_2},\frac{1}{n'_1},\frac{1}{n'_2}})
\nonumber\\
&=&
-I\otimes {\int} d k{\,}k_a|\bm k\rangle\langle \bm k|\otimes
\frac{1}{2}\Big(a^{\dag}_{\bf a}a^{\bf a}+a^{\bf a}a^{\dag}_{\bf a}\Big)
=P_a(1)+K_a
\ee
is the 4-momentum, vacuum contribution included, corresponding to $N=1$ reducible representation of the harmonic oscillator Lie algebra.

\end{document}